\documentclass[aps,showpacs,pra,onecolumn]{revtex4-2}

\usepackage{hyperref}
\hypersetup{
    colorlinks=true,
    linkcolor=black,
    filecolor=black,      
    urlcolor=purple,
    citecolor=purple,
}

\usepackage{exscale}
\usepackage{graphicx}
\usepackage{amsmath}
\usepackage{latexsym}
\usepackage{epstopdf}
\usepackage{amsfonts}
\usepackage{amssymb}
\usepackage{bbm}
\usepackage{times}
\usepackage[T1]{fontenc}
\usepackage{lipsum}
\usepackage{amsthm}
\usepackage{fancyhdr,txfonts,bbm}
\usepackage{appendix}
\usepackage{soul}
\usepackage[table]{xcolor}
\usepackage{epsfig,pstricks}
\usepackage{xspace}

\usepackage{epsfig,pstricks}
\usepackage{changes}

\newtheorem{definition}{Definition}

\newcommand{{\Cd}}{{\mathbb{C}^d}}
\newcommand{{\C}}{{\mathbb{C}}}

\newcommand{\MPBS}{\textsc{MPBS}\@\xspace}
\newcommand{\algo}[1]{\textsc{#1}\@\xspace}

\newcommand{\rcv}{\cellcolor{red!17}}
\newcommand{\rcd}{\cellcolor{red!22}}
\newcommand{\bcv}{\cellcolor{blue!17}}
\newcommand{\bcd}{\cellcolor{blue!22}}

\begin{document}

\title{Optimized QUBO formulation methods for quantum computing}





\author{Dario De Santis$^{1,}$}\email{dario.desantis@sns.it}\author{Salvatore Tirone$^{1,2,3}$, Stefano Marmi$^1$, Vittorio Giovannetti$^4$}
\affiliation{Scuola Normale Superiore, Piazza dei Cavalieri 7, I-56126 Pisa, Italy}
\affiliation{$^2$QuSoft, Science Park 123, 1098 XG Amsterdam, the Netherlands}
\affiliation{$^3$Korteweg--de Vries Institute for Mathematics, University of Amsterdam, Science Park 105-107, 1098 XG Amsterdam, the Netherlands}
\affiliation{$^4$NEST, Scuola Normale Superiore and Istituto Nanoscienze-CNR, I-56126 Pisa, Italy}


\begin{abstract}

Quantum computers have strict requirements for the problems that they can efficiently solve. One of the principal limiting factor for the performances of noisy intermediate-scale quantum (NISQ) devices is the number of qubits required by the running algorithm.
Several combinatorial optimization problems can be solved with NISQ devices once that a corresponding quadratic unconstrained binary optimization (QUBO) form is derived. Several techniques have been proposed to achieve such reformulations and, depending on the method chosen, the number of binary variables required, and therefore of qubits, can vary considerably.
The aim of this work is to drastically reduce the variables needed for these QUBO reformulations in order to unlock the possibility to efficiently obtain optimal solutions for a class of optimization problems with NISQ devices.
This goal is achieved by introducing novel tools that allow an efficient use of slack variables, even for problems with non-linear constraints, without the need to approximate the starting problem.
We divide our new techniques in two independent parts, called the iterative quadratic polynomial and the master-satellite methods.
Hence, we show how to apply our techniques in case of an NP-hard optimization problem inspired by a real-world financial scenario {called \algo{Max-Profit Balance Settlement} (MPBS)}.
We follow by submitting several instances of this problem to two D-wave quantum annealers,
comparing the performances of our novel approach with the standard methods used in these scenarios.
Moreover, this study allows to appreciate several performance differences between the D-wave Advantage and next-generation Advantage2 quantum annealers.
{We show that the adoption of our techniques in this context allows to obtain QUBO formulations with significantly fewer slack variables, i.e., around 90\% less, and D-wave annealers provide considerably higher correct solution rates, which moreover do not decrease with the input size as fast as when adopting standard techniques.}

\end{abstract}

\maketitle

\section{Introduction}

In recent years, many efforts have been devoted to understand to what extent modern noisy intermediate-scale quantum (NISQ) devices can help to solve optimization problems of any sort (see Refs.~\cite{Hauke,Yarkoni,Abbas} for recent reviews). 
Particular attention has been paid to their potential to solve NP-hard and NP-complete problems, which most of the times can be formulated as quadratic unconstrained binary optimizations (QUBO). For instance, all famous NP Karp's problems~\cite{Karp} can be casted in this form~\cite{Lucas}. 

The possibility to solve QUBO problems with quantum devices has been studied on several supports, especially on quantum annealers~\cite{Yarkoni,Abbas,Rajak,Hauke,Das, Farhi} but also on logical-gate quantum computers~\cite{farhi2014quantum,Khumalo,Chieza,QAOA} and Rydberg atom arrays~\cite{Ebadi,Nguyen,Cain}. 
Although these technologies are not expected to solve NP problems efficiently~\cite{Altshuler,Dickson}, namely providing exponential speed-ups compared to classical strategies, possible polynomial in-time advantages attract great interest.
Indeed, given the importance of the real-world applications involved with the solutions of these hard combinatorial problems~\cite{Yarkoni}, any achievable improvement is well-received. 
Exemplary scenarios come from finance~\cite{SamMugelrevORUS,SamMugel2,Venturelli,Herman,Buonaiuto,Egger_2020}, logistics~\cite{Weinberg,Sales,Hernandez}  and drug discovery~\cite{Zinner}.

Consider the situation where we aim to solve a generic constrained optimization problem with a QUBO solver. 
The first step is to reformulate it in a quadratic unconstrained form, namely as a QUBO. 
We call \textit{logical variables} those defining the initial problem. 
The constraints attached to the original problem can be enforced in a QUBO form by employing additional \textit{slack variables}.
The larger is the total number of variables, logical and slack, in a QUBO problem, the harder is in general to solve it. 

Whereas the well-established procedures to translate optimization problems into QUBOs~\cite{Glover} can be efficient in several scenarios, their implementation can be extremely inefficient at times.
To be more precise, we say that a method to obtain these formulations is inefficient for a problem whenever it requires too many slack variables, namely their number is comparable with that of logical variables.
In these cases, the practical usefulness of QUBO solvers can be highly limited. 
For instance, if we have a maximum size for the problems that our solver can receive, any reduction of slack variables allows to increase the total number of logical variables associated to the initial problem, namely larger problems can be tackled. Moreover, the more slack variables we implement, the higher is the connectivity that we require in our QUBO and therefore in the QUBO solver that we aim to use. This consequence can be a very limiting factor for the performances of modern NISQ devices.

The main motivation of this work is to unlock the possibility to solve certain classes of optimization problems with those quantum technologies specialized to solve QUBO problems.
For this purpose, we introduce the iterative quadratic polynomial and master-satellite methods.
Their goal is to provide QUBO forms requiring a minimal employment of slack variables.
The former consists in a new paradigm to translate problem constraints into corresponding quadratic penalty forms, while the latter allows the simultaneous enforcement of different constraints sharing the same restricted set of binary variables.
Noteworthy, our techniques can also be used to enforce non-linear equality and inequality constraints defined through integer or non-integer parameters, without the need to approximate the constraints or to employ slack variables solely to obtain corresponding linearizations. Hence, the same computational effort is required both for linear and non-linear constraints, where no approximate enforcing of the constraints is required. 
{Moreover, the QUBO variables employed by our methods for each constraint are not necessarily completely connected.
These features allow improving the output quality of NISQ devices used as QUBO solvers as fewer and less-connected qubits are in general required.}

We apply these techniques in the context of an NP-hard problem coming from finance, namely the \algo{Max-Profit Balance Settlement} (MPBS) problem~\cite{power}. We provide a drastic reduction of the slack variables necessary to enforce the corresponding inequality constraints. {We generate several instances  of this problem reflecting some main features of realistic datasets~\cite{power2020} and we show that our methods provide the corresponding QUBO forms by employing around 90\% less slack variables, if compared with the QUBO forms obtainable with a standard procedure~\cite{Glover}.}

We solve several instances of \algo{MPBS} with two different quantum annealers manufactured by D-Wave Systems, Inc.~\cite{Dwavereport}, namely the {D-wave} Advantage\_system4.1 and Advantage2\_prototype1.1. 
The annealer Advantage2\_prototype1.1, being a prototype of the next-generation D-wave annealers, has a reduced number of qubits but is characterized by higher connectivity and lower noise. 
Hence, not only do we compare the outputs obtained when the QUBO problems are generated with different approaches, but we also provide a comparison between the performances of these annealers. We show that, while by using the standard approach the solution quality drops quite fast with the input size, our methods unlock the possibility to consider much larger instances. {The number of successes that we obtain when the QUBO forms are generated with our methods are always significantly higher: the multiplicative factor between the successes obtainable with the two methods ranges from $7$ (smallest instances with Advantage\_system4.1) to $184$ (largest instances with Advantage2\_prototype1.1).}

{We point out the existence of other techniques attempting to reduce the slack variables employed for QUBO reformulations. For instance, the alternating direction method of multipliers~\cite{Boyd} and the augmented Lagrangian method~\cite{Hestenes,Powell} have recently been adopted 
in this context~\cite{Djidjev,otherADMM,Light}, {where preliminary QUBO problems have to be solved in order to get the final QUBO formulation of the original optimization problem.}
Another approach consists in enforcing constraints in an approximate fashion via a finely-tuned rescaling of the parameters defining the target linear inequalities~\cite{arxiv1st,solar,powernetwork}. 
{Differently from these strategies, neither we require to solve preliminary QUBOs nor we approximate the starting optimization problem. Rather, we change the paradigm under which constraints are enforced.}

The content of this work is structured as follows. In Section~\ref{MainTools}, we introduce the fundamental tools needed throughout this work, as the definition of graph-based problems, QUBO forms and the standard technique to achieve these quadratic forms. The first part of Section~\ref{SecIQPMS} is devoted to describe in an intuitive way the main ideas behind of our novel methods. Instead, the iterative quadratic polynomial and the master-satellite methods are respectively described in details in Sections~\ref{GQPmaster} and \ref{GQPsatellite}.
The latter method requires a tuning of the QUBO penalty multipliers that is radically different from those found in standard QUBO formulations. Different strategies for this tuning are presented in Section~\ref{tunepenalty}.
The last heuristic that we introduce is a generalization of the master-satellite approach, suitable for those cases where three or more constraints are defined over the same variables, which is proposed in Section~\ref{concatenation}. 
We identify a class of optimization problems, namely the locally-constrained problems, in Section~\ref{SECLOC}, which are particularly suitable for the application of our techniques.
The definition of the NP-hard \algo{Max-Profit Balance Settlement} (MPBS) problem is given in Section \ref{SecMPBSUC}, where we also analyse its QUBO formulation and the corresponding needs in terms of slack variables when using standard techniques. Instead,  we show how to apply  our novel techniques  to MPBS step-by-step in Section~\ref{IQPMSMPBSsec}. 
The corresponding improved performance obtainable in the context of quantum annealing is shown in Section~\ref{DwaveMPBS}, where the novel and the standard strategies are put in comparison.

\section{Main tools}\label{MainTools}
 QUBO problems are defined as optimizations of quadratic forms defined over a set of binary variables $\mathbf x = \{0,1\}^N$, without the possibility to attach any constraint~\cite{Glover}. Hence, we have the following equivalent notations:
\begin{eqnarray}
\mathop{\mbox{min/max}}\limits_{\substack{\mathbf x}}  \,\mathbf x^T Q \mathbf x = \mathop{\mbox{min/max}}\limits_{\substack{\mathbf x}}  \sum_{ i,j=1}^N Q_{ij} x_i x_j = \mathop{\mbox{min/max}}\limits_{\substack{\mathbf x}}  Q(\mathbf x) \, ,
\end{eqnarray}
where $Q$ is an upper-triangular $N\times N$ real-valued matrix and the corresponding optimization can either be a minimization or a maximization. Although these two cases are computationally equivalent, we consider the maximization case from now on.

Quantum annealers are the QUBO solvers that we adopt later in this work to solve these optimization problems, which in general are NP-hard~\cite{Lucas}. 
The explanation of how quantum annealers work in order to obtain optimal solutions of QUBO problems can be found, e.g., in Refs.~\cite{Hauke,Yarkoni,Abbas}.
{We point out that we do not consider other heuristics from classical computation to solve these problems. An example of such techniques is called simulated annealing~\cite{simannealing0,simannealing1,simannealing2}, the classical counterpart of quantum annealing, which searches into the space of possible solutions, while slowly decreasing the probability to accept worse alternatives of the current solution. A comparison between quantum and simulated annealing can be found in Ref.~\cite{qvscanneal}. A different heuristic designed for the solution of QUBO problems is called tabu search, which is put in contrast with simulated annealing in Ref.~\cite{tabusa}.}

We proceed by explaining how several optimizations can be translated into a QUBO problem and where does the necessity to introduce new methods for this translation come from.
Consider an optimization problem defined through the maximization of an objective function, which from now on we call~$W$, where some constraints on the possible solutions must be satisfied. 
Usually, from the very definition of the objective function, we obtain a natural set of {logical variables}~$\mathbf x=(x_1,x_2,\dots)$ associated to the units that describe the problem itself. 
For instance, if the problem requires to find a particular sub-graph of an input graph, the binary variables may be associated to its arcs and/or nodes, or, if we aim to subdivide a particular quantity among different parties, the binary variables may be used to describe the corresponding fractions. 
Whenever  $W(\mathbf x)$ can be written as a linear or quadratic form of the variables $(x_1,\,x_2,\,x_3,\dots)$, its QUBO form is obtained straightforwardly.
Nonetheless, in order to translate the constraints attached to our optimization problem in a QUBO form, we usually need additional binary variables~$\mathbf s=(s_1,s_2,\dots)$, called {slack variables}, which are employed solely to enforce constraints~\cite{Glover}.

This work is devoted to expose novel techniques to implement constraints in QUBOs.
We point out that, even if some constraints could be particularly demanding in terms of slack variables, others may be not. Hence, whenever we use the expression ``target constraints'', we refer to those on which we decide to apply our novel tools.

We study the generic situation of problems defined on directed, weighted and node-attributed multigraphs, namely:
\begin{definition}[{\sffamily{R}}-multigraph]\label{RMULTI}
An {\sffamily{R}}-multigraph consists of the quadruple $\mathcal{G}=(\mathcal V,\mathcal E, w,\mathbf f)$, where $\mathcal{V}$ is a set of nodes, $\mathcal{E}$ is a multiset of ordered pairs of nodes called arcs, $w:\mathcal E\rightarrow \mathbb{R}^+$ is a function assigning positive real numbers, called weights, to arcs and $\mathbf  f:\mathcal{V} \rightarrow \mathbb{R}^n$ is a function assigning real-valued $n$-dimensional vectors, called attributes, to nodes. 
\end{definition}
Hence, we allow multiple arcs $(u,v)\in \mathcal{E}$ between two nodes $u,v\in \mathcal V$, where each arc $(u,v)\in \mathcal{E}$ has an origin $u$, a target $v$ node and a positive weight $w(u,v)$. Moreover, each node $u\in \mathcal V$ may have attributes given by a real-valued $n$-dimensional vector $\mathbf{f}(u)=(f_1(u),\dots,f_n(u))$. Naturally, what follows can be applied to simpler graphs, e.g., undirected and/or unweighted. 

The majority of the optimization problems considered in the literature can be casted as follows: find the sub-graph $\mathcal{G}^*= ( \mathcal V^*,\mathcal E^*,w,\mathbf f)$ of a given {\sffamily{R}}-multigraph  $\mathcal{G}=(\mathcal V,\mathcal E, w,\mathbf f)$ that maximizes the quantity $W(\hat{\mathcal{G}}) $ under some given constraints, where $\hat{\mathcal{G}}$ is a generic sub-graph of $\mathcal{G}$.  In order to write these problems in a more compact form, we introduce a binary variable $x_i=\{0,1\}$ for each arc $(u,v)\in\mathcal{E}$. Hence, we obtain a bijection $(u,v)\,\leftrightarrow\, x_i$, where $i=1,2,\dots,N$ and $N=|\mathcal E|$.
Each sub-graph $\hat{\mathcal{E}}\subseteq \mathcal E$ is identified by a single combination of $\mathbf x = \{x_1,x_2,\dots,x_N\}\in\{0,1\}^N$, where $x_i=1$ if the corresponding arc is included in $\hat{\mathcal{E}}$ and $x_i=0$ otherwise. Given this bijection, each combination of $\mathbf x$ defines a different $\hat{\mathcal{E}}\subseteq \mathcal{E}$ and $\hat{\mathcal{V}}$, where $u\in \hat{\mathcal V}$ if there exists at least one arc in $\hat{\mathcal E}$ such that $u$ is either the origin or the target node. For instance, $ \mathcal E\,\leftrightarrow \,\mathbf x = \{1,1,\dots,1\}$.

We start by considering the class of optimization problems that have objective functions and constraints that are, respectively, (at most) quadratic and linear in the binary variables $x_i$. In case of $m_{EC}\geq 0$ equality constraints $EC_i(\mathbf x, w,\mathbf f) = 0$ and $m_{IC}\geq 0$ inequality constraints $IC_j(\mathbf x, w,\mathbf f)\leq 0$, an optimization problem of this kind can be written as:
\begin{equation}\label{optprob}
\begin{array}{cccc}
\mathcal E^* = \arg & \max_{\mathbf x} & W(\mathbf x, w,\mathbf f)  \\ 
    & \mbox{ s.t. } & EC_{i}(\mathbf x, w,\mathbf f) = 0   & \mbox{ for } i=1,\dots, m_{EC} \\ 
   &  & IC_j(\mathbf x, w,\mathbf f) \leq 0 &  \!\mbox{ for } j=1,\dots, m_{IC}
   \end{array} \, .
\end{equation}
Moreover, we assume the parameters defining the linear inequality constraints to be integer (positive or negative), so that $IC_j(\mathbf x)$ can only assume integer values. 
The linearity assumption of all the constraints and the requirement that the inequality ones are integer valued are necessary in order to introduce the standard method to obtain QUBO reformulations, as we show in  Section~\ref{QUBOintro}. Later, in Section~\ref{SecIQPMS}, we introduce novel methods that do not require the starting optimization problem to satisfy none of these assumptions: non-linear and non-integer constraints can be enforced without additional effort with respect to the linear integer-valued constraint cases.

\subsection{Standard QUBO formulation}\label{QUBOintro} 
The goal of a QUBO reformulation of the optimization problem (\ref{optprob}) is to provide a quadratic unconstrained binary optimization problem
\begin{equation}\label{QUBOprob00}
\mathcal E^* = \arg  \max_{\mathbf x,\mathbf s}  W(\mathbf x) + \lambda \left(\sum_{i=1}^{m_{EC}} P_i^{EC}(\mathbf x) + \sum_{j=1}^{m_{IC}} P_j^{IC}(\mathbf x,\mathbf s) \right) \, ,
\end{equation}
that provides the same optimal configurations provided by Eq. (\ref{optprob}), where $\mathbf s = (s_1,s_2,\dots, s_{\overline N})$ are $\overline N$ additional slack variables that are employed to enforce inequality constraints in a quadratic form.
For the sake of clarity, we dropped the notation that makes explicit the dependence of $W$, $EC_i$ and $IC_j$ from $w$ and $\mathbf f$. 
As we see later, linear equality constraints do not need any slack variables to be enforced in this form. In order for this reformulation to have the meaning that we required, we have to find quadratic forms $P$ that provide penalties, namely are smaller or equal than -1, whenever the corresponding equality or inequality constraint are violated by $\mathbf x$, otherwise, if satisfied, they have to be zero valued for at least one value of $\mathbf s$. For instance,  $P^{EC}_1(\mathbf x) \leq -1$ if and only if $\mathbf x$ violates the first equality constraint and $P^{EC}_1 (\mathbf x) =0$ if and only if $\mathbf x$ satisfies the same constraint. Here, $\lambda>0$ is a multiplier that has to be chosen large enough in order to make sure that all the combinations of $\mathbf x$ violating one or more constraints cannot be the optimal configurations selected by Eq.~(\ref{QUBOprob00}).  

The standard approach to enforce constraints in a quadratic form~\cite{Glover} reads as follows:
\begin{equation}\label{QUBOprob0}
\mathcal E^* = \arg  \max_{\mathbf x,\mathbf s}  W(\mathbf x) + \lambda \left(- \sum_{i=1}^{m_{EC}} EC^2_i(\mathbf x) - \sum_{j=1}^{m_{IC}} \left(IC_j(\mathbf x) + S_j(\mathbf s) \right)^2 \right) \, ,
\end{equation}
where $S_j(\mathbf s)$ are linear functions of the binary slack variables $\mathbf s = \{s_1, s_2,\dots, s_{\overline N}\}$.
{ The quadratic term that enforces the equality constraint $EC_i(\mathbf x)$ in the QUBO form is simply given by $ P^{EC}_i (\mathbf x) = - EC_i^2(\mathbf x)$. The reason of this choice is straightforward: any $\mathbf x$ that violates $EC_i(\mathbf x)$ provides a negative contribution. Therefore, as this optimization is formulated as a maximization, those $\mathbf x$ leading to a violation of an equality constraint are discouraged. Hence, if $\lambda > 0$ is set large enough, we are sure that the solution of Eq.~(\ref{QUBOprob0}) satisfies all the equality constraints. 

For what concerns the inequality constraints, we build some $S_j(\mathbf s)$ that help to translate these constraints into equality ones~\cite{Glover}. Before explaining how this goal is achieved, we notice that, if the inequality constraints are satisfied for at least one combination of $\mathbf x$, namely if there exists a solution of Eq.~(\ref{optprob}), then $\min_{\mathbf x} IC_j(\mathbf x) \leq 0$. Hence, if $S_j(\mathbf s)$ assumes all the values in the interval $[0,|\min_{\mathbf x} IC_j(\mathbf x)|]$, checking whether $\mathbf x$ satisfies the inequality constraint $IC_j(\mathbf x)\leq 0$ is equivalent to check whether $\mathbf x$ satisfies the equality constraint $IC_j(\mathbf x) + S_j(\mathbf s) = 0$ for at least one value of $\mathbf s$. Hence, the enforcement of this new equality constraint in the QUBO form can be made similarly to $EC_i(\mathbf x)$ by considering the quadratic term $ P^{IC}_j (\mathbf x,\mathbf s) = - (IC_j(\mathbf x) + S_j(\mathbf s))^2$. } Finally, in order for $S_j(\mathbf s)$ to have the properties described above, we use the definition:
\begin{eqnarray}\label{S}
S_j(\mathbf s) =  s_{\overline N_j} \overline S_j +  \sum_{i=1}^{\overline N_j -1} s_i 2^{i-1} \, ,
\end{eqnarray}
where $\overline N_j > 0$ and $\overline S_j> 0$ are such that $S_j(\mathbf s)$ assumes all the integer values in the interval $[0,|\min_{\mathbf x} IC_j(\mathbf x)|]$. Hence, $\overline N_j>0$ is the number of slack variables needed by this method to enforce $IC_j(\mathbf x)$ in a QUBO form, namely Eq.~(\ref{QUBOprob0}).
Notice that in Eq.~(\ref{S}) we implied a relabelling of the slack variable indices such that those associated to $IC_j(\mathbf x)$ are the first $\overline N_j >0$, where $\overline N=\sum_{j} \overline N_j$.
It is possible to apply similar techniques also in the case of non-integer-valued inequality constraints, but in general their enforcement requires more slack variables.
 
In summary, each term of the first and second sum of Eq.~(\ref{QUBOprob0}) is respectively associated to an equality or inequality constraint which provides a negative contribution, or penalty, if and only if  $\mathbf x$ violates the corresponding condition. More precisely, if $\mathbf x$ violates a constraint, the contribution of the corresponding constraint term is strictly negative for all $\mathbf s$.
On the other side, the maximizations over the same terms are zero-valued if and only if $\mathbf x$ satisfies the corresponding conditions.
Hence, a large-enough value of $\lambda>0$ guarantees that the optimal solution does not violate any constraint and therefore Eqs.~(\ref{optprob}) and~(\ref{QUBOprob0}) have the same solution.
Whereas a straightforward choice of this parameter is $\lambda>\max_{\mathbf x } W(\mathbf x )$, we provide other techniques later in this work for this tuning.

We underline that, in order to achieve this QUBO formulation, we have to make a disjoint use of slack variables, namely each slack variable appearing in $S_j(\mathbf s)$, does not appear in any other $S_{k}(\mathbf s)$ for $j\neq k$.
Moreover, the total number of slack variables $\overline N_j$ used to enforce $IC_j(\mathbf x)$ solely depends on $\min_{\mathbf x} IC_j(\mathbf x)$. As a result, the number of slack variables deployed is:
\begin{eqnarray}\label{overlineN}
\overline N  = \sum_{j=1}^{m_{IC}}\overline N_j = \sum_{j=1}^{m_{IC}} \left\lceil \log_2 \left(1 + \left|\min_{\mathbf x} IC_j(\mathbf x)\right| \right) \right\rceil \, .
\end{eqnarray}

\section{Iterative quadratic polynomial and master-satellite methods}\label{SecIQPMS}

The main goal of this work is to introduce a novel set of tools to obtain QUBO formulations of constrained binary optimization problems with the least possible amount of slack variables. This can help to reduce the number of qubits required to solve these problems on NISQ devices, potentially unlocking the possibility to tackle larger and more complex optimizations.
Our strategy consists of two partially independent methods: the iterative quadratic polynomial (IQP) method and the master-satellite (MS) method. While here we provide an intuitive explanation of these techniques, a more detailed description can be found in the next sections.

The IQP method provides a new way to enforce constraints in a QUBO form. This method is particularly well-suited for constraints defined on few binary variables.
Moreover, it can be easily generalized to enforce non-linear equality and inequality constraints into a QUBO form defined with integer or non-integer parameters. 
We summarize the main idea behind this method in Figure~\ref{IQPfig}, while its technical explanation is presented in Section \ref{GQPmaster}.

\begin{figure}
\begin{center}
\includegraphics[width=0.99\textwidth]{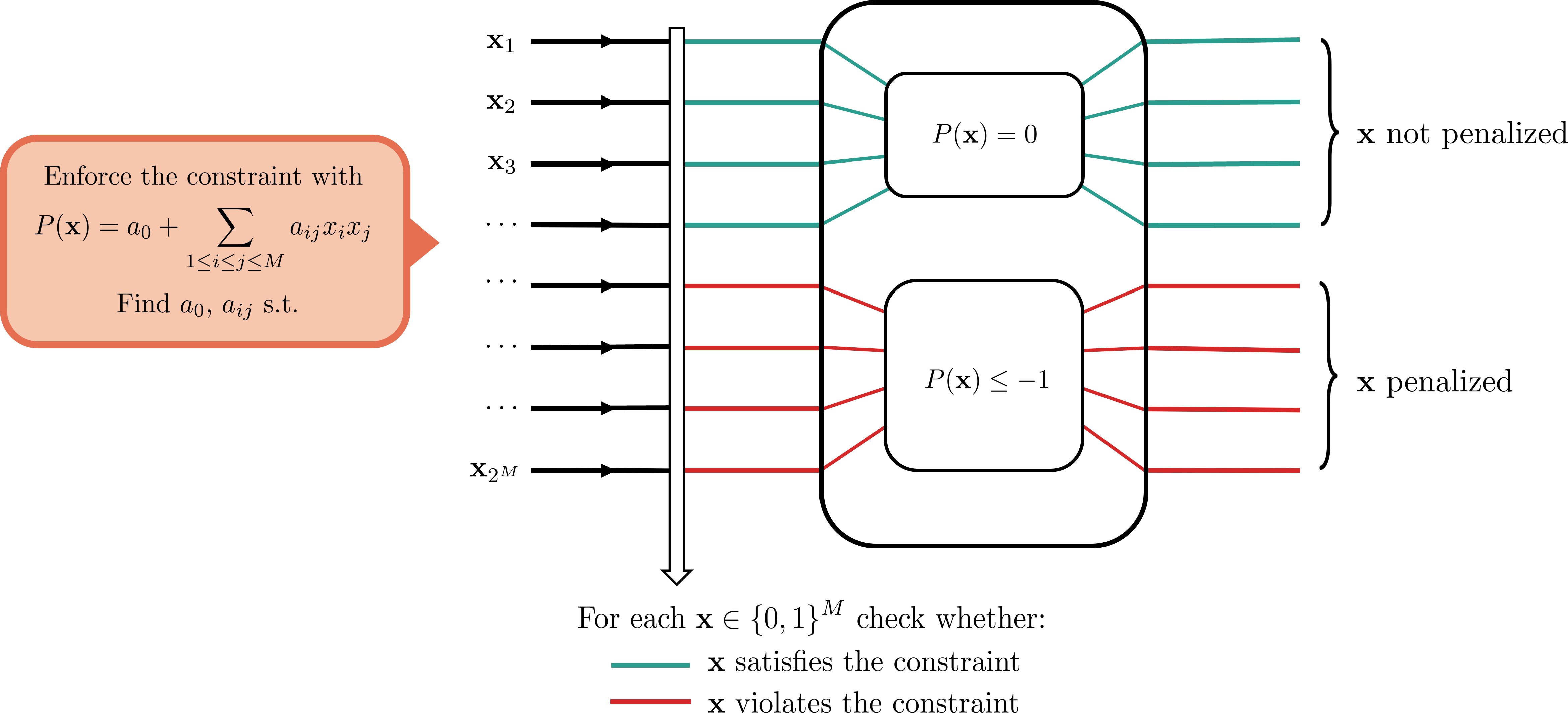}
\caption{{\bf Iterative quadratic polynomial method}:
Take a constraint defined over $M$ binary variables. Our goal is to enforce it in a QUBO form through the most generic quadratic polynomial $P(\mathbf x)$ in the variables $\mathbf x = (x_1,x_2,\dots, x_M)$.
We look for the values of its free parameters $a_0$ and $a_{ij}$ such that $P(\mathbf x)= 0$ for those combinations of $(x_1,x_2,\dots, x_M)$ satisfying the  constraint (green lines) and $P(\mathbf x)\leq -1 $ for those combinations violating the constraint (red lines). In case such a polynomial does not exists, we iteratively increase the number of slack variables in order to have more and more free parameters available to define the polynomial.
We give a detailed explanation of the IQP method (with and without slack variables) in Section~\ref{GQPmaster}.}\label{IQPfig}
\end{center}
\end{figure}

The second main contribution of this work is a novel interplay between the penalty quadratic forms of a QUBO problem that enforce constraints, called the MS method. 
While the standard approach enforces constraints independently from each other, our method exploits a synergy between different constraints sharing the same binary variables. In case of two or more such constraints, we can divide them into \textit{master} and \textit{satellite}.
Master constraints are enforced for each possible binary variable combination, while satellite constraints are enforced only for those combinations that \textit{already satisfy} the master constraints. 
As a result, the penalty quadratic forms associated with master constraints provide penalties/no-penalties whenever the original constraints are violated/satisfied, namely they perfectly enforce their corresponding constraints.
Instead, the penalty quadratic forms associated with satellite constraints are enforced only for those binary variables combinations that already satisfy the master constraints. A detailed description of the MS method is presented in Section~\ref{GQPsatellite}.

A consequence of this approach is that the quadratic forms associated to satellite constraints may provide \textit{accidental incentives}, namely a positive contribution, for those binary variables combinations violating the master constraints. 
Hence, in order to avoid the promotion of those combinations, we amplify the influence of master constraints penalty terms until we have a net penalty even for those combinations receiving accidental
incentives. 
This is done by a careful calibration of the QUBO penalty multipliers, namely some quantities analogous to the parameter $\lambda$ from Eq.~(\ref{QUBOprob0}). 
This tuning is explained in detail in  Section~\ref{tunepenalty}. 
We show that, through the adoption of this strategy, we are able to reduce the employment of slack variables.  We summarize the idea behind the MS method in Figure~\ref{MSfig}.

It is evident that, given a particular optimization problem, the task of deciding which constraints have to be enforced as master and which as satellite is not a trivial one. Remember that, this method requires the presence of at least one master constraint and the more master constraints we have, the more efficient the implementation of satellite constraints is going to be. We can either decide to consider as master those constraints that require the least amount of slack variables, with or without the IQP method, or maybe those that are the most regular among the graph, e.g., do not depend on $w$ and $\mathbf f$ but solely on the arcs and nodes in the multigraph. This situation is examined through an exemplary problem in Section~\ref{SecMPBSUC}.
We point out that, the separation of master and satellite constraints described here is called \textit{bipartite}, namely when constraint can either be master or satellite. We introduce a \textit{concatenated} version of the MS method in Section~\ref{concatenation}, which results to be even more efficient in terms of slack variables employed.

The implementation of the MS method is particularly straightforward when satellite constraints are enforced with the IQP method. Indeed, this method gives us the possibility to decide whether to associate a penalty or not to precise binary variables combinations. Moreover, when satellite constraints are enforced with the IQP method, even less  slack variables are necessary, if compared to the enforcement of the same constraints as master. Indeed, the slack variables needed by the IQP method increase with the number of binary variables combinations for which we require a penalty or not. Hence, we anticipate that the benefits of the conjunction between the IQP and the MS (IQPMS) methods are maximized when dealing with constraints defined on few common binary variables. 
Nonetheless, in Appendix~\ref{MSstandard} we show how to apply the MS method when constraints are enforced with the standard method explained in Section~\ref{QUBOintro}.

\begin{figure}
\begin{center}
\includegraphics[width=0.99\textwidth]{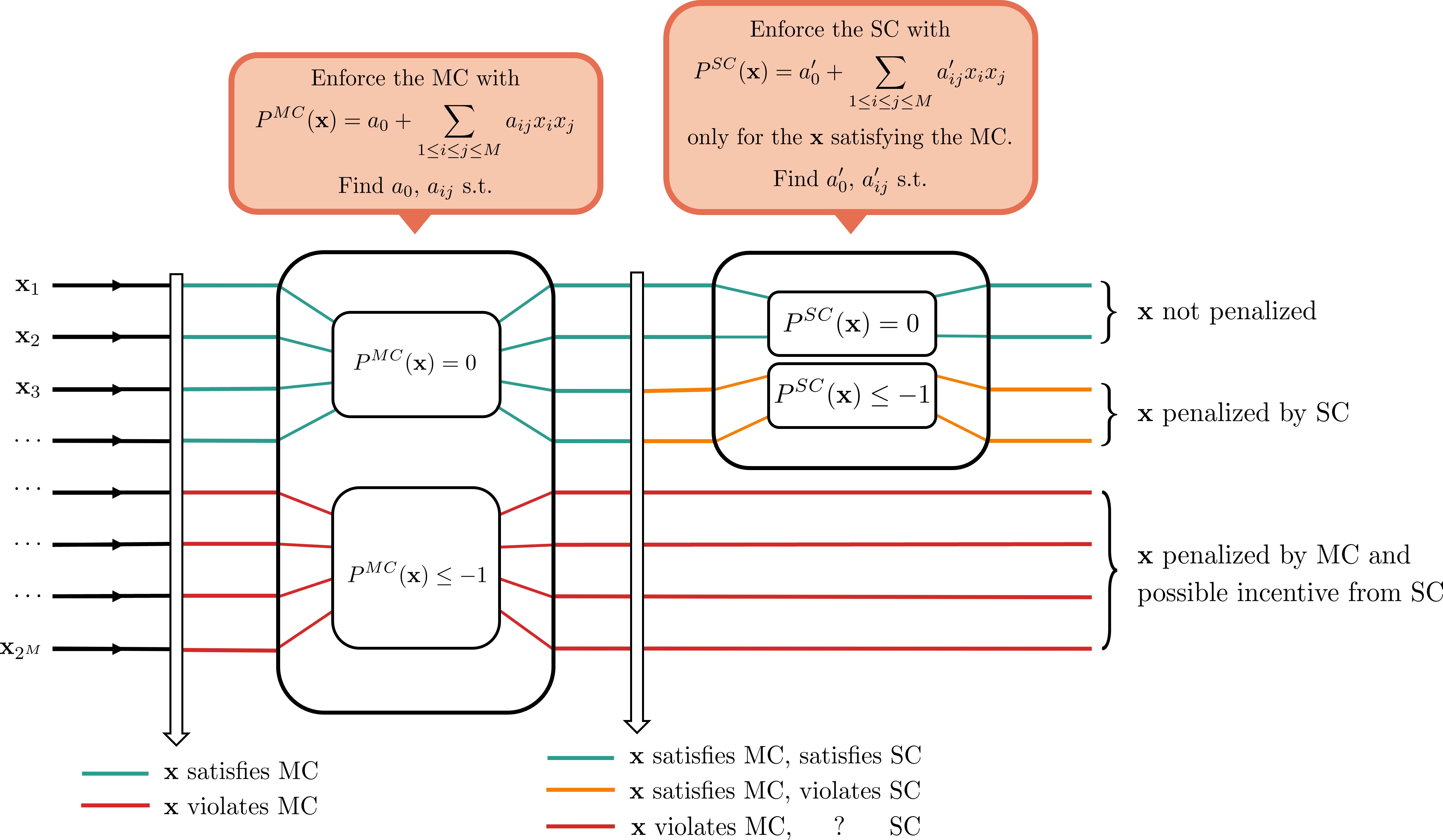}
\caption{{\bf Master-satellite method}:
Take two constraints defined over the same $M$ binary variables, where one is considered as master constraint (MC) and one as satellite constraint (SC). Our goal is to enforce them in a QUBO form through the most generic quadratic polynomials $P^{MC}(\mathbf x)$ and $P^{SC}(\mathbf x)$ in the variables $(x_1,x_2,\dots, x_M)$.
While the MC is enforced with the standard IQP method (see Figure \ref{IQPfig}), the SC is enforced only for those combinations already satisfying the MC (top left green lines). Compared to $P^{MC}(\mathbf x)$, $P^{SC}(\mathbf x)$ is defined through a reduced number of conditions: no requirements are applied for those combinations violating the MC. As a consequence, we may obtain accidental incentives $P^{SC}(\mathbf x)>0$ for these combinations (red lines). We solve this issue by amplifying the role of  $P^{MC}(\mathbf x)$ through a relative multiplier $\lambda^{MC}>1$ so that $\lambda^{MC} P^{MC}(\mathbf x) + P^{SC}(\mathbf x) \leq -1$ if either the MC or the SC is violated (orange and red lines) and is equal to 0 if they are both satisfied (top right green lines). In case one or both such polynomials do not exists, we iteratively increase the number of slack variables in order to have more and more free parameters available to define the polynomials.
We give a detailed explanation of the MS method in Section~\ref{GQPsatellite}.}\label{MSfig}
\end{center}
\end{figure} 

In summary, while the IQP method is more efficient for constraints defined over few variables, the MS approach can be applied for sets of constraints defined on the same subsets of binary variables. Depending on the features of the combinatorial optimization problem studied, either one or both methods can be applied. For the sake of clarity, we introduce the class of locally-constrained optimizations problems in Section \ref{SECLOC}, for which a simultaneous adoption of the IQPMS methods can be efficiently implemented, namely all constraints can be enforced with these new methods. Moreover, the IQPMS methods perform the best when the graphs defining these problems are lowly connected.

Finally, we point out that, while both the IQP and the MS methods offer several advantages, it is important to note their potential limitations. The IQP method can become computationally complex for constraints defined over several variables, and the effectiveness of the MS method may depend on the particular selection of master and satellite constraints.
Nonetheless, compared to existing approaches in the literature, such as the alternating direction method of multipliers~\cite{Boyd}, the augmented Lagrangian method~\cite{Hestenes,Powell} or approximate QUBO enforcing, which have recently been adopted in Refs.~\cite{Djidjev,otherADMM,Light,arxiv1st,solar,powernetwork}, our methods do not require solving preliminary QUBOs or approximating the problem constraints.

\subsection{IQP method}\label{GQPmaster}

In this section we introduce a novel technique to enforce constraints in the QUBO form. We start by focusing on the construction of a quadratic polynomial $P(\mathbf x,\mathbf s)$ that enforces an inequality constraint $IC(\mathbf x) \leq 0$ of the target optimization problem, see Eq.~(\ref{optprob}). As we explain later, this technique can also be used for more general constraints, but for the sake of simplicity we begin with the linear inequality case.
We consider the case where the constraint $IC(\mathbf x) = IC(\mathbf y)$ is defined over $M\leq N$ binary variables, namely over a subset $\mathbf y \subseteq \mathbf x$ of all the $N$ binary variables $\mathbf x = (x_1,x_2,\dots, x_N)$ defining the starting optimization problem. Hence, differently from the previous section (see Figure~\ref{IQPfig}), we relabel the binary variables associated to the target inequality constraint by using $\mathbf y = (x_1,x_2,\dots, x_M)$.

Consider a generic quadratic polynomial (QP) in the following $M+ \overline M$ variables $\{\mathbf y, \mathbf s\}$, where $\mathbf y = (x_1,x_2,\dots, x_M)$ are the binary variables associated to the studied constraint $IC(\mathbf y)$ and $\mathbf s = \{s_1,s_2,\dots,s_{\overline M}\}$ are $\overline M$ slack variables:
\begin{equation}\label{Pquad0}
P(\mathbf y, \mathbf s) =a_0 + \sum_{1\leq i\leq j\leq M} a_{ij} x_i x_j+\sum_{1\leq k\leq l\leq \overline M} \overline a_{k l} s_k s_l +\sum_{\substack{1\leq i\leq M \\ 1\leq k \leq \overline M}}  b_{i k} x_i s_k \, .
\end{equation}
The number of free parameters in $P(\mathbf y, \mathbf s) $ are:
\begin{equation}\label{degreesslack0}
\# \{a_0,a_{ij}, \overline a_{kl} , b_{ik}\} = 1 + \frac{\left(M+\overline M\,\right)^2 + M+ \overline M}{2}  \, .
\end{equation}
The slack variables used to define $P(\mathbf y, \mathbf s)$ are disjoint from those considered to define QPs associated to other constraints~\footnote{If $IC_{1}(\mathbf y)\leq 0$ and $IC_{2}(\mathbf y') \leq 0$ are enforced by $P_{1}(\mathbf y,\mathbf s)$ and $P_{2}(\mathbf y',\mathbf s')$, respectively, they would make use of $\overline M$ and $\overline M'$ independent slack variables.}.

{\bf QP with no-slack variables:} We start by considering the scenario without slack variables, namely $\overline M=0$. In order to obtain a quadratic form $P(\mathbf y)$ that enforces the constraint $IC(\mathbf y)\leq 0$, we have to solve the following linear system in the parameters $a_{ij}$:
\begin{equation}\label{genconds0} 
\left\{
 \,\,\, \begin{array}{ccc}
P(\mathbf y)= 0 & \mbox{ if }  \mathbf y \mbox{ satisfies } IC(\mathbf y)\leq 0  \\ 
P(\mathbf y) \leq - 1 & \mbox{ if }  \mathbf y \mbox{ violates } IC(\mathbf y)\leq 0  \\ 
\end{array}\right. \, .
\end{equation}
If such a solution exists, then we managed to translate  the constraint $IC(\mathbf y)\leq 0$ into a QUBO form without using slack variables.
We say that each value of $\mathbf y$ identifies a different \textit{sub-constraint}, which can be either assigned to a \textit{no-penalty} $P(\mathbf y)=0$ or to a \textit{penalty} $P(\mathbf y)\leq -1$. Hence, in order for $P(\mathbf y)$ to represent $IC(\mathbf y)\leq 0$, it has to satisfy $2^M$ sub-constraints, one for each value of $\mathbf y \in \{0,1\}^{M}$. It may sound that we can use no slack variables only when $\# a_{ij} = 1+ (M^2+M)/2$ is not smaller than the number of sub-constraint $2^M$. Nonetheless, as we see later, it is not uncommon to find quadratic polynomials that enforce a number of sub-constraints larger than the number of its free parameters. {This is because inequality sub-constraints, namely $P(\mathbf y)\leq -1$, are less restrictive than equality ones. Finally, notice that, if the constraint is satisfied for all $\mathbf y \in \{0,1\}^M$, then we do not even need to enforce this constraint via a penalty polynomial: it is not violated for any combination of $\mathbf y$ and therefore no penalty has to be applied. Indeed, in this case we would simply obtain the null polynomial, namely $a_0=a_{ij}=0$ for all $i$ and $j$}.

{\bf QP with slack variables:} The implementation of slack variables into this scenario cannot be made by simply considering the linear system~(\ref{genconds0}), where $P(\mathbf y)$ is replaced by $P(\mathbf y,\mathbf s)$.
Indeed, even if $P(\mathbf y,\mathbf s)$ has several more degrees of freedom coming from the presence of slack variables, these are not exploited in the system as there is no reference of them in the linear system that we should solve. Consider for instance a value of $\mathbf y$ associated to a no-penalty sub-constraint. In this case, we would require $P(\mathbf y,\mathbf s) = 0$ for all the values of $\mathbf s$. Hence, even if we have more degrees of freedom compared to the no-slack variable method, in this way we would solely have more conditions in the linear system. 
Therefore, it is easy to convince ourselves that this approach provides no improvement with respect to the scenario of a QP with no slack variables. 

In order to exploit the potential of the slack variables inside a generic QP in this context, we start by noting the following. If we replace $P(\mathbf y)$ with $P(\mathbf y,\mathbf s)$ in Eq.~(\ref{genconds0}), the penalty (inequality) sub-constraints  are less restrictive than the no-penalty (equality) ones. Indeed, the former type allows $P (\mathbf y,\mathbf s) $ to assume any value smaller or equal to -1 and for different values of $\mathbf s$ and the same $\mathbf y$ the polynomial $P (\mathbf y,\mathbf s) $ can assume different penalties. Instead, for no-penalty sub-constraints we require $P (\mathbf y,\mathbf s)$ to be exactly~0, for all $\mathbf y$ and $\mathbf s$. Nonetheless, if $\mathbf y$ is associated to a no-penalty, we do not need  $P (\mathbf y,\mathbf s) =0$ to be true \textit{for all the values} of $\mathbf s$, it is enough to require it for \textit{at least one} $\mathbf s$, while at the same time no incentives are provided for the other values of $\mathbf s$. Indeed, $P (\mathbf y,\mathbf s)$ is one component of the QUBO formulation of an optimization problem, which undergoes a maximization over $\mathbf x$ (which contains all the variables in $\mathbf y$) and $\mathbf s$. Hence, our minimal requirements on the parameters $a_{ij}$, $\overline a_{kl}$ and $b_{ik}$  defining $P (\mathbf y,\mathbf s)$ are: 
\begin{equation}\label{genconds00}
\left\{
\begin{array}{ccc}
P(\mathbf y,\mathbf s)\leq - 1 &  \mbox{ if }  \mathbf y \mbox{ violates } IC(\mathbf y)\leq 0 &\hspace{-1.35cm} \mbox{for all } \mathbf s \\ 
 P(\mathbf y,\mathbf s)\leq  0 &  \mbox{ if }  \mathbf y \mbox{ satisfies } IC(\mathbf y)\leq 0& \hspace{-1.35cm} \mbox{for all } \mathbf s  \\  
 P(\mathbf y,\mathbf s)=  0 &  \mbox{ if }  \mathbf y \mbox{ satisfies } IC(\mathbf y)\leq 0&\hspace{-0.1cm} \mbox{for at least one } \mathbf s
\end{array}\right.
\end{equation}
or, equivalently,
\begin{equation}\label{genconds000}
\left\{
\begin{array}{ccc}
P(\mathbf y,\mathbf s)\leq - 1 &  \mbox{ if }  \mathbf y \mbox{ violates } IC(\mathbf y)\leq 0 & \hspace{-0.2cm}  \mbox{ for all } \mathbf s \\ 
\max_{\mathbf{s}} P(\mathbf y,\mathbf s) = 0 &  \mbox{ if }  \mathbf y \mbox{ satisfies } IC(\mathbf y)\leq 0  \\ 
\end{array}\right.  \, .
\end{equation}

The enforcement of the sub-constraints proposed in Eqs.~(\ref{genconds00}) and~(\ref{genconds000}) can be casted as a \textit{feasibility problem}. In case of a single slack variable $\mathbf s = s_1$, namely for $\overline M = 1$, it assumes the form:
\begin{equation}\label{hybrid1slack0}
\begin{array}{cccccc}
 \arg & \min_{a_0,a_{ij}, \overline a_{kl} , b_{ik}} & 1   &  \\ 
 & \hspace{-9mm} \mbox{s.t.} &  P(\mathbf{y},0) \leq  -1   \land P(\mathbf{y},1) \leq  -1   & \mbox{ if }  \mathbf y \mbox{ violates } IC(\mathbf y)\leq 0  \\ 
 & &  \hspace{-8mm} ( P(\mathbf{y},0) = 0   \land P(\mathbf{y},1) \leq  0 )  \lor ( P(\mathbf{y},0) \leq  0   \land P(\mathbf{y},1) =  0 ) &   \mbox{ if }  \mathbf y \mbox{ satisfies } IC(\mathbf y)\leq 0\\ 
\end{array} \, , 
\end{equation}
where the symbols $\land$ and $\lor$ represent, respectively, the {\sffamily{AND}} and {\sffamily{OR}} logical operations. Hence, the solution of this feasibility problem provides the parameters $a_{ij}, \overline a_{kl}$ and $ b_{ik}$ that defines the quadratic polynomial $P(\mathbf y, s_1)$ that enforces our constraint.

{\bf IQP method:} Consider the following iterative procedure to enforce constraints in a QUBO form, where in each step we add a slack variable to the QP used, until this enforcement is possible.
Hence, given a target constraint, we execute the following step-by-step algorithm:
\begin{itemize}
\item STEP 0: Apply the method with no-slack variables given in Eq.~(\ref{genconds0}). If it provides a solution, halt.
\item STEP 1: Apply the method with $\overline M=1$ slack variable given in Eqs.~(\ref{genconds00}) and (\ref{genconds000}). If it provides a solution, halt.
\item STEP 2: Apply the method with $\overline M=2$ slack variables given in Eqs.~(\ref{genconds00}) and (\ref{genconds000}).
If it provides a solution, halt.
\item \,\,\,\,\,\,\dots
\item STEP $\overline M$:  Apply the method with $\overline M$ slack variables given in Eqs.~(\ref{genconds00}) and (\ref{genconds000}), with the following feasibility form:
\begin{equation}\label{hybridgeneral}
\begin{array}{ccccc}
\arg & \min_{a_0,a_{ij}, \overline a_{kl} , b_{ik}} &  1 & & \\
 &\mbox{s.t.} &  \bigwedge_{\mathbf s} P(\mathbf y, \mathbf s ) \leq -1 & \mbox{ if }  \mathbf y \mbox{ violates } IC(\mathbf y)\leq 0  \\ 
& &\bigvee_{\mathbf s} \left( P(\mathbf y, \mathbf s ) = 0 \land P(\mathbf y, \mathbf s' \neq \mathbf s) \leq 0 \right)   & \mbox{ if }  \mathbf y \mbox{ satisfies } IC(\mathbf y)\leq 0 
\end{array} \, ,
\end{equation}
\end{itemize}
where the {\sffamily{AND}} and {\sffamily{OR}} logical operations are performed over $\mathbf s\in \{0,1\}^{\overline M}$.

In the following, we use the expression ``IQP method'' to refer to this algorithm. Notice that, if this procedure halts at a certain step, then the same amount of slack variables are enough for the IQP method to encode the target constraint as a QUBO penalty term. Remember that we start from STEP 0, which corresponds to the no-slack variables scenario. Hence, this method is designed to enforce our constraint via a QP (\ref{Pquad0}) employing the least number of slack variables. In the following, the QP obtained through this method could be simply called IQP.

We point out that, as we increase the number of slack variables throughout these steps, $P(\mathbf y,\mathbf s)$ gets more and more free-parameters at disposal (see Eq.~(\ref{degreesslack0})), while at the same time we apply the same amount of ``strong'' equality conditions and more ``soft'' inequality conditions (see Eq.~(\ref{genconds00})) in the corresponding linear system. An indicative estimate of the slack variables necessary for this kind of method is given by requiring that the free parameters in $P(\mathbf y,\mathbf s)$ (see Eq.~(\ref{degreesslack0})) is larger than the number of sub-constraints, which in the general case are $2^M$, namely the conditions in the linear system~(\ref{genconds0}) considered with no slack variables. Nonetheless, as we show it later with an example, a lower number of slack variables is often sufficient. Hence, it should already be clear that this method is particularly useful for small $M$. Indeed, while for the standard method the number of slack variables do not change with $M$, in general the number of slack variables needed by the IQP method increases with $M$.

Finally, we underline two more qualities of this technique. First, {this method can be extended to any linear/non-linear equality/inequality constraint defined with integer/non-integer parameters.} Indeed, consider the linear systems~(\ref{genconds00}) and~(\ref{hybridgeneral}) defining the parameters of the quadratic polynomial enforcing the constraint. The proposed procedure makes no reference to the form of $IC(\mathbf y)$ and, more importantly, its linearity is never exploited. The only operation that we have to perform is to verify whether $IC(\mathbf y)$ is violated or not for each $\mathbf y$. It follows that, in case we need to enforce any other type of constraint $C(\mathbf y) =\mbox{True}$, the same procedure described in this section can be adopted, where we simply replace ``$\mathbf y$ satisfies/violates $IC(\mathbf y)\leq 0$'' with ``$\mathbf y$ satisfies/violates $C(\mathbf y)$''. {A second merit of the IQP method is that, differently from the standard approach, it does not require all the variables to be coupled to each other: some of the parameters $a_{ij},\, \overline{a}_{kl},\, b_{ik}$ can be null. We encounter this scenario for several constraints defining the optimization problem studied later in this work, namely the \algo{MPBS} problem (see Section~\ref{IQPMSMPBSsec}).  
This scenario can improve the performance of NISQ devices used as QUBO solvers as sparser formulations require fewer and less-connected qubits.}

\subsection{MS methods}\label{GQPsatellite}

Consider the scenario where two (or more) constraints are defined over the same subset of variables $\mathbf y \subseteq\mathbf x  $, where the first has been enforced in a quadratic form either with the IQP method or any other method, e.g., as in Section \ref{QUBOintro}. We call this constraint the master constraint, while the second, which has yet to be enforced, as satellite. We follow by presenting a variation of the IQP method suited to enforce satellite constraints. This technique, as we show below, employs even less slack variables, if compared with the method presented in Section \ref{GQPmaster}, which instead is better suited for master constraints.

The main difference between the standard IQP method and its variation for satellite constraints is that we do not impose sub-constraints for all the possible combinations of the binary variables $\mathbf y=(x_1,x_2,\dots, x_M)$, but only for the strictly smaller subset of combinations satisfying the master constraints. The final discussion of the previous section should already convince the reader that having less than $2^M$ sub-constraints allows to implement satellite constraints with even fewer slack variables.

We start by considering two constraints sharing the same variables $\mathbf y$, where we call ``$MC(\mathbf y)=$ True'' the constraint considered as master, and ``$SC(\mathbf y)=$ True'' the satellite one, where both can either be {linear/non-linear equality/inequality constraints defined with integer/non-integer parameters}.
Requiring $P^{SC}(\mathbf y,\mathbf s)$, obtained form Eq.~(\ref{Pquad0}), to enforce the satellite constraint with $\overline M$ slack variables corresponds to verify whether there exists a solution of the following linear system in the parameters $a_{ij}$, $\overline a_{kl}$ and $b_{ik}$:
\begin{equation}\label{gencondssat}
\left\{
\begin{array}{ccc}
P^{SC}(\mathbf y,\mathbf s)\leq - 1 &  \mbox{ if }  \mathbf y \mbox{ satisfies } MC(\mathbf y) \mbox{ and violates } SC(\mathbf y) & \hspace{-1.35cm} \mbox{for all } \mathbf s \\ 
P^{SC}(\mathbf y,\mathbf s) \leq  0 &  \mbox{ if }  \mathbf y \mbox{ satisfies } MC(\mathbf y) \mbox{ and satisfies } SC(\mathbf y)&  \hspace{-1.35cm} \mbox{for all } \mathbf s \\ 
P^{SC}(\mathbf y,\mathbf s) =  0 &  \mbox{ if }  \mathbf y \mbox{ satisfies } MC(\mathbf y) \mbox{ and satisfies } SC(\mathbf y)& \hspace{-0.1cm} \mbox{for at least one } \mathbf s
\end{array}\right.
\end{equation}
or, equivalently,
\begin{equation}\label{gencondssat2}
\left\{
\begin{array}{ccc}
 P^{SC}(\mathbf y,\mathbf s)\leq - 1 &  \mbox{ if }  \mathbf y \mbox{ satisfies } MC(\mathbf y) \mbox{ and violates } SC(\mathbf y) & \hspace{-0.2cm}  \mbox{ for all } \mathbf s \\ 
\max_{\mathbf{s}} P^{SC}(\mathbf y,\mathbf s) = 0 &  \mbox{ if }  \mathbf y \mbox{ satisfies } MC(\mathbf y) \mbox{ and satisfies } SC(\mathbf y)  \\ 
\end{array}\right.  \, .
\end{equation}

The step-by-step IQP method described in the previous section can be generalized to the satellite scenario, where we simply restrict the corresponding conditions for $P^{SC}(\mathbf y,\mathbf s)$ to the $2^M - n(MC)<2^M$ sub-constraints that satisfy the master constraints, as shown in Eqs.~(\ref{gencondssat}) and~(\ref{gencondssat2}), where $n(MC) \geq 1$ is the number of combinations of  $\mathbf y$ that violate the master constraint. Hence, all the corresponding steps can be formulated similarly. For instance, the feasibility problem formulation (\ref{hybridgeneral}) reads:
\begin{equation}\label{feasMS}
\begin{array}{ccccc}
\arg & \min_{a_{ij}, \overline a_{kl} , b_{ik}} &  1 & & \\
 &\mbox{s.t.} &  \bigwedge_{\mathbf s} P^{SC}(\mathbf y, \mathbf s ) \leq -1 & \mbox{ if }  \mathbf y \mbox{ satisfies } MC(\mathbf y) \mbox{ and violates } SC(\mathbf y)  \\ 
& &\bigvee_{\mathbf s} P^{SC}(\mathbf y, \mathbf s ) = 0 \land P^{SC}(\mathbf y, \mathbf s' \neq \mathbf s) \leq 0  & \mbox{ if }  \mathbf y \mbox{ satisfies } MC(\mathbf y) \mbox{ and satisfies } SC(\mathbf y) 
\end{array} \, ,
\end{equation}
where the {\sffamily{AND}} and {\sffamily{OR}} logical operations are performed over $\mathbf s\in \{0,1\}^{\overline M}$.
Remember that this procedure can be straightforwardly applied to constraints that are non-linear or not defined through integer parameters, as described in Sec. \ref{GQPmaster}.

{Notice that, in case all the combinations of $\mathbf y$ satisfying the master constraint satisfy the satellite constraint too, then there is no need to enforce the satellite constraint with a quadratic polynomial at all. Indeed, in this case a solution of Eq. (\ref{feasMS}) would be given (without employing any slack variable) by $a_0=a_{ij}= \overline a_{kl} = b_{ik}= 0$ for all $i$, $j$, $k$ and $l$.}

In order to make the notation clearer, we considered the case of one master and one satellite constraint. In case of multiple master constraints, we have to require sub-constraints only for those $\mathbf y$ satisfying \textit{all} the master constraints. Hence, in such a scenario we would impose even less sub-constraints, namely $2^M - n(MC)<2^M$, where now $n(MC) \geq 1$ is the number of combinations that violate at least one master constraint.
Naturally, if we have more than one satellite constraint, the solution of this linear system has to be performed independently for each one of these constraints.

{ The natural consequence of not imposing any sub-constraint for those $\mathbf y$ violating the master constraints is that we may obtain an accidental incentive $P^{SC}(\mathbf y,\mathbf s)>0$ for some $(\mathbf y,\mathbf s)$, where $\mathbf y$ violates the master constraint. In this situation such a combination would be promoted, and therefore we need a strategy to avoid that the optimal solution of the corresponding QUBO problem violates one or more constraints. This goal is achieved by amplifying the weight of master constraints as shown in Section \ref{tunepenalty}.}

We say that this MS method adopts a \textit{bipartite} approach as, even in case of more than two constraints involved, each constraint is either considered as a master or a satellite constraint. Instead, in case of three or more constraints, we can either adopt the strategy introduced in this section or its \textit{concatenated} version, presented in Section~\ref{concatenation}.

\subsection{Tuning of the relative multipliers}\label{tunepenalty}

Consider the QUBO form (\ref{QUBOprob00}) of a generic optimization problem (\ref{optprob}), where a single constraint has been enforced with the IQP method presented in Section~\ref{GQPmaster}. Since the corresponding quadratic form $P(\mathbf x,\mathbf s)= P(\mathbf y,\mathbf s)$ provides penalties/no-penalties if and only if $\mathbf y$ violates/satisfies the corresponding constraint, no adjustments have to be applied to the form given in Eq.~(\ref{QUBOprob00}). The same applies if multiple constraints are enforced with the standard IQP method, namely when the MS method is not exploited.

In case two or more constraints are enforced with the MS method, some adjustments to Eq.~(\ref{QUBOprob00}) have to be applied. In practice, we need to amplify the role of the master constraint quadratic form as follows:
\begin{equation}\label{MStrans}
P^{MC}(\mathbf y,\mathbf s) + P^{SC}(\mathbf y,\mathbf s)\,\, \longrightarrow \,\,  \lambda^{MC} P^{MC}(\mathbf y,\mathbf s) + P^{SC}(\mathbf y,\mathbf s') \, ,
\end{equation} 
where the master and satellite constraints considered here are two constraints picked from the inequality and equality constraints of Eq.~(\ref{optprob}) that we decided to enforce with the MS approach, hence sharing the same subset of binary variables $\mathbf y$. Hence, $\lambda^{MC} P^{MC}(\mathbf y,\mathbf s) + P^{SC}(\mathbf y,\mathbf s')$ replaces two of the quadratic polynomials on the r.h.s. of Eq.~(\ref{QUBOprob00}).
 The multiplier  $\lambda^{MC} \geq 1$ can be defined as follows:
\begin{equation}\label{lambdaMC0}
\lambda^{MC} = 1 + \gamma \max \left\{ 0, \max_{\mathbf y,\mathbf s'} P^{SC}(\mathbf y,\mathbf s') \right\} \, ,
\end{equation} 
where $\gamma>1$ and, if the maximization on the r.h.s. is greater than zero, we say that at least one of the satellite constraints provide an \textit{accidental incentive}. In case of no accidental incentives, $\lambda^{MC}=1$.

Thanks to this transformation, we finally achieve a QUBO form of these two constraints that faithfully represents the corresponding constraints. Indeed, the r.h.s. of Eq.~(\ref{MStrans}) is smaller or equal than -1 if $\mathbf y$ violates either the master or the satellite constraint and is zero-valued for at least one value of $\mathbf s$ if both constraints are satisfied:
\begin{equation}
\left\{
\begin{array}{ccc}
\lambda^{MC} P^{MC}(\mathbf y,\mathbf s) + P^{SC}(\mathbf y,\mathbf s')\leq - 1 &  \mbox{ for all } \mathbf s & \mbox{ if }  \mathbf y \mbox{ violates } MC(\mathbf s) \vee  SC(\mathbf s)   \\ 
\lambda^{MC} P^{MC}(\mathbf y,\mathbf s) + P^{SC}(\mathbf y,\mathbf s') \leq 0 & \mbox{ for all } \mathbf s & \mbox{ if }  \mathbf y \mbox{ satisfies } MC(\mathbf s) \land  SC(\mathbf s)  \\
\lambda^{MC} P^{MC}(\mathbf y,\mathbf s) + P^{SC}(\mathbf y,\mathbf s') = 0 & \mbox{ for at least one } \mathbf s & \mbox{ if }  \mathbf y \mbox{ satisfies } MC(\mathbf s) \land  SC(\mathbf s)   
\end{array}\right.  \, .
\end{equation}

Consider the case where we enforce multiple constraints as master and satellite (either with the IQP method or not) defined over the same shared set of variables $\mathbf y \subseteq \mathbf x$. Both master and satellite constraints can be chosen among the equality and inequality constraints. We apply the following transformation of the QUBO form:
\begin{equation}\label{MStrans2}
\sum_{i=1}^{m_{MC}}P_i^{MC}(\mathbf y,\mathbf s) +\sum_{j=1}^{m_{SC}} P_j^{SC}(\mathbf y,\mathbf s)\,\, \longrightarrow \,\, \lambda^{MC}  \sum_{i=1}^{m_{MC}}P_i^{MC}(\mathbf y,\mathbf s) +\sum_{j=1}^{m_{SC}} P_j^{SC}(\mathbf y,\mathbf s) \, ,
\end{equation} 
where, even if not made explicit, the slack variables employed for the master and satellite quadratic polynomials are independent. Therefore, this transformation involves $m_{MC}+m_{IC}$ quadratic polynomials picked from Eq.~(\ref{QUBOprob00}).
Similarly to Eq.~(\ref{lambdaMC0}), we use:
\begin{equation}
\lambda^{MC} = 1 + \gamma \max \left\{ 0, \max_{\mathbf y,\mathbf s'} \sum_{j=1}^{m_{SC}} P_j^{SC}(\mathbf y,\mathbf s') \right\} \, .
\end{equation}

For what concerns the multiplier $\lambda$ that appears in Eq.~(\ref{QUBOprob00}), we can apply the ordinary strategies that can be found in the literature. Several exemplary problems for which this parameter is calculated are given in Ref.~\cite{Lucas}, where at times a different parameter $\lambda_i$ is associated to each quadratic penalty term $P_i(\mathbf x,\mathbf s)$. In what follows, we define a class of optimization problems, called locally-constrained, for which we propose a different tuning of these parameters.

\subsection{Master-satellite concatenation}\label{concatenation}

We briefly describe a method to concatenate the MS approach in order to reduce even more the number of considered sub-constraints, and therefore of slack variables. 
This approach can be applied solely for those cases where three or more constraints share the same subset of variables $\mathbf y \subseteq \mathbf x$. For instance, consider the four constraints $C_{1}(\mathbf y)$, $C_{2}(\mathbf y)$, $C_{3}(\mathbf y)$ and $C_{4}(\mathbf y)$ (we omit the ``$=\mbox{True}$'' notation) defined over the same binary variables. 

The approach of the previous sections requires to identify two groups of constraints, where all the constraints in the second group are satellite of the constraints in the first group. For instance, we could choose the master constraints to be $C_{1}(\mathbf y)$ and $C_{2}(\mathbf y)$, while their satellite constraints are respectively  $C_{3}(\mathbf y)$ and $C_{4}(\mathbf y)$.
Each constraint in the satellite group is actively enforced solely for those combinations of $\mathbf y$ that satisfy the constraints from the master group. On the other side, all the master constraints have to be enforced for all the $2^{M}$ combinations of $\mathbf y=(x_1,x_2,\dots,x_M)$.

We concatenate the MS method among all the constraints defined over $\mathbf y\subseteq \mathbf x$. In particular, we start by choosing two constraints, e.g., $C_{1}(\mathbf y)$ and $C_{2}(\mathbf y)$, and enforce $C_{1}(\mathbf y)$ \textit{as the master constraint of} $C_{2}(\mathbf y)$. Hence, $C_{1}(\mathbf y)$ would be enforced for all the combinations of $\mathbf y$, while $C_{2}(\mathbf y)$ only for those $\mathbf y$ satisfying $C_{1}(\mathbf y)$ (as the master-satellite approach described in the previous sections in the case of two constraints). Nonetheless, differently from before, we continue this concatenation of constraints by considering $C_{1}(\mathbf y) \land C_{2}(\mathbf y)$ as the master constraint of $C_{3}(\mathbf y)$. Hence, $C_{3}(\mathbf y)$ would be enforced only for those $\mathbf y$ satisfying $C_{1}(\mathbf y) \land C_{2}(\mathbf y)$. Finally, $C_{4}(\mathbf y)$ would be enforced as a satellite constraint of all the other constraints, namely with $C_{1}(\mathbf y) \land C_{2}(\mathbf y) \land C_{3}(\mathbf y)$ being its master constraint. Therefore, $C_{4}(\mathbf y)$ has to be enforced for those $\mathbf y$ already satisfying all the other constraints.
This hierarchy of constraint enforcement allows to consider fewer sub-constraints if compared with the non-concatenated case, and therefore, as show before, less slack variables.

This approach requires the introduction of multiple relative multipliers that have the role to counterbalance the accidental incentives coming from the enforcement of $C_{2}(\mathbf y)$, $C_{3}(\mathbf y)$ and $C_{4}(\mathbf y)$. If we consider the previously described example of four constraints, we would need three different multipliers similar to $\lambda^{MC}$ from Eq.~(\ref{lambdaMC0}) via the following replacement of the quadratic polynomials enforcing $C_{1,2,3,4}(\mathbf y)$ in Eq.~(\ref{QUBOprob00}):
\begin{equation}
P_1(\mathbf y,\mathbf s) + P_2(\mathbf y,\mathbf s)+ P_3(\mathbf y,\mathbf s)+ P_4(\mathbf y,\mathbf s) \,\longrightarrow \, \lambda_{1} P_{1}(\mathbf y,\mathbf s) +\lambda_{2} P_{2}(\mathbf y,\mathbf s) +\lambda_{3} P_{3}(\mathbf y,\mathbf s) + P_{4}(\mathbf y,\mathbf s)  \, ,
\end{equation}
where $P_{i}(\mathbf y,\mathbf s)$ is the IQP obtained while considering the constraint $C_{i}(\mathbf y)$ in this concatenated scheme, namely as a satellite of all the constraints $C_{j}(\mathbf y)$ with $j<i$ and master of those with $j>i$. Remember that each $C_i(\mathbf y)$ can either be an equality or an inequality constraint. The tuning of the parameters $\lambda_{i}$ is given by the following generalization of  Eq.~(\ref{lambdaMC0}):
\begin{equation}\label{relativemultiplierconc}
\lambda_{i} =  1 +\gamma \max\left\{ 0 , \max_{\mathbf y,\mathbf s}\sum_{j>i} P_{j}(\mathbf y,\mathbf s) \right\}  \, ,
\end{equation}
where $\gamma >1$.

\section{Locally-constrained optimization problems}\label{SECLOC}

A class of optimization problems that is particularly suited for the application of our techniques is given by the {locally-constrained} ones, namely those having constraints that are defined node-by-node and depend solely on the variables $x_i$ associated to the arcs incident to each node (see Definition~\ref{RMULTI}).
The optimization problems of this type assume the form:
\begin{equation}\label{locoptprob}  
\begin{array}{cccc}
\mathcal E^* = \arg & \max_{\mathbf x} & W(\mathbf x) & \\ 
    & \mbox{ s.t. } & EC_{u,i}(\mathbf  x) = 0 & \mbox{ for } u\in \mathcal{V(\mathbf x)} \,\,\mbox{ and }\,\, i=1,\dots, m_{EC}  \\ 
   &  & IC_{u,j}(\mathbf  x) \leq 0&  \mbox{ for } u\in \mathcal{V(\mathbf x)} \,\,\mbox{ and  }\,\, j=1,\dots, m_{IC}
   \end{array} 	\, ,
\end{equation}
where the equality and inequality constraints $EC_{u,i}(\mathbf  x)$ and $IC_{u,j}(\mathbf  x)$ are defined solely over those binary variables $\mathbf y_u \subseteq \mathbf x = \{x_1,x_2,\dots, x_N\}$ associated to the arcs incident to the corresponding node $u$ and $\mathcal{V}(\mathbf x) \subseteq \mathcal V$ is the subset of nodes identified by the arcs selected by $\mathbf x$. For instance, if $\mathbf x  = \{ 1, 0 , 0 , \dots, 0\}$, where $x_1=1$ corresponds to an arc connecting $u=1$ and $u=2$, $\mathcal{V}(\mathbf x) $ would consists of the nodes $u=1$ and $u=2$.
Hence, for each node in $\mathcal{V}(\mathbf x) $ we impose $m_{EC}$ equality constraints and $m_{IC}$ inequality constraints~\footnote{Extending our approach to node-dependent $m_{IC,u}$ and $m_{EC,u}$ is straightforward but, for the sake of simplicity, we consider each node having the same number of equality and inequality constraints.}. Notice that, while in Eq.~(\ref{optprob}) $m_{EC}$ and $m_{IC}$ referred to the total number of equality and inequality constraints in the whole graph, respectively, in Eq.~(\ref{locoptprob}) they refer to the number of equality/inequality constraints per node~\footnote{If Eq.~(\ref{optprob}) is locally-constrained, it can be written in the form of Eq.~(\ref{locoptprob}) via the rescaling $m_{EC/IC} \rightarrow m_{EC/IC} / |\mathcal V|$.}.

Locally-constrained optimization problems are particularly appropriate for the application of the IQPMS methods. These problems present a natural subdivision of the constraints defined over the same set of binary variables $\mathbf y$. Indeed, each node $u$ has constraints defined solely over the variables $\mathbf y_u \subseteq \mathbf x$ associated to the arcs incident to $u$. Therefore, the application of the IQPMS methods is extremely direct.

Consider a locally-constrained problem with at least two constraints per node and divide them into $m_{MC}\geq 1$ master constraints $MC_{u,i}(\mathbf x)$ and $m_{SC}\geq 1$ {satellite} constraints $SC_{u,j}(\mathbf x)$, where $m_{MC}+m_{SC}= m_{EQ} + m_{IQ}$. All the linear equality constraints, since they do not need the employment of slack variables, are promoted to master constraints, while the remaining ones can either be satellite or master. 
We can adopt the IQPMS methods introduced in Sections~\ref{GQPmaster} and~\ref{GQPsatellite} to enforce the master and satellite constraints as quadratic penalty terms $P^{MC}_{u,i}(\mathbf y_u,\mathbf s)$ and $P^{SC}_{u,j}(\mathbf y_u,\mathbf s)$, which are defined over the variables $\mathbf y_u\subseteq \mathbf x$ associated to the arcs incident to the node $u$.
Therefore, the QUBO formulation of the optimization problems in this class, thanks to the IQPMS methods, assume the form: 
\begin{equation}\label{QUBOprob}
\mathcal E^* = \arg  \max_{\mathbf x,\mathbf s}  W(\mathbf x)  + \sum_{u=1}^{|\mathcal{V}|} \lambda_u \left( \lambda^{MC}_{u} \sum_{i=1}^{m_{MC}} P^{MC}_{u,i}(\mathbf y_u,\mathbf s) + \sum_{j=1}^{m_{SC}} P^{SC}_{u,j} (\mathbf y_u,\mathbf s)   \right) \, ,
\end{equation}
where the quantities $\lambda^{MC}_{u,i}\geq 1$ and $\lambda_u\geq 1$ have two different roles. The relative multipliers $\lambda^{MC}_{u,i}$ ensure that, for each combination of the binary variables associated to the node $u$, if $\mathbf y_u$ violates one of the constraints, the net contribution of the term inside the parenthesis is smaller than or equal to -1. Indeed, we need to amplify the effect of master constraints accordingly to the accidental incentives that $P^{SC}_{u,j}(\mathbf y_u,\mathbf s)$ provide. Instead, if $\mathbf y_u$ satisfies the constraints, the term inside the parenthesis has to be zero for at least one value of $\mathbf s$ and non-positive for the remaining combinations of $\mathbf s$. These quantities have the same role of the parameter $\lambda^{MC}$ defined in Eqs.~(\ref{lambdaMC0}) and~(\ref{MStrans2}). 
Instead, the parameters $\lambda_u$ replace the role of $\lambda$ in Eq.~(\ref{QUBOprob00}). Hence, they have to amplify enough the penalties in order to suppress those $\mathbf y_u$ that violate one or more constraints.

\subsection{Tuning of the relative multipliers and the MS concatenation}
We first discuss the tuning of the relative multipliers $\lambda_{u}^{MC}$ and later we focus on $\lambda_u$.
A straightforward application of the results presented in Section~\ref{tunepenalty} allows us to consider the following definition
\begin{equation}\label{relativemultiplier}
\lambda_{u}^{MC} =  1 +\gamma \max\left\{ 0 , \max_{\mathbf y_u,\mathbf s} \sum_{j=1}^{m_{SC}} P^{SC}_{u,j}(\mathbf y_u,\mathbf s) \right\} \, ,
\end{equation}
where $\gamma>1$. 
Notice that all the relative multipliers of the same node $u$ have the same value.

In what follows, we describe three proposals for the tuning of $\lambda_u$, which we call \textit{local, neighbour} and \textit{global}. Our first approach is to choose a value of $\lambda_u$ such that the minimum penalty applied when at least one constraint in $u$ is violated is (in modulo) greater than the maximum contribution provided by the arcs incident in $u$ to $W(\mathbf x)$. In order to do this, we define $W_u(\mathbf x)$ to be the restriction of $W(\mathbf x)$ to the node $u$. More precisely, if we write $W(\mathbf x) = \sum_{1\leq i \leq j \leq N} W_{ij} x_i x_j$, in $W_u(\mathbf x)$ we include only those terms where either $x_i$, $x_j$ or both belong to $\mathbf y_u$, namely corresponds to arcs in $\mathcal E(u)$. Hence, we define:
\begin{equation}\label{lambdalocal}
\lambda_u^{local}  = \gamma \max_{\mathbf x} W_u(\mathbf x) \, ,
\end{equation}
where $\gamma>1$. 

The next two approaches that we define naturally provide larger values of $\lambda_u$. 
On the one side it helps preventing the ``non-local'' violation of constraints that may occur in presence of a local tuning. In Appendix~\ref{mpbsmulti} we describe this phenomenon in the context of the {\algo{Max-Profit Balanced Settlement}} optimization problem: an NP-hard problem that we analyse in Section~\ref{SecMPBSUC}. Nonetheless, the presence of penalty terms that are too large compared to those found in $W(\mathbf x)$ can suppress the output quality of several QUBO solvers, especially quantum ones. 
Moreover, depending on the structure of $W(\mathbf x)$, the evaluation of these two extra methods may be more demanding (this is not the case for diagonal  $W(\mathbf x)$). Therefore, in this work we privilege the use of the local tuning, and, in case of necessity, we pick larger values of $\gamma$.

The second possible tuning that we discuss extends the previous approach to those nodes neighbouring $u$, namely those sharing an arc with $u$. Similarly to $W_u(\mathbf x)$, we now define $W_u^{neigh}(\mathbf x)$ to be the restriction of $W(\mathbf x)$ to the binary variables associated to arcs incident to $u$ and its neighbouring nodes. Hence, we define:
\begin{equation}\label{lambdaneigh}
\lambda_u^{neigh}  = \gamma \max_{\mathbf x} W^{neigh}_u(\mathbf x) \, ,
\end{equation}
where $\gamma>1$.

The last discussed tuning is labelled as global because it is not node-dependent and it makes sure that the penalty applied is always large enough whenever any constraint is violated. Indeed, it consists in setting:
\begin{equation}\label{lambdaglobal}
\lambda_u^{global}  = \gamma \max_{\mathbf x} W(\mathbf x) \, ,
\end{equation}
where $\gamma>1$. Whereas this is a sure choice to make our QUBO formulation exactly encoding the starting optimization problem, namely they provide the same solution, it has some downturns. Depending on the structure of $W(\mathbf x)$, the optimization $\max_{\mathbf x} W(\mathbf x)$ may have a computational cost that is similar to the whole optimization problem, and therefore evaluating $\lambda_u^{global}$ may become excessively demanding. More importantly, this choice can easily produce a QUBO problem where the penalty terms are (in modulo) extremely larger than the objective function terms. This scenario can deeply affect the solution quality of QUBO solvers and, moreover, we expect that in many common situations this brute-force approach is in fact unnecessary.

To conclude this section, we give an exemplary reformulation of the results regarding the concatenated IQPMS methods to the case of locally-constrained problems. Imagine to have four constraints per node, hence defined over  the same binary variables $\mathbf y_u\subseteq \mathbf x$, and concatenate them as in Section~\ref{concatenation}. We can reformulate the QUBO form~(\ref{QUBOprob}) of this locally-constrained optimization problem as:
\begin{equation}\label{QUBOprobconc}
\mathcal E^* = \arg  \max_{\mathbf x,\mathbf s}  W(\mathbf x)  + \sum_{u=1}^{|\mathcal{V}|} \lambda_u \left({\color{white} \Gamma^\Gamma}\!\!\!\!\!\! \lambda_{u,1} P_{u,1}(\mathbf y_u,\mathbf s) +\lambda_{u,2} P_{u,2}(\mathbf y_u,\mathbf s) +\lambda_{u,3} P_{u,3}(\mathbf y_u,\mathbf s) + P_{u,4}(\mathbf y_u,\mathbf s)  \right) \, ,
\end{equation}
where $P_{u,i}(\mathbf x,\mathbf s)$ is the IQP obtained while considering the constraint $C_{u,i}(\mathbf x)$ in this concatenated scheme, namely as a satellite of all the constraints $C_{u,j}(\mathbf x)$ with $j<i$ and master of those with $j>i$, for $i,j=1,2,3,4$. The tuning of the parameters $\lambda_{u,i}$ is given by the following generalization of  Eq.~(\ref{relativemultiplierconc}):
\begin{equation}\label{relativemultiplierconcc}
\lambda_{u,i} =  1 +\gamma\max\left\{ 0 , \max_{\mathbf x,\mathbf s} \sum_{j>i} P_{u,j}(\mathbf y_u,\mathbf s) \right\}  \, .
\end{equation}
Regarding the multipliers $\lambda_u$ appearing in Eq.~(\ref{QUBOprobconc}), we can adopt the same techniques discussed above in this section.

\section{Case of study: \MPBS}\label{SecMPBSUC}

We present the locally-constrained optimization problem that we focus on in this work, which is inspired by a real-world financial scenario. Consider a network of users having a set of pending receivables that still have to be paid for. Each receivable indicates the details of a precise pending payment and therefore it is defined by its value, the corresponding creditor and debtor users and some additional temporal features, such as the dates the receivable has been submitted and when the corresponding payment falls due. These scenarios are often handled by \textit{receivable funders} who have the role of anticipating the resolution of receivables: users are requested to pay a fee in order to have instant access to a lump sum of capital, which significantly eases the cash-ﬂow issues associated with receivables. Nonetheless, a different strategy for receivable resolution has been proposed in Ref.~\cite{power}, where a \textit{do-ut-des} mechanism is exploited. 
The main goal in the design of this network-based receivable funding is to select the largest-amount subset of receivables for which autonomous payment among users is possible, without involving the funder. This optimization has to be performed under two main constraints.
First, the account balance of each user has to be upper- and lower-bounded by some pre-fixed values before and after the resolution of the chosen receivables. We call this constraint CAP/FLOOR, where the CAP (FLOOR) value corresponds to the upper (lower) bound of the account balance.
Secondly, each user accepts to realize at least one anticipated payment for which they are debtor if and only if they also receive at least one payment corresponding to a receivable for which they are creditor. Hence, this constraint, called IN/OUT, prevents users from only receiving or only executing the payment of receivables.

To reformulate this optimization problem in a more mathematical framework, we say that each day, we have a different set of receivables $\mathcal{R}$ that constitutes the pending transactions, namely those not resolved the day before together with those added in the meantime between the two days. Therefore, each receivable is defined by its amount $w(u,v) > 0$, the corresponding debtor $u$ and creditor $v$, together with several temporal features which we do not exploit in this analysis~\cite{power}. Indeed, we focus on the single-day resolution problem defined by the {\sffamily{R}}-multigraph induced by $\mathcal{R}$, formulated as:

\begin{figure}
\begin{center}
\includegraphics[width=0.99\textwidth]{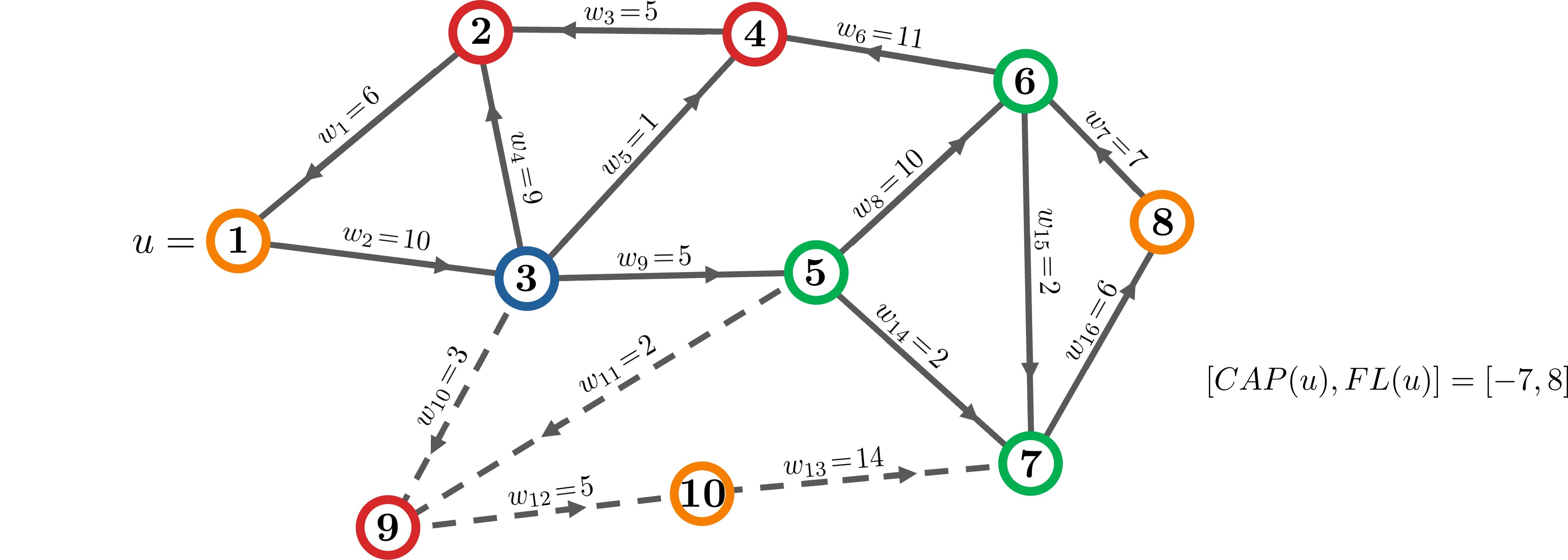}
\caption{Instance of an {\sffamily{R}}-multigraph with $|\mathcal V|=10$ nodes and $N=|\mathcal E|=16$ arcs, where the arrows points from the debtor to the creditor. We have nodes with $N(u)=2$ (orange), 3 (red), 4 (green) and 5 (blue) incident arcs. The goal of \MPBS is to maximize the value exchanged, namely the objective function $W(\mathbf x)=\sum_i x_i w_i$ where the binary variable $x_i$ associated to the $i$-th arc is equal to 0/1 if the corresponding arc is not activated/activated, under two constraints. The first, called CAP/FLOOR, requires each node to exchange a net value inside the $[CAP(u),FL(u)]$ range, which in this case is $[-7,8]$ for each node. For instance, if we activate all the arcs incident to the node $u=1$, we obtain a net value of $w_1-w_2=-4$. The second constraint, called IN/OUT, requires each node to either activate at least one incoming and one outgoing arc or no arcs can be activated for that node. Bold arcs indicate the solution of this \MPBS instance, while the activation of any dashed arc leads to a violation of CAP/FLOOR and/or IN/OUT. Indeed, node $u=10$ cannot activate its incident arcs without violating CAP/FLOOR, as $w_{12}-w_{13}=-9$. As a consequence, the arc $(u,v)=(9,10)$ with value $w_{12}=5$ cannot be activated. Hence, node $u=9$ cannot activate any of its remaining incoming arcs without violating IN/OUT. The solution subgraph $\mathcal E^*$ of this example is given by the combination of binary variables $\mathbf x^*$ such that $x_i^* = 0$ for $i=10,11,12,13$ and $x_i^*=1$ otherwise.}\label{MPBSlore}
\end{center}
\end{figure} 

\begin{definition}[Receivables {\sffamily{R}}-multigraph]
The R-multigraph induced by the set of receivables $\mathcal{R}$ consists of the quadruple~$\mathcal{G}=(\mathcal V,\mathcal E, w,\mathbf f)$, where $\mathcal{V}$ is a set of nodes, $\mathcal{E}$ is a multiset of ordered pairs of nodes called arcs, and $w:\mathcal E\rightarrow \mathbb{R}^+$ assigns positive numerical values to arcs and $\mathbf f:\mathcal V \rightarrow \mathbb R^4$ assigns four attributes to each node. 
The nodes $u\in \mathcal{V}$ represent the users in $\mathcal{R}$, the arcs $(u,v)\in \mathcal E$ represent the receivables in $\mathcal R$, where $u$ represents the corresponding debtor and $v$ the creditor, and $w(u,v)$ is the value of the receivable $(u,v)$. Each node $u\in\mathcal V$ is assigned attributes $\mathbf f(u) = (bl_r(u), bl_a(u), cap(u), fl(u))$, respectively the corresponding receivable balance, actual balance, cap and floor values.
\end{definition}

{Hence, given a set of receivables $\mathcal{R}$, we can identify a corresponding {\sffamily{R}}-multigraph. The aim of the optimization problem that we are about to define is to maximize the overall value of the receivables paid under the condition that the constraints CAP/FLOOR and IN/OUT described above are satisfied by all users. More precisely:}

\begin{definition}[{\algo{Max-Profit Balanced Settlement}}]\label{defMPBS} Given the {\sffamily{R}}-multigraph $\mathcal{G}=(\mathcal V,\mathcal E, w,\mathbf f)$ induced by the receivables $\mathcal R$, find
\begin{eqnarray}
\mathcal{E}^*&=&\arg \max_{\hat{\mathcal{E}}\subseteq \mathcal{E}} \sum_{(u,v)\in\hat{\mathcal{E}}} w(u,v)  \, , \label{MAXPROFIT} \\ 
&\mbox{s.t.}& \,\, {CAP/FLOOR(u)},\,\,\,  \forall u \in \mathcal{V}(\hat{\mathcal{E}}),  \label{capfloor} \\
&& \,\, \mbox{IN/OUT(u)},\,\,\,  \forall u \in \mathcal{V}(\hat{\mathcal{E}}) . \label{inout} 
\end{eqnarray}
where $\hat{\mathcal{E}}$ is a generic subset of all the arcs/transactions $\mathcal E$ and $\mathcal{V}(\hat{\mathcal{E}}) = \{u \in \mathcal{V} \,|\, (u,v)\lor (v,u) \in \hat{\mathcal E} \}$.
Condition~(\ref{capfloor}) confines the balance of the users that take part to the solution multigraph $\mathcal{E}^*$ to be inside a finite region, so that floor and cap conditions are not violated after the realization of the selected transactions:
\begin{equation}\label{capflooru}
CAP/FLOOR (u): \,\, \sum_{v: (v,u)\in \hat{\mathcal{E}} } w(v,u)- \sum_{v: (u,v)\in \hat{\mathcal{E}} } w(u,v) \in [fl(u)-bl_a(u) , cap(u)-bl_r(u)].
\end{equation}
Condition~(\ref{inout}) imposes that each user taking part to the solution multigraph $\mathcal{E}^*$ has to be debtor and creditor for at least one transaction:
\begin{equation}\label{inoutu}
IN/OUT (u): |\{(u,v)\,|\, (u,v) \in \hat{\mathcal{E}}\}| \geq 1, \mbox{ and } |\{(v,u)\,|\, (v,u)\in \hat{\mathcal{E}}\}|\geq 1 \, .
\end{equation}
\end{definition}

{We start by underlying the locally-constrained nature of this optimization problem: each constraint is defined node-by-node and only depends on the arcs incident to that same node.} Secondly, the \algo{Max-Profit Balanced Settlement} (\algo{MPBS}) problem is NP-hard as it can be reduced from the \algo{SubsetSum} problem~\cite{power}, which is a special case of the \algo{Knapsack} problem. 
{This problem presents the same structure of other well-known minimum-cost circulation problems, which have been extensively studied for decades (see, e.g.,~\cite{Goldenberg}).}
In the following, we show how the IQPMS methods can be used to reduce the amount of slack variables needed for a QUBO formulation of \MPBS with respect to the standard procedure of Section \ref{QUBOintro}. A simple example of this problem is depicted in Figure \ref{MPBSlore}.

It is obvious that, if a node $u$ has either no incoming or outgoing arc in the input {\sffamily{R}}-multigraph, the optimal solution of the corresponding \MPBS problem cannot include any arc incident to $u$, otherwise IN/OUT($u$) would be violated. Moreover, the exclusion of all the arcs incident to $u$ may cause another node $v$ to not have any incoming or outgoing arc. In the following, in order to make our exposition clearer, we assume that all the nodes of the considered input {\sffamily{R}}-multigraphs have at least one incoming and one outgoing arc.

\subsection{QUBO formulation of \MPBS}\label{MPBSQUBOzero}

In order to cast \algo{MPBS} into a QUBO form, we proceed as follows. First, we associate to each arc $(u,v)\in \mathcal{E}$ a binary variable $x_i\in\{0,1\}$. Hence, we obtain a string of $N$ binary variables $\mathbf{x}=(x_1,x_2,\dots,x_N)$, where $|\mathcal{E}|=N$ is the total number of arcs in $\mathcal E$. 
Given the one-to-one relation $(u,v) \leftrightarrow i$ we can rewrite any expression in Definition \ref{defMPBS} with the binary variables $x_i$. For instance, if we use the definition $W=\mbox{diag} (w_1,w_2,\dots , w_{N})$, the maximization of the objective function of \MPBS can be formulated as:
\begin{equation}\label{targetQUBO}
\max_{\hat{\mathcal{E}}\subseteq \mathcal{E}} \sum_{(u,v)\in\hat{\mathcal{E}}} w(u,v)  = \max_{\mathbf x\in \{0,1\}^{N} } \sum_i x_i w_i \, = \max_{\mathbf x\in \{0,1\}^{N} } \mathbf{x}^T W \mathbf{x}= \max_{\mathbf x\in \{0,1\}^{N} } W(\mathbf x),
\end{equation} 
where we forced the previous notation by defining: $W(\mathbf x)=  \mathbf{x}^T W \mathbf{x}$. From now on, we solely consider the quadratic polynomial form of QUBO elements, e.g., $W(\mathbf x)$, instead of the corresponding matricial form, e.g., $\mathbf{x}^T W \mathbf{x}$.

Our goal is to achieve a QUBO form of \algo{MPBS} defined over the $N$ logical variables $\mathbf x = \{x_1,\dots,x_N\}$ and $\overline N$ slack binary variables $\mathbf s=\{s_1,\dots,s_{\overline N}\}$, hence considering the optimization of a quadratic form $Q(\mathcal G,\mathbf x,\mathbf s)$ defined over $N+\overline N$ variables $\{\mathbf x,\mathbf s\}$ that corresponds to the instance of \algo{MPBS} identified by $\mathcal G$:
\begin{equation}\label{QUBOtutorial}
 \arg\, \max_{\mathbf \mathbf x,\mathbf s } Q(\mathcal G,\mathbf x,\mathbf s)  \, .
\end{equation}
 Given the one-to-one relation between subsets of arcs $\hat{\mathcal{E}} \subseteq \mathcal{E}$ and binary variables combinations $\mathbf{x}\in\{0,1\}^N$, we define:
\begin{equation}\label{solution}
\mathcal{E}^* \, \longleftrightarrow \, \mathbf{x}^* \, ,
\end{equation}
where $\mathcal{E}^*$ is the solution of \MPBS. 
We say that the quadratic polynomial $Q(\mathcal{G},\mathbf x,\mathbf s)$ provides a QUBO formulation of \MPBS if
\begin{equation}\label{solution0}
 \arg\, \max_{\mathbf x,\mathbf s} Q(\mathcal{G},\mathbf x,\mathbf s) = \{\mathbf{x}^*,\mathbf{s}^*\} \, .
\end{equation}
Hence, in order to achieve this formulation, we must translate the constraints of \MPBS, namely CAP/FLOOR and IN/OUT, through quadratic penalty terms $P^{CF}(\mathbf x,\mathbf s)$ and $P^{IO}(\mathbf x,\mathbf s)$, respectively, in order to obtain:
\begin{equation}\label{QUBOMPBS} 
\arg\, \max_{\mathbf x,\mathbf s }  Q(\mathcal{G}, \mathbf x,\mathbf s) 
=\arg\, \max_{\mathbf x,\mathbf s} W (\mathbf x) + P^{CF}(\mathbf x,\mathbf s) + P^{IO} (\mathbf x,\mathbf s ) = \{\mathbf x^*,\mathbf s^*\}.
\end{equation}

 \MPBS is locally-constrained. Indeed, constraints are clearly defined node-by-node. As we show later, we can achieve the following QUBO structure similar to Eq.~(\ref{QUBOprob}):
\begin{eqnarray}\label{QUBOprobMPBSx}
Q(\mathcal G,\mathbf x,\mathbf s) =    W(\mathbf x)  + \sum_{u=1}^{|\mathcal{V}|} \lambda_u \left( \lambda^{IO}_{u} P^{IO}_{u}(\mathbf x,\mathbf s) +  P^{CF}_{u} (\mathbf x,\mathbf s)   \right) \, ,
\end{eqnarray} 
where IN/OUT($u$) is considered the master constraint of the node $u$, while CAP/FLOOR($u$) as the corresponding satellite constraint.

We define $\mathcal E(u)=\mathcal{E}^+(u) \cup\mathcal{E}^-(u)$ to be the subset of arcs incident to the node $u$, where $\mathcal{E}^+(u)$ is the subset of incoming arcs $(v,u)$ where $u$ represents the creditor and $\mathcal{E}^-(u)$ is the subset of outgoing arcs $(u,v)$ where $u$ represents the debtor. Hence, $|\mathcal{E}(u)|=|\mathcal{E}^+(u)|+|\mathcal{E}^-(u)|$. We use the notation $N(u)=|\mathcal E (u)|$ and $N^{\pm}(u)=|\mathcal{E}^{\pm}(u)|$.
Moreover, we use $\mathcal E(u)$ and $\mathcal{E}^{\pm}(u)$ to indicate specific subsets of the binary variables $\mathbf x$. For instance, ``$i\in \mathcal{E}^+(u)$'' stands for all the values of $i$ such that $x_i$ corresponds to a $(v,u)$ arc. Instead, $\overline N(u)$ corresponds to the amount of slack variable needed to enforce a constraint associated to the node $u$, where the specific constraint will be clear from the context. We remember that each constraint is enforced with independent slack variables.

We conclude this section by expressing the CAP/FLOOR  and IN/OUT  conditions with the binary variables $x_i$ associated to the arcs/pending transactions $(u,v)\in\mathcal E$. First, it is straightforward to check that Eq.~(\ref{capflooru})  corresponds to:
\begin{equation}\label{capfloori}
\mbox{CAP/FLOOR($u$)}:\,\, FL(u) \leq \sum_{i\in \mathcal{E}^+(u)} x_i w_i  - \sum_{i\in \mathcal{E}^- (u)} x_i w_i \leq CAP(u) \, ,
\end{equation}
where $FL(u)=fl(u) - bl_a(u)$ and $CAP(u)=cap(u)-bl_r(u)$. Moreover, the IN/OUT  condition for the node $u$ given in Eq.~(\ref{inoutu}) is equivalent to the following pair of inequality constraints:
\begin{equation}\label{inouti}
\mbox{IN/OUT($u$)}:\,\, \left\{
\begin{array}{ccc}
IC^{IO}_{u,1}(\mathbf x)= -\left( |\mathcal{E}^-(u)| \sum_{i\in \mathcal{E}^+(u)} x_i \right) + \sum_{i\in \mathcal{E}^-(u)} x_i \leq 0 \, , \\ 
IC^{IO}_{u,2}(\mathbf x)= -\left( |\mathcal{E}^+(u)| \sum_{i\in \mathcal{E}^-(u)} x_i \right)+ \sum_{i\in \mathcal{E}^+(u)} x_i \leq 0 \, .
\end{array}
\right.
\end{equation}

\subsection{Standard QUBO formulation of IN/OUT}\label{inoutslack}

\begin{table}
$$
\begin{array}{|c|c|c|}
\hline\,\,\,\,\, \mbox{ Arcs }\,\,\,\,\, & \mbox{ Scenario } &  \,\,\,\mbox{ IN/OUT (standard) }\,\,\, \\ \hline
N(u)=2 & \mbox{1vs1}  & \overline N(u) =0  \\ \hline
N(u)=3 & \mbox{1vs2}  & \overline N(u) =2  \\ \hline
N(u)=4 & \begin{array}{c} \mbox{1vs3} \\ \mbox{2vs2} \end{array}  & \begin{array}{c} \overline N(u) =4 \\ \overline N(u) =4 \end{array}  \\ \hline
N(u)=5 & \begin{array}{c} \mbox{1vs4} \\ \mbox{2vs3} \end{array}  & \begin{array}{c} \overline N(u) =4 \\ \overline N(u) =6 \end{array}  \\ \hline
  \end{array}
$$
\caption{Amount  of slack variables $\overline N(u)$ necessary to the standard method to enforce IN/OUT($u$), where $N(u)$ is the number of arcs. The $X$vs$Y$ scenario occurs when on the target node there are $X$ arcs of one type (incoming/outgoing) and $Y$ of the opposite type.}\label{TABELLONEIO}
\end{table} 

We start by showing how the standard approach discussed in Section~\ref{QUBOintro} applies to cast the IN/OUT constraints into a QUBO form. 
Each IN/OUT constraint in Eq.~(\ref{inouti}) can be considered as a different inequality constraint $IC^{IO}_{u,j}(\mathbf x) \leq 0$, with $j=1,2$, of a locally-constrained problem (see Eq.~(\ref{locoptprob})). 

It is easy to check that, for each $u$, the IN/OUT inequality constraints have the same minimum  $\min_{\mathbf x} IC_{u,j} (\mathbf x)= - |\mathcal{E}^+(u)|\, |\mathcal{E}^-(u)|$.
{A straight application of the standard procedure described in Section \ref{QUBOintro} would involve the introduction of a quantity analogous to $S_j(\mathbf s)$ (see Eq. (\ref{QUBOprob0})), where the amount $\overline N(u)$ of slack variables $\mathbf s = \{s_1,s_2,\dots,s_{\overline N(u)}\}$ is chosen such that $S_j(\mathbf s)$ can assume all the integer values in the interval $[0, |\min_{\mathbf x} IC^{IO}_{u,j}(\mathbf x)| ] = [0, |\mathcal{E}^+(u)|\, |\mathcal{E}^-(u)|]$. In general, the larger is this interval, the more slack variables are necessary   (see Eq.~(\ref{overlineN})). For instance, we could enforce  $IC^{IO}_{u,1}(\mathbf x) \leq 0$ into a QUBO form through:
\begin{equation}\label{IOslack}
 P^{IO,1}_u(\mathbf x,\mathbf s) =- \left(IC^{IO}_{u,1}(\mathbf x) + S_1(\mathbf s) \right)^2 = - \left( -\left( |\mathcal{E}^+(u)| \sum_{i\in \mathcal{E}^-(u)} x_i \right) + \sum_{i\in \mathcal{E}^+(u)} x_i +  S_1(\mathbf{s}) \right)^2 ,
\end{equation}
where $S_1(\mathbf{s})$ is given by Eq.~(\ref{S}) and assumes all the integer values in $[0,|\mathcal{E}^+(u)|\, |\mathcal{E}^-(u)| ]$. Nonetheless, this path would employ more slack variables than necessary.}

{
Indeed, the combinations of $\mathbf x$ for which $IC^{IO}_{u,1}(\mathbf x)= - |\mathcal{E}^+(u)|\, |\mathcal{E}^-(u)|$ violate $IC^{IO}_{u,2}(\mathbf x)\leq 0$ and, similarly,  the combinations of $\mathbf x$ for which $IC^{IO}_{u,2}(\mathbf x)= - |\mathcal{E}^+(u)|\, |\mathcal{E}^-(u)|$ violate $IC^{IO}_{u,1}(\mathbf x) \leq 0$. Hence, requiring  $S_1(\mathbf s)$ to be able to assume the value $|\mathcal{E}^+(u)|\, |\mathcal{E}^-(u)|$, so that Eq. (\ref{IOslack}) can be zero-valued when $IC^{IO}_{u,1}(\mathbf x)  = -|\mathcal{E}^+(u)|\, |\mathcal{E}^-(u)|$, is not necessary, as there would be anyway a violation of the other inequality constraint $IC^{IO}_{u,2}(\mathbf x)\leq 0$.
 Instead, the same is not true for those $\mathbf x$ such that $IC^{IO}_{u,1}(\mathbf x)$ or $IC^{IO}_{u,2}(\mathbf x)$ are equal to the second-lowest possible value, namely $1 - |\mathcal{E}^+(u)|\, |\mathcal{E}^-(u)|$. 
 
 As a consequence, we consider the QUBO quadratic penalty that corresponds to $IC^{IO}_{u,1}(\mathbf x) \leq 0$ as Eq. (\ref{IOslack}), 
}
where $S_1(\mathbf{s})$ is given by Eq.~(\ref{S}) and assumes all the integer values in $[0,|\mathcal{E}^+(u)|\, |\mathcal{E}^-(u)| - 1]$.
Hence, the implementation of the first IN/OUT inequality constraint requires $\lceil\log_2 ( |\mathcal{E}^+(u)|\, |\mathcal{E}^-(u)| )\rceil$
slack variables (see Eq. (\ref{overlineN})).

In summary, in order to implement a quadratic penalty term that enforces IN/OUT($u$) in a QUBO formulation of \MPBS, we can use the standard technique described in Section~\ref{QUBOintro} and construct $P^{IO}_u(\mathbf x,\mathbf s)= P^{IO,1}_u(\mathbf x,\mathbf s) + P^{IO,2}_u(\mathbf x,\mathbf s)$, where $P^{IO,1}_u(\mathbf x,\mathbf s)$ is given in Eq.~(\ref{IOslack}) and $P^{IO,2}_u(\mathbf x,\mathbf s)$ can be obtained similarly. We underline that, while these two addendum share the same logical binary variables, the slack variables are disjoint, namely those involved in $P^{IO,1}_u(\mathbf x,\mathbf s)$ do not enter in the definition of $P^{IO,2}_u(\mathbf x,\mathbf s)$ and vice-versa.
The slack variable cost of this implementation for each node is:
$$
\overline N^{IO}(u)= 2 \left\lceil\log_2 \left( \left|\mathcal{E}^+(u)\right|\, \left|\mathcal{E}^-(u)\right| \right)\right\rceil \, .
$$ 
Hence, the number of slack variables needed to implement the IN/OUT constraint in the whole {\sffamily{R}}-multigraph is:
\begin{equation}\label{NIO}
\overline N^{IO}=2 \sum_{u\in \mathcal{V}(u)} \left\lceil \log_2 \left|\mathcal{E}^+(u)\right|\, \left|\mathcal{E}^-(u)\right|  \right\rceil \, .
\end{equation}
In Table~\ref{TABELLONEIO} we report the amount of slack variables required by the standard method to enforce IN/OUT($u$) in a QUBO.

{We notice that the amount of slack variables needed for the QUBO formulation of this constraint strictly depends on the topology of the multigraph considered. Indeed, the amount of slack variables given by Eq.~(\ref{NIO}) is a function of the number of incoming and outgoing arcs of each node.  }

\subsection{Standard QUBO formulation of  CAP/FLOOR}\label{capfloorslack}
We proceed by applying the standard method discussed in Section~\ref{QUBOintro} to convert CAP/FLOOR($u$) into a QUBO form.
This constraint can be considered as one of the linear inequality constraints in Eq.~(\ref{optprob}), where now $S(\mathbf s)$ has to assume all the integer values in $[0,CAP(u)-FL(u)]$.
Hence, to obtain the corresponding QUBO form, we use Eq.~(\ref{S}) to define the matrix $P^{CF}_u$ that appears in Eq.~(\ref{QUBOMPBS}) and we obtain:
\begin{equation}\label{CFslack}
P^{CF}_u(\mathbf x,\mathbf s)=- \left( \sum_{i\in \mathcal{E}^+(u)} w_i x_i -\sum_{i\in \mathcal{E}^-(u)} w_i x_i  - CAP(u) +  S(\mathbf{s}) \right)^2 .
\end{equation}
Hence, the implementation of this quadratic penalty term requires 
$\overline N\!{\color{white}.}^{CF}(u) = \lceil \log_2 (1+ CAP(u) -FL(u))\rceil $
slack variables for each node (see Eq.~(\ref{overlineN})) and therefore a total of
\begin{equation}\label{NCF}
\overline N^{CF}= \sum_{u\in \mathcal{V}} \left\lceil \log_2 (1+ CAP(u) -FL(u))\right\rceil
\end{equation}
  slack variables to implement the CAP/FLOOR constraint in the whole  {\sffamily{R}}-multigraph. 

{Hence, when adopting the standard method presented in Section \ref{QUBOintro}, differently from the IN/OUT case, the slack variable cost for this constraint does not depend on the topology of the graph, i.e., the amount of incoming and outgoing arcs per node, but it only depends on the values of $CAP(u)$ and $FL(u)$. Notice that, in case of multigraphs taken from real-world financial scenarios, the order of magnitude of $CAP(u) - FL(u)$ can easily be around $10^5 \sim 10^6$. Therefore, the amount of slack variables $\overline N\!{\color{white}.}^{CF}(u)$ needed per node, even in case of few incoming and outgoing transactions, can be larger than $16$. Later, when considering real instances of \MPBS, we set $CAP(u) -FL(u)=15$ in order to need 4 slack variables per node for the enforcement of this constraint.}

\subsection{\MPBS on quantum annealers}

We can plug the newly formulated quadratic penalties for CAP/FLOOR and IN/OUT given respectively in Eqs.~(\ref{CFslack}) and~(\ref{IOslack}), into Eq.~(\ref{QUBOprobMPBSx}) and obtain the standard QUBO formulation of \MPBS. 
As a result, the number of binary variables over which we optimize is given by $N=|\mathcal{E}|$ variables $x_i$, which are associated to the arcs of the input {\sffamily{R}}-multigraph, ${\overline N}\!{\color{white}.}^{IO}$ slack variables to enforce the IN/OUT (see Eq.~(\ref{NIO})) and $\overline N\!{\color{white}.}^{CF}$ for CAP/FLOOR (see Eq.~(\ref{NCF})).

One possibility to solve a QUBO problem with quantum technologies is via a quantum annealer, e.g., those commercially produced by D-wave \cite{Dwavereport}. In the best-case scenario, namely when each QUBO variable is not coupled with too many variables and the topology of multigraph of arcs $\mathcal{E}$ is particularly fortunate, D-wave annealers employ around one qubit per QUBO variable. 
Hence, by using the standard technique described above, the qubits employed to include CAP/FLOOR and IN/OUT are in general much more than those representing the actual transactions. As a result, any attempt to use classical or quantum QUBO solvers in order to solve \MPBS for larger and larger {\sffamily{R}}-multigraphs is highly limited by the number of slack variables required by this procedure.


While in the near future quantum processing units with more qubits and higher connectivities may be introduced and therefore better computing performances may be achieved, any reduction of the slack variables needed to implement \MPBS as a QUBO makes the use of quantum annealers more suitable to solve instances of this problem. In case of more advantageous strategies, such reductions could give the opportunity to approach real-world problems with a larger input size.
An approach that could be adopted together with the slack variable reduction in order to efficiently use quantum annealers is given by splitting large QUBO problems in smaller sub-QUBOs, e.g., by considering the algorithm \algo{qbsolv}~\cite{qbsolv}. This method allows to provide sub-optimal solutions to QUBO problems that would not fit entirely on a quantum chip.

In the following sections we first analyse the features of the realistic {\sffamily{R}}-multigraphs studied in Ref.~\cite{power2020} and later we show why they are particularly suited  for the application of our newly introduced techniques, namely the IQPMS methods. 
Real-world instances of \MPBS are indeed characterized by a low number of arcs per node and therefore, as discussed in Section~\ref{GQPmaster}, the adoption of IQPMS is particularly effective.
Step-by-step, we show how a drastic slack variables reduction is possible. Thanks to our approach, all the nodes with 3 or less arcs require no slack variables in order for IN/OUT($u$) and CAP/FLOOR($u$) to be imposed in a QUBO form. Moreover, for 4-arc nodes, we need only 1 slack variable to implement IN/OUT($u$). Instead, for nodes with 5 arcs, we need either 1 or 2 slack variables for IN/OUT($u$) and from 0 to 2 slack variables for CAP/FLOOR($u$). {We summarize these slack variable costs in Table \ref{TABELLONE}. Hence, compared to the standard method presented in Section \ref{QUBOintro}, our methods implement significantly less slack variables and this amount depends only on the multigraph topology: no dependency from the problem parameters occurs. We remind that the standard method enforces CAP/FLOOR($u$) in a QUBO form with a slack variable cost that directly depends on $CAP(u) - FL(u)$ (see Eq.~(\ref{NCF})).}
 Finally, we provide various approaches to tune the corresponding penalty multipliers.

\subsection{Realistic datasets}\label{dataset}

In~\cite{power,power2020} the authors show how to obtain sub-optimal solutions of \MPBS for real scenarios obtained by the pan-European bank company UniCredit. The authors consider time-dependent {\sffamily{R}}-multigraph, which vary day-by-day depending on the transactions resolved each day and the newly submitted ones. The average number of arcs per node obtained by averaging over a  time span of two years is (see Table 2 from~\cite{power2020}):
\begin{equation}\label{arcsxnode}
 2 \frac{\langle |\mathcal{E}|\rangle_{DATA}}{\langle |\mathcal{V}|\rangle_{DATA}}\in [2.04,\, 2.77],
\end{equation}
where these averages vary depending on the particular problem parameters chosen, e.g., cap($u$), fl($u$) and the lifetime that each receivable is allowed to stay in the {\sffamily{R}}-multigraph.
Instead, the single days with the largest ratios are such that:
\begin{equation}\label{arcsxnodemax}
2 \frac{\langle |\mathcal{E}|\rangle_{MAX}}{\langle |\mathcal{V}|\rangle_{MAX}}\in [2.80,\, 4.59].
\end{equation}
The authors of~\cite{power2020} also analyse the cases where the CAP constraint is dropped (while FLOOR is still enforced), namely for cap($u$)$=+\infty$. Nonetheless, the corresponding values of Eqs.~(\ref{arcsxnode}) and (\ref{arcsxnodemax}) do not change significantly.

It is evident that a technique that allows a reduction of the number of slack variables needed to enforce \MPBS constraints for nodes with a number of arcs inside the ranges described here would be of fundamental importance in order to use QUBO solvers. As anticipated, we indeed realize such a protocol which works nicely for nodes with 5 or less arcs. All the nodes with 6 or more arcs should be treated separately and would in general need more slack variables, but the techniques introduced in this work, namely the IQPMS methods, can still be exploited.

 \section{IQPMS methods for \MPBS}\label{IQPMSMPBSsec}
 
 We show how the techniques introduced in this work, namely the IQPMS methods, apply to obtain a QUBO representation of \MPBS. We apply the IQPMS methods, where IN/OUT($u$) is enforced as master constraint and CAP/FLOOR($u$) as satellite. 
We summarize the results of this section in Table~\ref{TABELLONE}, where we provide the number of slack variables needed by our techniques to enforce the constraints of \MPBS in the QUBO form.
{As we show, the quadratic polynomials found by the IQPMS methods enforcing IN/OUT($u$) and CAP/FLOOR($u$) can be non-completely connected, namely some of the parameters $a_{ij},\, \overline{a}_{kl},\, b_{ik}$ defining the corresponding polynomials~(\ref{Pquad0}) can be null. Hence, if the quadratic forms enforcing IN/OUT($u$) and CAP/FLOOR($u$) do not couple the same two or more variables for one or more node $u$, we get a QUBO where the logical variables are not completely connected and therefore solving such a problem is in general easier than in the completely-connected case. Similarly, the same applies to the slack variables, which are not completely connected to each other and to the logical variables.
Remember that the standard method in general employs several more slack variables and it completely connects all the logical and slack variables to each other.}

\begin{table} 
$$
\begin{array}{|c|c|c|c|} 
\hline\,\,\,\,\, \mbox{ Arcs }\,\,\,\,\, & \mbox{ Scenario } &  \,\,\,\mbox{ IN/OUT (IQPMS) }\,\,\, &\,\,\mbox{ CAP/FLOOR (IQPMS) }\,\, \\ \hline
N(u)=2 & \mbox{1vs1}  & \overline N(u) =0 & \overline N(u) =0 \\ \hline
N(u)=3 & \mbox{1vs2}  & \overline N(u) =0 & \overline N(u) =0 \\ \hline
N(u)=4 & \begin{array}{c} \mbox{1vs3} \\ \mbox{2vs2} \end{array}  & \begin{array}{c} \overline N(u) =1 \\ \overline N (u)=1 \end{array} & \begin{array}{c}  \overline N (u)=0 \\  \overline N(u) =0 \end{array} \\ \hline
N(u)=5 & \begin{array}{c} \mbox{1vs4} \\ \mbox{2vs3} \end{array}  & \begin{array}{c} \overline N(u) =1 \\ \overline N (u)=2 \end{array} & \begin{array}{c}  \overline N (u)=0,1 \\  \overline N(u) =0,1,2 \end{array} \\ \hline
  \end{array}
$$
\caption{Amount  of slack variables $\overline N(u)$ necessary to enforce IN/OUT($u$) and CAP/FLOOR($u$) with the IQPMS methods, where $N(u)$ is the number of arcs. IN/OUT($u$) is considered as a master constraint and CAP/FLOOR($u$) as its satellite. The $X$vs$Y$ scenario occurs when on the target node there are $X$ arcs of one type (incoming/outgoing) and $Y$ of the opposite type. {Comparing the slack variable cost of the IQPMS methods with the standard method, the slack variable cost of our methods is significantly lower and solely depends on the topology of the problem multigraph (see Table \ref{TABELLONEIO} and Section \ref{capfloorslack} for a comparison). Indeed, differently from the case of CAP/FLOOR($u$) enforced in a QUBO form with the standard method (see Eq. (\ref{NCF})), $\overline N(u)$ never depends on the values of the problem parameters. }}\label{TABELLONE}
\end{table}

 \subsection{Reduction of slack variables for IN/OUT}\label{INOUTnoslack}

An interesting feature of this constraint is that it is homogenous among the graph, namely, modulo a relabelling of the binary variables involved in $u$, all the nodes having the same values of $|\mathcal E^+(u)|$ and $|\mathcal E^-(u)|$ have the same $P^{IO}_u(\mathbf x, \mathbf s)$. Moreover, the same is true if we swap the values of $|\mathcal E^+(u)|$ and $|\mathcal E^-(u)|$: the IN/OUT penalty term for a node with $|\mathcal E^+(u)|=2$ and $|\mathcal E^-(u)|=1$ can be simply obtained from the IN/OUT penalty given to another node having $|\mathcal E^+(u)|=1$ and $|\mathcal E^-(u)|=2$.
Indeed, this penalty requires each node to have either none or at least one incoming and one outgoing arc, while the values associated to each arc $w_i$ and the attributes $CAP(u)$ and $FL(u)$ do not influence this constraint. This regularity is the main reason why we pick IN/OUT as master: once that we enforce this constraint for a given pair of $|\mathcal E^+(u)|$ and $|\mathcal E^-(u)|$, we automatically obtain the IN/OUT($u$) penalty term for all the nodes with either the same $|\mathcal E^+(u)|$ and $|\mathcal E^-(u)|$ or swapped values of $|\mathcal E^+(u)|$ and $|\mathcal E^-(u)|$.

We proceed by exposing how the IQP method for master constraints applies to IN/OUT. As anticipated, this method is particularly effective when $N(u)$ is not too large and therefore we fix our attention to those nodes with $N(u)\leq 5$ arcs.

{\bf  Nodes with 2 arcs: }
Consider a node with $N(u)=2$. Without loss of generality, we associate the binary variable $x_1$ to the incoming arc and $x_2$ to the outgoing one.
The IQP method provides the following solution with no slack variables
\begin{equation}
P^{IO}_u(\mathbf x)=-x_1-x_2+2x_1x_2 = \left\{
\begin{array}{cccc}
0 &  \mbox{ if } &  \mathbf x= \{0,0\}, \, \{1,1\} \\ 
-1 &  \mbox{ if } &  \mathbf x= \{1,0\}, \, \{0,1\} 
\end{array}
\right. \, ,
\end{equation}
where $\{x_1,x_2\}= \{0,0\} , \{1,1\}$ are the combinations that satisfy IN/OUT, while  $\{x_1,x_2\}= \{1,0\} , \{0,1\}$ violate IN/OUT.   \\

{ \bf Nodes with 3 arcs: } Consider a node with $N(u)=3$. We either have $|\mathcal E^+(u)|=1$ and $|\mathcal E^-(u)|=2$ or  $|\mathcal E^+(u)|=2$ and $|\mathcal E^-(u)|=1$. 
Hence, we simply say that we have a 1vs2 scenario and associate $x_1$ to the type of arc that appears only once. 
More precisely, if $|\mathcal E^+(u) |= 1$, we associate $x_1$ to the incoming arc and $x_{2,3}$ to the outgoing ones. Otherwise, we associate $x_1$ to the outgoing arc and $x_{2,3}$ to the incoming ones when $|\mathcal E^-(u) |= 1$.

Interestingly, even if the free parameters in the IQP with $N=3$ are fewer than the independent conditions that have to be applied on $P^{IO}_u(\mathbf x)$, respectively $\#a_{ij} = 7 < M(u) = 2^{N(u)} = 8$, the IQP method provides infinite many solutions with no slack variables. Remember that the standard method required 2 slack variables to enforce the same constraint (see Section~\ref{inoutslack}). Hence, we replace some inequality sub-constraints with equality sub-constraints in Eq.~(\ref{genconds0})  in order to have a larger penalty for the string that violates IN/OUT and is likely to provide a large incentive to the objective function $\sum_i w_i x_i$ of the QUBO formulation of \MPBS, namely for $\{x_1,x_2,x_3\} = \{0,1,1\}$.
Hence, the following linear system of 8 equations in 7 variables
\begin{equation}\label{sysforIO3}
P^{IO}_u(\mathbf x)=a_0+ \sum_{1 \leq  i\leq j \leq 3} a_{ij} x_i x_j 
 \hspace{0.4cm} \mbox{ s.t. }\hspace{0.4cm} \left\{
\begin{array}{cccccc}
P^{IO}_u(\mathbf x) =  0 & \mbox{ if } &  \mathbf x_u= \{0,0,0\},& \!\!\{1,1,0\},&\!\!\{1,0,1\},&\!\!\{1,1,1\}  \\ 
P^{IO}_u(\mathbf x)  = -1 &  \mbox{ if } &  \mathbf x_u= \{1,0,0\},&\!\!\{0,1,0\},&\!\! \{0,0,1\} \\ 
P^{IO}_u(\mathbf x) \leq  -1 & \mbox{ if } &  \!\!\mathbf x_u= \{0,1,1\} \\ 
\end{array}
\right. \, , 
\end{equation}
provides the following unique solution:
\begin{equation}\label{IO3}
P^{IO}_u(\mathbf x)= -x_1-x_2-x_3+2 x_1 x_2 +2 x_1 x_3 - x_2 x_3=\left\{
\begin{array}{cccccc}
 0 & \mbox{ if } &  \mathbf x= \{0,0,0\},& \!\!\{1,1,0\},&\!\!\{1,0,1\},&\!\!\{1,1,1\} \\ 
 -1 &  \mbox{ if } &  \mathbf x= \{1,0,0\},&\!\!\{0,1,0\},&\!\! \{0,0,1\} \\ 
 -3 & \mbox{ if } & \mathbf x= \{0,1,1\} \\ 
\end{array}
\right. \, .
\end{equation}

{ \bf Nodes with 4 arcs: } 
Consider a node with $N(u)=4$. In the following, the 1vs3 case refers to the $|\mathcal E^+(u)|=1$ and $|\mathcal E^-(u)|=3$ scenario and  the $|\mathcal E^+(u)|=3$ and $|\mathcal E^-(u)|=1$ scenario. Instead, the 2vs2 case refers to the $|\mathcal E^+(u)|=2$ and $|\mathcal E^-(u)|=2$ scenario.

The IQP method needs 1 slack variable in order to enforce IN/OUT as a QUBO penalty term for this type of nodes, which we call $s_1$. Remember that the standard method required 4 slack variables to enforce the same constraint (see Section~\ref{inoutslack}). Before providing the explicit solutions obtained for $P^{IO}_u(\mathbf x, s_1)$, we explain the ordering chosen for the binary variables.  Concerning the 1vs3 case, we associate $x_1$ to the type of arc that appears only once in $u$, while $x_{2,3,4}$ are associated to the other type of arcs~\footnote{For instance, if $|\mathcal E^+(u)|=3$ and $|\mathcal E^-(u)|=1$, we associate $x_1$ to the only outgoing arc of the node.}. Instead, for the 2vs2 case, $x_{1,2}$ are associated to the incoming arcs and $x_{3,4}$ to the outgoing ones.

Finally, the corresponding quadratic forms are:
\begin{eqnarray}
&\mbox{ (1vs3)}& \hspace{1cm} P^{IO}_u(\mathbf x,s_1) = -x_1 -x_2 -x_3 -x_4 +x_1 x_2 + 2 x_1 x_3 + x_1 x_4 - x_2 x_4 - s_1 + s_1 x_1  + s_1 x_2  - s_1 x_3  + s_1 x_4 \, , \\ 
&\mbox{ (2vs2)}& \hspace{1cm} P^{IO}_u(\mathbf x,s_1) = -x_1 - x_2 - x_3 - x_4 - x_1 x_2  - x_3 x_4 + s_1 + 2 s_1 x_1  + 2 s_1 x_2  + 2 s_1 x_3  + 2 s_1 x_4  \, .  
\end{eqnarray}
Concerning the sparsity of these penalty terms, we notice that the zero-valued parameters in the solution IQPs~(\ref{Pquad0}) are $a_{2,3}=a_{3,4}=a_0=0$ in the 1vs3 scenario and  $a_{1,3}=a_{1,4}= a_{2,3}=a_{2,4}=a_0=0$ in the 2vs2 scenario.

{ \bf Nodes with 5 arcs: } Consider a node with $N(u)=5$ arcs. In the following, the 1vs4 case refers to the $|\mathcal E^+(u)|=1$ and $|\mathcal E^-(u)|=4$ scenario and the $|\mathcal E^+(u)|=4$ and $|\mathcal E^-(u)|=1$ scenario. Similarly, the 2vs3 case refers to the $|\mathcal E^+(u)|=2$ and $|\mathcal E^-(u)|=3$ scenario and the $|\mathcal E^+(u)|=3$ and $|\mathcal E^-(u)|=2$ scenario.

The IQP method applied to the 1vs4 case allows a solution with 1 slack variable, which we call $s_1$, while for the 2vs3 case the IQP method requires 2 slack variables, which we call $s_1$ and $s_2$. 
Remember that the standard method required 4 and 6 slack variables, respectively, to enforce the same constraint (see Section~\ref{inoutslack}).
Before providing the explicit solutions obtained for $P^{IO}_u(\mathbf x, \mathbf s)$, we explain the ordering chosen for the binary variables.  
Concerning the 1vs4 case, we associate $x_1$ to the type of arc that appears only once in $u$, while $x_{2,3,4,5}$ are associated to the other type of arcs. 
Instead, for the 2vs3 case, $x_{1,2}$ are associated to the type of arcs that appears twice in $u$, while $x_{3,4,5}$ are associated to the other type of arcs. 

Finally, the corresponding quadratic forms are:
\begin{eqnarray}
&\hspace{-0.45cm} \mbox{ (1vs4)} \hspace{0.7cm} P^{IO}_u(\mathbf x,s_1) =& \hspace{-0.45cm} -x_1 -x_2 -x_3 -x_4 - x_5+2x_1 x_2 +2 x_1 x_3 + 2 x_1 x_4 + 2x_1 x_5 -x_2 x_3 - x_2 x_4 - x_2 x_5 - x_3 x_4    \\
& & \hspace{-0.45cm} - x_3 x_5 - x_4 x_5 - 5 s_1 + 2 s_1 x_2 + 2 s_1 x_3 + 2 s_1 x_4 + 2 s_1 x_5 \, , \nonumber \\  
&\mbox{ (2vs3)} \hspace{0.7cm} P^{IO}_u(\mathbf x,s_1,s_2) =& -x_1 -x_2 -x_3 -x_4 - x_5 - x_1 x_2 - x_3 x_4 - x_3 x_5 - x_4 x_5 -2 s_1 -2 s_2 +2 s_1 x_1 +2 s_1 x_2  \\ 
& &  +2 s_1 x_3 +2 s_1 x_4 +2 s_1 x_5 + s_2 x_3 + s_2 x_4 + s_2 x_5  \nonumber   \, .  
\end{eqnarray}
Concerning the sparsity of these penalty terms, we notice that the zero-valued parameters in the solution IQPs~(\ref{Pquad0}) are $b_{11}=0$ in the 1vs4 scenario and $a_{13}=a_{14}=a_{15}=a_{23}=a_{24}=a_{25}=\overline a_{12} = b_{12} = b_{22} = 0$ in the 2vs3 scenario.


\subsection{Reduction of slack variables for CAP/FLOOR}\label{CAPFLOORnoslack}

We follow by enforcing CAP/FLOOR in the QUBO form as a satellite constraint of IN/OUT, which has already been treated as a master constraint. In this section, we construct  the quadratic penalty terms $P^{CF}_u(\mathbf x,\mathbf s)$ with the IQP method by requiring that it provides penalties if $\mathbf x$ satisfies IN/OUT($u$) and violates CAP/FLOOR($u$). Moreover, this quadratic form has to give no-penalties for at least one combination of $\mathbf s$ whenever $\mathbf x$ satisfies IN/OUT($u$) and CAP/FLOOR($u$). In other terms, we require $P^{CF}_u(\mathbf x,\mathbf s)$ to penalize the combinations violating CAP/FLOOR($u$) solely when they already satisfy IN/OUT($u$) (see Section~\ref{GQPsatellite}). It follows that such a $P^{CF}_u(\mathbf x,\mathbf s)$ could provide accidental incentives when IN/OUT($u$) is violated and therefore we have to proceed as described in Section~\ref{tunepenalty}.
 Hence, we look for a IQP~(\ref{Pquad0}) that satisfies the linear system~(\ref{gencondssat}), which now assumes the form:
 \begin{equation}\label{gencondssatX}
\left\{
\begin{array}{ccc}
P^{CF}_u(\mathbf x,\mathbf s)\leq - 1 &  \mbox{ if }  \mathbf x \mbox{ satisfies IN/OUT($u$) and violates CAP/FLOOR($u$) } & \hspace{-1.4cm} \mbox{for all } \mathbf s \\ 
P^{CF}_u(\mathbf x,\mathbf s) \leq  0 &  \mbox{ if }  \mathbf x \mbox{ satisfies IN/OUT($u$)  and satisfies CAP/FLOOR($u$) } &  \hspace{-1.4cm} \mbox{for all } \mathbf s \\ 
P^{CF}_u(\mathbf x,\mathbf s) =  0 &  \mbox{ if }  \mathbf x \mbox{ satisfies IN/OUT($u$)  and satisfies CAP/FLOOR($u$) } & \hspace{-0.2cm} \mbox{for at least one } \mathbf s
\end{array}\right. \, 
\end{equation}

Before proceeding, we briefly analyse the slack variables cost expected for IQP method in this scenario. The total number of free parameters inside a IQP grows quadratically with $N(u)+\overline N(u)$ (see Eq.~(\ref{degreesslack0})), where $\overline N(u)$ is the number of slack variables employed. In case of master constraints, the total number of combinations of $\mathbf x$ for which we have to enforce sub-constraints are $2^{N(u)}$. The advantage of considering CAP/FLOOR($u$) a satellite constraint of IN/OUT($u$) is that we can ignore those sub-constraints violating IN/OUT($u$), which are $2^{N^+(u)} + 2^{N^-(u)}-2$~\footnote{The number of combinations of $\mathbf x$ such that there are no outgoing (incoming) arcs activated are $2^{N^+(u)}$. The number of combinations of $\mathbf x$ such that there are no incoming arcs activated are $2^{N^-(u)}$. Since the null combination $\mathbf x = \{0,0,\dots,0\}$ is allowed by IN/OUT($u$), we obtain that the number of combinations violating IN/OUT($u$) are $2^{N^+(u)} + 2^{N^-(u)}-2$.}. Hence, the total number of sub-constraints that have to be imposed for CAP/FLOOR($u$) are:
\begin{equation}\label{numbersubCF}
\# cond (u) = 2^{N(u)} - 2^{N^+(u)} - 2^{N^-(u)} +2 \, .
\end{equation}

As discussed in Section~\ref{GQPmaster}, an indicative estimate for the minimum number of slack variables necessary to IQP method to enforce a given constraint is given by that $\overline N(u)$ such that the number of free parameters (see Eq.~(\ref{degreesslack0})) is larger than the number of sub-constraints (see Eq.~(\ref{numbersubCF})), namely for:
$$
\# \{a_{ij}, \overline a_{kl} , b_{ik}\} = 1 + \frac{(N(u)+\overline N(u))^2 + N(u)+ \overline N(u)}{2}   > \#cond(u) = 2^{N(u)} - 2^{N^+(u)} - 2^{N^-(u)} +2 \, .
$$
From the values of $\#\{a_{ij}, \overline a_{kl} , b_{ik}\}$ and $\# cond(u)$ reported in Table~\ref{2Naij}, we expect that one or more slack variables are necessary to enforce CAP/FLOOR($u$) in a quadratic form with the IQP method only for $N(u)\geq 5$. As we show later, this is indeed true. Moreover, by generating several random graphs, a major part of the 5-arc nodes studied required either zero or one slack variable and very few needed two slack variables.

\begin{table}[t]
$$
\begin{array}{|c|c|c|} \hline 
 \mbox{ Arcs } &  \#\{a_{ij},\overline a_{kl},b_{ik}\}  &  \# cond   \\ \hline
N(u) = 2 & (\overline N(u)=0 ) \,\,\, 4 & 2 \\ \hline
N(u) = 3 & (\overline N(u)=0 ) \,\,\, 7 & 4 \\ \hline
N(u) = 4 & (\overline N(u)=0 ) \,\,\, 11 & \begin{array}{c} \!\mbox{(1vs3)}\,\,\,8 \\ \mbox{(2vs2)}\,10 \end{array} \\ \hline
N(u) = 5 & \begin{array}{c}  (\overline N(u)=0 ) \,\,\, 16 \\  (\overline N(u)=1 ) \,\,\, 22 \\  (\overline N(u)=2 ) \,\,\, 29 \end{array} \,\,\, & \begin{array}{c} \mbox{(1vs4)}\,\,\,16 \\ \mbox{(2vs3)}\,\,\,22 \end{array} \\ \hline
  \end{array}
$$
\caption{Comparison between the number of free parameters $\#\{a_{ij},\overline a_{kl},b_{ik}\}$ in an IQP for different number of arcs $N(u)$ and slack variables $\overline N(u)$ with the number of sub-constraints $\#cond(u)$ required by CAP/FLOOR($u$) when enforced as a satellite constraint of IN/OUT($u$). The indicative minimum amount of slack variables necessary is given when $\#\{a_{ij},\overline a_{kl},b_{ik}\}>\# cond(u)$. No slack variables are indeed necessary for $N(u)\leq 4$, while for $N(u)=5$ one or two may be needed. The $X$vs$Y$ scenario occurs when on the target node there are $X$ arcs of one type (incoming/outgoing) and $Y$ of the opposite type.}\label{2Naij}
\end{table}

In case of $N(u)=2,3,4$ we need no slack variables to enforce CAP/FLOOR($u$) in the QUBO form with our methods and moreover we have a closed form for the corresponding polynomials $P^{CF}_u(\mathbf x)$. 
Moreover, since the linear systems~(\ref{gencondssatX}) leave free many parameters of the IQP with no slack variables, we can insert back some sub-constraints violating IN/OUT($u$) while still obtaining a solution in a closed form. In particular, for $N(u)=2$ we can add 2 conditions back, and therefore we do not need to assume IN/OUT($u$) to be satisfied, for $N(u)=3$ we can add 3 conditions back and  for $N(u)=4$ we can add 3 conditions back in the 1vs3 scenario and 1 in the 2vs2 scenario.
Instead, for what concerns the $N(u)=5$ case, we could not find a closed form of $P^{CF}_u(\mathbf x,\mathbf s)$. Therefore, in the numerical simulations presented below, the corresponding IQPs have been calculated node by node. In addition, while looking for these polynomials, we incentivised solutions with fewer slack variables and with sparser representative matrices.

In the following, we provide the quadratic forms $P^{CF}_u(\mathbf x)$ that are the solutions of the linear system~(\ref{gencondssatX}) for $N(u)=2,3,4$. These penalties solely depend on the particular combinations $\mathbf x$ that satisfy IN/OUT($u$) and violate CAP/FLOOR($u$). Hence, given a node $u$, we simply have to check which are these combinations and make a simple substitution inside a parametrized polynomial, without the need to solve any linear system. We remember that for this constraint the standard method required $\overline N\!{\color{white}.}^{CF}(u) = \lceil \log_2 (1+ CAP(u) -FL(u))\rceil $ slack variables for each node. Consider that, if we consider realistic scenarios as those analysed in~\cite{power2020}, $CAP(u) -FL(u)$ can easily be greater than $10^6$.

{ \bf Nodes with 2 arcs: } In this case CAP/FLOOR($u$) and IN/OUT($u$) can be enforced at the same time with a single polynomial. It is enough to find an IQP in two variables satisfying the following linear system:
$$ \left\{
\begin{array}{ccc}
P^{IO,CF}_u (0,0) =  0   \\
P^{IO,CF}_u (x_1,x_2) =  -1 & \mbox{ for } & \,\{x_1,x_2\}=  \{1,0\}, \{0,1\}   \\ 
P^{IO,CF}_u (1,1) =   \sigma(1,1)  \\ 
\end{array}
\right. 
$$
where $\sigma(1,1) = 0 $ if $\{x_1,x_2\}=\{1,1\}$ satisfies CAP/FLOOR($u$) and $\sigma(1,1) = -1 $ otherwise.
Remember that $\{x_1,x_2\}=\{0,0\}$ always satisfies both CAP/FLOOR($u$) and IN/OUT($u$) while $\{x_1,x_2\}=\{1,0\}$ and $\{x_1,x_2\}=\{0,1\}$ always violate IN/OUT($u$). 
If $\{x_1,x_2\}=\{1,1\}$ satisfies CAP/FLOOR($u$), we obtain $P^{IO,CF}_u (x_1,x_2)=-x_1-x_2+2x_1 x_2$. Instead, if $\{x_1,x_2\}=\{1,1\}$ violates CAP/FLOOR, we have $P^{IO,CF}_u (x_1,x_2)=-x_1-x_2+x_1 x_2$. Indeed, we can express the solution IQP as follows:
\begin{equation}
P^{IO,CF}_u(\mathbf x) = -x_1-x_2 + (2+\sigma_{11}) x_1 x_2 \, .
\end{equation}

{ \bf Nodes with 3 arcs: } We consider the ordering introduced in Section~\ref{INOUTnoslack} for which $x_1$ is associated to the type of arc that appears only once in $u$, while $x_{2,3}$ correspond to the other type of arcs.
Similarly to the $N(u)=2$ case, we express the linear system~(\ref{gencondssatX}) as a function of the CAP/FLOOR($u$) penalties/no-penalties $\sigma(x_1,x_2,x_3)$ that have to be assigned to those combinations $\{x_1,x_2,x_3\}$ satisfying IN/OUT($u$). In particular, we consider the system:
\begin{equation}\label{SISTEMONE3} \left\{
\begin{array}{ccc}
P^{CF}_u (0,0,0) =  0  \\
P^{CF}_u (x_1,x_2,x_3) =  -1 & \mbox{ for } & \{x_1,x_2,x_3\}=  \{0,1,1\},  \{0,0,1\}, \{1,0,0\}   \\ 
P^{CF}_u (x_1,x_2,x_3) =   \sigma(x_1,x_2,x_3) & \mbox{ for } & \{x_1,x_2,x_3\}=\{1,1,0\},\{1,0,1\},\{1,1,1\} 
\end{array}
\right. 
\end{equation}
where $\sigma(x_1,x_2,x_3) = 0 $ if $\{x_1,x_2,x_3\}$ satisfies CAP/FLOOR($u$) and $\sigma(x_1,x_2,x_3) = -1 $ otherwise. Notice that, as anticipated, we added three extra sub-constraints in order to completely determine all the free parameters of $P^{CF}_u(\mathbf x)$. Indeed, we required penalties for $\{x_1,x_2,x_3\}=  \{0,1,1\},  \{0,0,1\}, \{1,0,0\}$, which violate IN/OUT($u$). The solution of such system is:
\begin{eqnarray}
& P^{CF}_u(\mathbf x) =& -x_1 + (1+\sigma(1,0,1)+ \sigma(1,1,0)+ \sigma(1,1,1)) x_2 - x_3 + (\sigma(1,1,1)- \sigma(1,0,1)) x_1 x_2 + (2+\sigma(1,0,1)) x_1x_3  \nonumber \\
& & + ( \sigma(1,1,1)-\sigma(1,0,1)-\sigma(1,1,0)-1 ) x_2 x_3 \, .
\end{eqnarray}

{ \bf Nodes with 4 arcs: } We consider the ordering introduced in Section~\ref{INOUTnoslack} for which, in the 1vs3 scenario $x_1$ is associated to the type of arc that appears only once in $u$, while $x_{2,3,4}$ correspond to the other type of arcs. Instead, for the 2vs2 scenario we consider $x_{1,2}$ associated to the incoming arcs, while $x_{3,4}$ are associated to the outgoing ones. 
Similarly to the $N(u)=2$ case, we express the linear system~(\ref{gencondssatX}) as a function of the CAP/FLOOR($u$) penalties/no-penalties $\sigma(x_1,x_2,x_3,x_4)$ that have to be assigned to those combinations $\{x_1,x_2,x_3,x_4\}$ satisfying IN/OUT($u$). In particular, for the 1vs3 case we consider the system:
\begin{equation}\label{SISTEMONE4} \left\{
\begin{array}{ccc}
P^{CF}_u (0,0,0,0) =  0   \\
P^{CF}_u (x_1,x_2,x_3,x_4) =  -1 & \mbox{ for } & \,\,\{x_1,x_2,x_3,x_4\}=  \{0,1,1,0\}, \{0,0,1,1\}, \{0,1,1,1\}  \\
P^{CF}_u (x_1,x_2,x_3,x_4) =   \sigma(x_1,x_2,x_3,x_4) & \mbox{ for } & \begin{array}{c} \{x_1,x_2,x_3,x_4\}=\{1,1,0,0\},\{1,0,1,0\},\{1,1,1,0\},\\ 
  \{1,0,0,1\},\{1,1,0,1\},\{1,0,1,1\},\{1,1,1,1\} \end{array}
\end{array}
\right. 
\end{equation}
where $\sigma(x_1,x_2,x_3,x_4) = 0 $ if $\{x_1,x_2,x_3,x_4\}$ satisfies CAP/FLOOR($u$) and $\sigma(x_1,x_2,x_3,x_4) = -1 $ otherwise. Notice that we added three extra penalty sub-constraints for $\{x_1,x_2,x_3,x_4\}=  \{0,1,1,0\},  \{0,0,1,1\}, \{0,1,1,1\}$, which violate IN/OUT($u$). The solution of such system is given in Appendix~\ref{FORMULOZZI}.

Instead, for the 2vs2 case we consider the system:
\begin{equation}\label{SISTEMONE42} \left\{
\begin{array}{ccc}
P^{CF}_u (0,0,0,0) =  0   \\
P^{CF}_u (0,0,1,1) =  -1  \\
P^{CF}_u (x_1,x_2,x_3,x_4) =   \sigma(x_1,x_2,x_3,x_4) & \mbox{ for } & \begin{array}{c} \{x_1,x_2,x_3,x_4\}=\{1,0,1,0\},\{0,1,1,0\},\{1,1,1,0\}, \{1,0,0,1\}, \\  \{0,1,0,1\},\{1,1,0,1\}, \{1,0,1,1\},\{0,1,1,1\},\{1,1,1,1\} \end{array} 
\end{array}
\right. 
\end{equation}
where $\sigma(x_1,x_2,x_3,x_4) $ is defined analogously. Notice that we added the extra penalty sub-constraint for $\{x_1,x_2,x_3,x_4\}=  \{0,0,1,1\}$, which violates IN/OUT($u$). The solution of such system is given in Appendix~\ref{FORMULOZZI}. 

{\bf Nodes with 5 arcs: } We randomly generated multigraphs and, by using the IQP method for satellite constraints, we never needed more than 1 slack variable to enforce CAP/FLOOR($u$) in the 1vs4 scenario and 2 slack variables for the 2vs3 scenario. Moreover, we often encountered instances where even less slack variables were needed, namely 0 slack variables in the 1vs4 scenario and 0 or 1 slack variables in the 2vs3 scenario. Differently from the $N(u)=2,3,4$ cases, we could not achieve a closed form for $P^{CF}_u(\mathbf x,\mathbf s)$ parametrized by some quantities analogous to the $\sigma(\mathbf x)$ considered above. Hence, node by node, we had to solve the linear system~(\ref{gencondssatX}).

 \subsection{Tuning of \MPBS penalty multipliers}\label{lambdas}
 
 We showed how to implement CAP/FLOOR($u$) and IN/OUT($u$) conditions with a reduced number of slack variables, if compared to the standard method, for nodes $u\in \mathcal{V}$ with $N(u)\leq 5$ arcs. In this section we show how to set the multipliers $\lambda_u$ and $\lambda^{IO}_u$ in order to be sure that the combinations $\mathbf{x}$ that solve the maximization in the QUBO encoding the \MPBS problem satisfy CAP/FLOOR($u$) and IN/OUT($u$). In other terms, we want $Q(\mathcal{G},\mathbf x, \mathbf s)$ to faithfully represent \MPBS, and therefore to satisfy Eq.~(\ref{QUBOMPBS}). In the following, we use the corresponding single-node contribution:
 \begin{equation}\label{uQUBO}
 Q_u(\mathcal{G},\mathbf x,\mathbf s) =   \left( \sum_{i\in \mathcal{E}(u)} w_i x_i \right) + \lambda_u \left( \lambda^{IO}_u P^{IO}_u(\mathbf x,\mathbf s) + P^{CF}_u(\mathbf x,\mathbf s) \right) \, .
 \end{equation}
First, we show how to set $\lambda_u$ and then we proceed with $\lambda^{IO}_u$. From now on, we assume that our QUBO formulation of \MPBS is made with the IQPMS methods.

The tuning of these multipliers has to take into account the possibility of particularly unfortunate scenarios that may occur. Hence, $\lambda_u$ and $\lambda^{IO}_u$ have to be set large enough so that also these scenarios get the right penalty factors that allow to obtain the correct optimal result $\mathbf x^*$ (see Eq.~(\ref{solution})). We explained a possible approach to this problem in Section~\ref{tunepenalty}, where we did not assume any particular structure for the target locally-constrained problem. By studying a particular problem, such as \MPBS, it may occur that its particular features allow to set smaller multipliers. While in the following we limit to apply the formulas obtained in Section~\ref{tunepenalty}, in Appendix \ref{mpbsmulti} we prove that the worst-case requirements that could be find within instances of \MPBS require these same multiplier bounds.

First of all, we can set the IN/OUT($u$) relative multiplier $\lambda^{IO}_u$ by using Eq.~(\ref{relativemultiplier}).  The role of the satellite master and satellite constraints are now played by IN/OUT and CAP/FLOOR, respectively. Therefore, we obtain:
\begin{equation}
\lambda^{IO}_u=  1 + \gamma \max\left\{0, \max_{\mathbf x,\mathbf s} P^{CF}_{u}(\mathbf x,\mathbf s)\right\} \, 
\end{equation}
where $\gamma>1$.
For what concerns $\lambda_u$, the local, neighbour and global tunings described in Section~\ref{tunepenalty} leads, respectively, to: 
\begin{eqnarray}\label{lambdalocal}
&\lambda_u^{local}=&\gamma \, \sum_{i\in \mathcal{E}(u)} w_i = \lambda w(u) \, , \\ 
\label{lambdaneigh}&\lambda_u^{niegh}=&\gamma \, \sum_{v\in \mbox{\scriptsize neigh}(u)} \sum_{i\in \mathcal{E}(v)} w_i \, , \\ 
&\lambda_u^{global}=&\gamma \, \sum_{i \in \mathcal E} w_i \, ,
\end{eqnarray}
 where $\gamma>1$ and neigh($u$) is the set of nodes sharing at least one arc with $u$. We defined the total amount of the transactions associated to $u$ as $ w(u) = \sum_{i\in \mathcal{E}(u)} w_i$. Instead, $\lambda_u^{neigh}$ is proportional to the sum of the transactions in $u$ and its neighbouring nodes. Finally, $\lambda^{global}_u$ is node-independent and is proportional to the total amount of transactions in the graph. 
 
Concerning the possibility to consider a global tuning of $\lambda_u$, this choice surely imposes CAP/FLOOR($u$) and IN/OUT($u$) for all the nodes $u$ of the  {\sffamily{R}}-multigraph and therefore, if we modify $Q(\mathcal G,\mathbf x,\mathbf s)$ accordingly to these new multipliers, we get  $\{\mathbf x^* ,\mathbf  s^*\} = \arg (\max_{\mathbf x,\mathbf s} Q(\mathcal G,\mathbf x,\mathbf s))$. Indeed, as soon as one node violates either CAP/FLOOR($u$) or IN/OUT($u$), a penalty greater (in modulo) than the value of all the transactions is applied and therefore we are ensured that  CAP/FLOOR($u$) or IN/OUT($u$) are satisfied for all the graph nodes.

Nonetheless, this approach generates very large penalties which may compromise the quality of the solution due to the evident unbalance between the weight of the incentives and the penalties for each $\mathbf x$. Indeed, large unbalances between the terms of a QUBO may be problematic when using quantum annealers, e.g., D-wave.
For these reasons, we believe that Eqs.~(\ref{lambdalocal}) and~(\ref{lambdaneigh}) are the best pick to start studying the \MPBS problem within this QUBO framework.

\section{\MPBS simulations on D-wave quantum annealers}\label{DwaveMPBS}

We tested how the standard approach to translate constraints in a quadratic form compares to the IQPMS methods when they are adopted to solve instances of \MPBS with D-wave quantum annealers, which are mainly designed to solve QUBO problems. 
We generated several random {\sffamily{R}}-multigraphs $\mathcal G$ with different numbers of arcs and nodes. 
Each $\mathcal G$ identified a different \MPBS problem, which we solved exactly by exhaustion, namely by checking all the possible combinations of $\mathbf x$ in Eq.~(\ref{optprob}).
Hence, we used the standard and the IQPMS methods to realize a QUBO formulation of each \MPBS problem and solved them with two different D-wave quantum annealers, Advantage\_system4.1 (Adv1) and the prototype Advantage2\_prototype1.1 (Adv2), respectively with 5627 and 563 operating qubits. {The corresponding connectivities, namely the maximum number of qubits interacting with each qubit, are equal to 15 for Adv1 (Pegasus topology) and 20 for Adv2 (Zephyr topology) \cite{Dwavereport}. }Finally, we compared the rates of success for which these annealers were able to find the corresponding optimal solutions together with the analysis of other performance figure of merits. Since all the instances have been solved on both quantum annealers, our results allow to compare their performances. For another comparative study of D-wave annealers, see, e.g., Ref.~\cite{Djidjev2}.

We follow by describing the features of the {\sffamily{R}}-multigraphs generated. 
We considered 5 different arc-settings, each one with a different number of arcs, which were $N= 10,12,14,16,18$. 
We made this choice because, how we see later, within this range we can appreciate how the standard method goes from being satisfactory to almost useless. 
For each setting, we generated 4 instances of {\sffamily{R}}-multigraphs with different numbers of nodes, for a total of 20 instances.
Hence, fixed a setting, or number of arcs, the different instances had different arcs/nodes ratios, ranging from 1.5 to 2.2. Hence, the average number of arcs incident to each node was $\langle N(u)\rangle \in[3,4.4]$. 
In order to generate an {\sffamily{R}}-multigraph with a given number of arcs and nodes, we first randomly distributed the minimal amount of arcs that allowed each node to have at least one incoming and one outgoing arc. The remaining arcs were then distributed randomly until the desired number of arcs was achieved, where we prevented this procedure to generate nodes with $N(u)\geq 6$. 
We assigned integer values $w_i$ to arcs which were randomly sampled from the interval $[1,18]$. 
Finally, we chose $FL(u)=-7$ and $CAP(u)=8$ for all the generated instances.

\begin{table} 
$$
\begin{array}{>{\centering\arraybackslash$} p{1.5cm} <{$}>{\centering\arraybackslash$} p{2.2cm} <{$}>{\centering\arraybackslash$} p{2.55cm} <{$}>{\centering\arraybackslash$} p{2.55cm} <{$}>{\centering\arraybackslash$} p{2.55cm} <{$}}  
 \mbox{Arcs}& \begin{array}{>{\centering\arraybackslash$} p{2.5cm} <{$}}\mbox{Instance} \end{array} &  \begin{array}{>{\centering\arraybackslash$} p{2.8cm} <{$}} \begin{array}{c}  \mathbf x\mbox{ satisfying} \\ \mbox{\MPBS constraints} \end{array}\end{array} &    \begin{array}{>{\centering\arraybackslash$} p{2.8cm} <{$}} \begin{array}{c} \mbox{QUBO variables} \\ \mbox{\,\, (standard)} \end{array}\end{array}  &   \begin{array}{>{\centering\arraybackslash$} p{2.8cm} <{$}} \begin{array}{c} \mbox{QUBO variables} \\ \mbox{\,\, (IQPMS)} \end{array} \end{array} \\   \\ 
  N = 10 &  \begin{array}{>{\centering\arraybackslash$} p{2.5cm} <{$}} \bcv\,\,\, |\mathcal{V}|= 6 \,\,\, \\ \bcd |\mathcal{V}|= 6 \\ \bcv |\mathcal{V}|= 5 \\ \bcd |\mathcal{V}|= 5 \end{array} & \begin{array}{>{\centering\arraybackslash$} p{2.8cm} <{$}} \bcv 18\,\,\, (1.758\%) \\ \bcd 16\,\,\, (1.563\%) \\ \bcv 7\,\,\, (0.684\%) \\ \bcd 9\,\,\, (0.879\%) \end{array} & \begin{array}{>{\centering\arraybackslash$} p{2.8cm} <{$} } \bcv 50  \\ \bcd 50 \\ \bcv 50 \\ \bcd 50 \end{array} & \begin{array}{>{\centering\arraybackslash$} p{2.8cm} <{$}} \bcv 14  \\ \bcd 15 \\ \bcv 17 \\ \bcd 15 \end{array}  \\  
 N = 12 & \begin{array}{>{\centering\arraybackslash$} p{2.5cm} <{$}} \rcv |\mathcal{V}|= 8  \\ \rcd |\mathcal{V}|= 7 \\ \rcv |\mathcal{V}|= 7 \\ \rcd |\mathcal{V}|= 6 \end{array}& \begin{array}{>{\centering\arraybackslash$} p{2.8cm} <{$}} \rcv 7\,\,\, (0.171\%) \\ \rcd 35\,\,\, (0.854\%) \\ \rcv 5\,\,\, (0.122\%) \\ \rcd 17\,\,\, (0.415\%) \end{array} & \begin{array}{>{\centering\arraybackslash$} p{2.8cm} <{$}} \rcv 60  \\ \rcd 60 \\ \rcv 60 \\ \rcd 60 \end{array} & \begin{array}{>{\centering\arraybackslash$} p{2.8cm} <{$}} \rcv 15  \\ \rcd 16 \\ \rcv 16 \\ \rcd 18 \end{array}  \\  
 N = 14 & \begin{array}{>{\centering\arraybackslash$} p{2.5cm} <{$}} \bcv |\mathcal{V}|= 9  \\ \bcd |\mathcal{V}|= 8 \\ \bcv |\mathcal{V}|= 8 \\ \bcd |\mathcal{V}|= 7 \end{array} & \begin{array}{>{\centering\arraybackslash$} p{2.8cm} <{$}} \bcv26\,\,\, (0.159\%) \\ \bcd 8\,\,\, (0.049\%) \\ \bcv 7\,\,\, (0.043\%) \\ \bcd 41\,\,\, (0.250\%) \end{array}& \begin{array}{>{\centering\arraybackslash$} p{2.8cm} <{$}} \bcv 70  \\ \bcd 70 \\ \bcv 70 \\ \bcd 68 \end{array} & \begin{array}{>{\centering\arraybackslash$} p{2.8cm} <{$}} \bcv 18  \\ \bcd 20 \\ \bcv  22 \\ \bcd 21 \end{array}  \\  
 N = 16 & \begin{array}{>{\centering\arraybackslash$} p{2.5cm} <{$}} \rcv |\mathcal{V}|= 10  \\ \rcd|\mathcal{V}|= 9 \\ \rcv |\mathcal{V}|= 9 \\\rcd |\mathcal{V}|= 8 \end{array} & \begin{array}{>{\centering\arraybackslash$} p{2.8cm} <{$}} \rcv 31\,\,\, (0.047\%) \\ \rcd 32\,\,\, (0.049\%) \\ \rcv 16\,\,\, (0.024\%) \\ \rcd 38\,\,\, (0.060\%) \end{array}& \begin{array}{>{\centering\arraybackslash$} p{2.8cm} <{$}} \rcv 80  \\ \rcd 78 \\ \rcv 80 \\ \rcd 80 \end{array} & \begin{array}{>{\centering\arraybackslash$} p{2.8cm} <{$}} \rcv 20  \\ \rcd 23 \\ \rcv 23 \\ \rcd 27 \end{array}  \\  
 N = 18 & \begin{array}{>{\centering\arraybackslash$} p{2.5cm} <{$}} \bcv |\mathcal{V}|= 12  \\ \bcd |\mathcal{V}|= 11 \\ \bcv |\mathcal{V}|= 10 \\ \bcd|\mathcal{V}|= 9 \end{array} & \begin{array}{>{\centering\arraybackslash$} p{2.8cm} <{$}} \bcv 32\,\,\, (0.012\%) \\ \bcd 28\,\,\, (0.011\%) \\ \bcv 8\,\,\, (0.003\%) \\ \bcd 14\,\,\, (0.005\%) \end{array}& \begin{array}{>{\centering\arraybackslash$} p{2.8cm} <{$}} \bcv 86  \\ \bcd 90 \\ \bcv 90 \\ \bcd 90 \end{array} & \begin{array}{>{\centering\arraybackslash$} p{2.8cm} <{$}} \bcv  21  \\ \bcd 24 \\ \bcv  26 \\ \bcd 28 \end{array}  \\  
  \end{array}
$$
\caption{Main features of the generated \MPBS instances, labelled by the corresponding number of arcs $N$ and nodes $|\mathcal V|$. 
For each instance, we provide the number of combinations $\mathbf x$ that satisfy the \MPBS constraints and the corresponding percentage with respect the total number of possible combinations, namely $2^N$. Finally, we report the number of variables $N+\overline N$ (logical and slack) that were needed to obtain a QUBO representation with the standard and the IQPMS methods. {Hence, the total number of slack variables employed by the standard and the IQPMS methods is obtained by subtracting from the corresponding QUBO variables column the number of arcs $N$ of the corresponding instance. By considering all the generated instances, IQPMS needed from 82.5\% (first instance with $N=10$ and $|\mathcal V|=5$) to 95.6\% ($N=18$ and $|V|=12$) less slack variables than the standard method.} }\label{satcomb}
\end{table}

After the generation of these \MPBS problems, we exactly solved them by exhaustion. 
Hence, for each instance we collected the corresponding optimal  combination $\mathbf x^*$ and the optimal value $W^* = \sum_i w_i x^*_i$ (see Section~\ref{MPBSQUBOzero}). Moreover, in order to estimate how restrictive the constraints of this problem are and what is the probability to randomly sample a combination of $\mathbf x$ that simply satisfies our  constraints, we explicitly count how many combinations satisfy IN/OUT($u$) and CAP/FLOOR($u$) for all~$u$.
While the possible combinations grow exponentially with $N$, the amount of $\mathbf x$ that do not lead to a constraint violation did not grow appreciably. After this analysis, we generated the QUBO representations of all the problems both with the standard and the IQPMS methods. Finally, the tuning of the penalty multipliers was executed as described in the previous sections, where we adopted the local approach and  we chose $\gamma=2$. The main features of the instances generated together with the corresponding amount of combinations $\mathbf x$ satisfying the constraints and the number of variables needed to generate the described QUBO representations can be found in Table~\ref{satcomb}. Moreover, in Figure~\ref{grafico_statistiche}(left) we show how the average number of QUBO variables needed changes when we increase the number of arcs in the instances considered. We show that the standard method needs approximately 4.88 variables per arc added to the graph, while the IQPMS methods need only 1.30 variables.

\begin{table} 
$$
\hspace{-0.5cm} \begin{array}{>{\centering\arraybackslash$} p{1cm} <{$}>{\centering\arraybackslash$} p{1.6cm} <{$}>{\centering\arraybackslash$} p{1.6cm} <{$}>{\centering\arraybackslash$} p{2.5cm} <{$}>{\centering\arraybackslash$} p{2.5cm} <{$}>{\centering\arraybackslash$} p{1.6cm} <{$}>{\centering\arraybackslash$} p{2.5cm} <{$}>{\centering\arraybackslash$} p{2.5cm} <{$}}
\mbox{Arcs} &
\begin{array}{>{\centering\arraybackslash$} p{2cm} <{$}} \mbox{Annealer} \end{array} &  
\begin{array}{>{\centering\arraybackslash$} p{2cm} <{$}} 	\langle\mbox{qubits}\rangle \\ \mbox{(standard)} \end{array} &
\begin{array}{>{\centering\arraybackslash$} p{2.9cm} <{$}} \begin{array}{c} \langle\mbox{qubits/variable}\rangle \\ \mbox{\,\, (standard)} \end{array}\end{array} &
\begin{array}{>{\centering\arraybackslash$} p{2.9cm} <{$}} \langle\mbox{max. chain length}\rangle \\ \mbox{(standard)} \end{array} &
\begin{array}{>{\centering\arraybackslash$} p{2cm} <{$}} \langle\mbox{qubits}\rangle \\ \mbox{(IQPMS)} \end{array} &  
 \begin{array}{>{\centering\arraybackslash$} p{2.9cm} <{$}} \begin{array}{c} \langle\mbox{qubits/variable}\rangle \\ \mbox{\,\, (IQPMS)} \end{array} \end{array} &
\begin{array}{>{\centering\arraybackslash$} p{2.9cm} <{$}}  \langle\mbox{max. chain length}\rangle \\ \mbox{(IQPMS)} \end{array} \\  \\ 
N = 10 & 
\begin{array}{>{\centering\arraybackslash$} p{2cm} <{$}} \bcv \mbox{Adv1} \\ \bcd \mbox{Adv2} \end{array} &  
\begin{array}{|>{\centering\arraybackslash$} p{2cm} <{$}} 101.75 \bcv  \\ 94.38 \bcd  \end{array} &
\begin{array}{>{\centering\arraybackslash$} p{2.9cm} <{$}} \bcv 2.04 \\ \bcd 1.89 \end{array} &
\begin{array}{>{\centering\arraybackslash$} p{2.9cm} <{$}} 5.88\bcv  \\ 4.63 \bcd  \end{array} &
\begin{array}{|>{\centering\arraybackslash$} p{2cm} <{$}} 23.00 \bcv  \\20.50 \bcd  \end{array} &
\begin{array}{>{\centering\arraybackslash$} p{2.9cm} <{$}}1.51 \bcv  \\ 1.34\bcd  \end{array} &
\begin{array}{>{\centering\arraybackslash$} p{2.9cm} <{$}}2.25 \bcv  \\ 2\bcd  \end{array} \\ 
N = 12 & 
\begin{array}{>{\centering\arraybackslash$} p{2cm} <{$}} \rcv \mbox{Adv1} \\ \rcd \mbox{Adv2} \end{array} &  
\begin{array}{|>{\centering\arraybackslash$} p{2cm} <{$}} 112.13\rcv  \\ \rcd 106.63 \end{array} &
\begin{array}{>{\centering\arraybackslash$} p{2.9cm} <{$}} 1.87 \rcv  \\ 1.78\rcd  \end{array} &
\begin{array}{>{\centering\arraybackslash$} p{2.9cm} <{$}} 5.38\rcv  \\ 4.63\rcd  \end{array} &
\begin{array}{|>{\centering\arraybackslash$} p{2cm} <{$}} 21.75\rcv  \\20.50 \rcd  \end{array} &
\begin{array}{>{\centering\arraybackslash$} p{2.9cm} <{$}} \rcv 1.34 \\ 1.26\rcd  \end{array} &
\begin{array}{>{\centering\arraybackslash$} p{2.9cm} <{$}} 2.13 \rcv  \\ 2\rcd  \end{array} \\
N = 14& 
\begin{array}{>{\centering\arraybackslash$} p{2cm} <{$}} \bcv \mbox{Adv1} \\ \bcd \mbox{Adv2} \end{array} &  
\begin{array}{|>{\centering\arraybackslash$} p{2cm} <{$}} \bcv 148.63 \\ 134.63 \bcd  \end{array} &
\begin{array}{>{\centering\arraybackslash$} p{2.9cm} <{$}} \bcv  2.14\\ 1.94\bcd  \end{array} &
\begin{array}{>{\centering\arraybackslash$} p{2.9cm} <{$}} \bcv 6.38 \\ 6.25 \bcd  \end{array} &
\begin{array}{|>{\centering\arraybackslash$} p{2cm} <{$}} \bcv 28.50 \\ 26.88 \bcd  \end{array} &
\begin{array}{>{\centering\arraybackslash$} p{2.9cm} <{$}} 1.41\bcv  \\ 1.33\bcd  \end{array} &
\begin{array}{>{\centering\arraybackslash$} p{2.9cm} <{$}} \bcv 2.38 \\ 2 \bcd  \end{array}  \\ 
N = 16& 
\begin{array}{>{\centering\arraybackslash$} p{2cm} <{$}} \rcv \mbox{Adv1} \\ \rcd \mbox{Adv2} \end{array} &  
\begin{array}{|>{\centering\arraybackslash$} p{2cm} <{$}} \rcv 165.00 \\ 152.86\rcd  \end{array} &
\begin{array}{>{\centering\arraybackslash$} p{2.9cm} <{$}} \rcv  2.08 \\1.92 \rcd  \end{array} &
\begin{array}{>{\centering\arraybackslash$} p{2.9cm} <{$}} \rcv 6.75 \\ 5.13\rcd  \end{array} &
\begin{array}{|>{\centering\arraybackslash$} p{2cm} <{$}} \rcv 33.50 \\ 31.5\rcd  \end{array} &
\begin{array}{>{\centering\arraybackslash$} p{2.9cm} <{$}} \rcv 1.44 \\ 1.35\rcd  \end{array} &
\begin{array}{>{\centering\arraybackslash$} p{2.9cm} <{$}} \rcv  2.38\\ 2.13\rcd  \end{array}  \\ 
N = 18& 
\begin{array}{>{\centering\arraybackslash$} p{2cm} <{$}} \bcv \mbox{Adv1} \\ \bcd \mbox{Adv2} \end{array} &  
\begin{array}{|>{\centering\arraybackslash$} p{2cm} <{$}} \bcv 183.75 \\ \bcd 174.38 \end{array} &
\begin{array}{>{\centering\arraybackslash$} p{2.9cm} <{$}} \bcv 2.06 \\1.96 \bcd  \end{array} &
\begin{array}{>{\centering\arraybackslash$} p{2.9cm} <{$}} \bcv 6.63 \\ 5.88\bcd  \end{array} &
\begin{array}{|>{\centering\arraybackslash$} p{2cm} <{$}} \bcv 37.38 \\34.63 \bcd  \end{array} &
\begin{array}{>{\centering\arraybackslash$} p{2.9cm} <{$}} \bcv  1.51\\ 1.34\bcd  \end{array} &
\begin{array}{>{\centering\arraybackslash$} p{2.9cm} <{$}} \bcv 2.63 \\ 2.13 \bcd  \end{array} 
\end{array}
$$
\caption{Several details about the D-wave embeddings of our instances of \MPBS, both for the standard and the IQPMS methods. These statistics have been performed for the embeddings generated on Adv1 and Adv2 by averaging the data over all the instances of \MPBS with the same number of arcs. In $\langle$qubits$\rangle$ we report the average number of qubits needed to embed our \MPBS instances. In $\langle$qubits/variable$\rangle$ we report the ratio between $\langle$qubits$\rangle$ and the average number of variables needed to provide a QUBO formuation of our \MPBS instances, which can be obtained from Table~\ref{satcomb}. In $\langle$max. chain length$\rangle$ we report the average maximum chain length encountered in the embeddings.}\label{Dwavestatshard} 
\end{table}

\begin{figure}
\begin{center}
\includegraphics[width=1.01\textwidth]{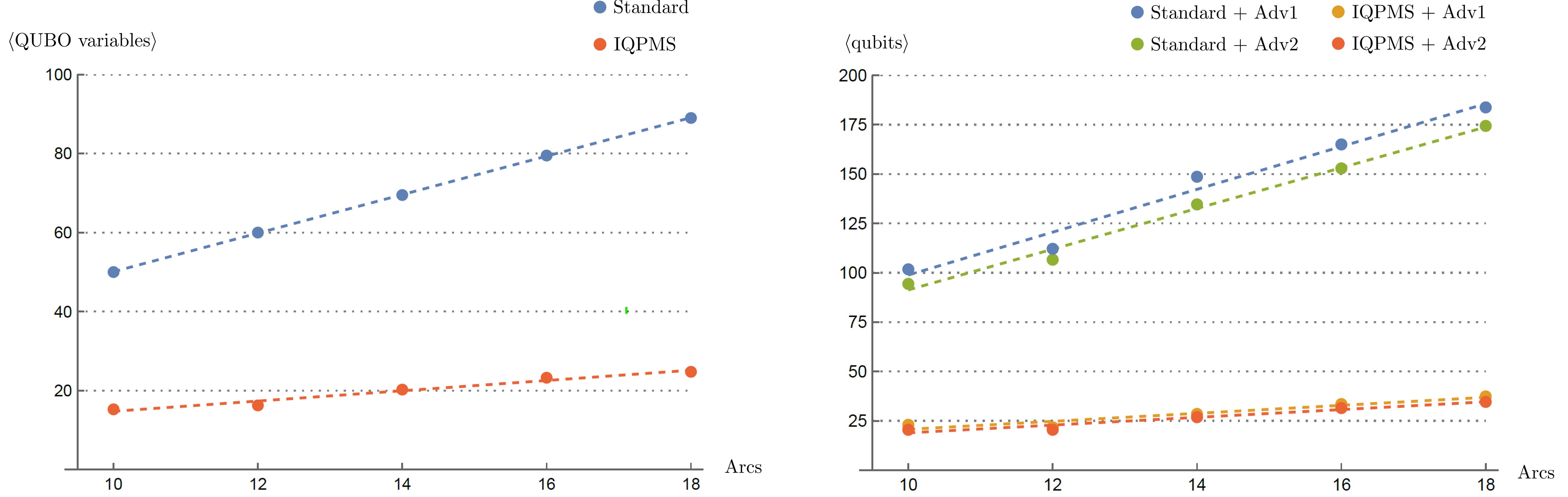}
\caption{(Left) Average number of variables needed by the standard and IQPMS methods to provide a QUBO form of the instances generated for different number of arcs. The averages have been performed over the four instances with the same number of arcs. 
From the linear interpolation of these data (dashed lines), we infer that the IQPMS methods requires in average $m(\mbox{IQPMS})= 1.30$ QUBO variables per arc added to the graph, while the standard method needs $m(\mbox{standard})= 4.88$ QUBO variables.
(Right) Average number of qubits needed to embed the QUBO instances generated with the standard and IQPMS methods into the quantum annealers Adv1 and Adv2 for different number of arcs. The averages have been performed over the four instances with the same number of arcs. 
From the linear interpolation of these data (dashed lines), we infer that the qubits employed per arc added to the \MPBS instance to obtain an embedding on a D-wave annealer when the IQPMS and the standard methods are adopted are $m(\mbox{IQPMS + Adv2})= 1.96$, $m(\mbox{IQPMS + Adv1})= 2.03$, $m(\mbox{standard + Adv2})= 10.31$ and $m(\mbox{standard + Adv1})= 10.84$.
}\label{grafico_statistiche}
\end{center}
\end{figure}

The number of variables needed by the IQPMS methods are much fewer, in general less than a third. Moreover, the ratio between the variables needed with the standard and the IQPMS methods increases with $N(u)$. One could object that these results are highly influenced by our arbitrary choice of $[FL(u),CAP(u)]=[-7,8]$. This is in true, for better or for worse. Indeed, this choice caused the standard method to need 4 slack variables per node, as described in Section~\ref{capfloorslack}. Nonetheless, as discussed above, in realistic scenarios $CAP(u)-FL(u)$ could easily assume values in the order of $10^5$ and therefore the standard method would need more than 15 slack variables per node only to implement CAP/FLOOR. Notice that our methods do not depend on the particular values of $FL(u)$ and $CAP(u)$, as shown in Table~\ref{TABELLONE}, and therefore its performances would not change in realistic scenarios. Hence, our pick tried to find a compromise between the possibility to obtain QUBO formulations of \MPBS with the standard method that were solvable by D-wave, namely $CAP(u)-FL(u)$ not too large, and the idea of being somehow faithful with the possible scenarios encountered in realistic scenarios, namely $CAP(u)-FL(u)$ not too small. Nonetheless, even if we would pick smaller interval $[FL(u),CAP(u)]$, and therefore requiring even less slack variables for the standard QUBO implementation of CAP/FLOOR, we remind that the IQPMS methods always require zero slack variables for this constraint when we have $N(u)\leq 4$ arcs. Moreover, the IQPMS methods need anyway less slack variables to implement the master constraint IN/OUT (compare Tables~\ref{TABELLONEIO} and \ref{TABELLONE}). 

We followed by submitting these QUBO problems to the D-wave quantum annealers Advantage\_system4.1 and  Advantage2\_prototype1.1, where the latter is the latest quantum annealer prototype having enhanced performance. 
In the following, we simply call these two annealers Adv1 and Adv2, respectively. 
For each one of the 20 instances of \MPBS described above, we performed 4k annealing runs with Adv1 and 4k annealing runs with Adv2. These runs were performed both for the QUBO forms obtained with the standard and the IQPMS methods.
Moreover, the 4k runs dedicated to each instance, annealer and method were subdivided in 16 blocks of 250 runs. 
The motivation of this choice is the following. Every time we ask the D-wave online services to perform a given number of runs, they generate a particular embedding that maps our QUBO problem into an initialization of their quantum chip and all the runs are then performed with that particular embedding.
Hence, our motivation to split the runs in 16 different blocks was to average out the possible advantages/disadvantages coming from particularly fortunate/unfortunate embeddings generated by D-wave.
As a result, we performed a total 320k runs on D-wave quantum annealers. The annealing times have not been tuned and therefore all the runs had the same default duration of $20\mu s$.
\begin{table} 
$$
\hspace{-0.5cm} \begin{array}{>{\centering\arraybackslash$} p{1.1cm} <{$}>{\centering\arraybackslash$} p{1.6cm} <{$}>{\centering\arraybackslash$} p{2.5cm} <{$}>{\centering\arraybackslash$} p{2.5cm} <{$}>{\centering\arraybackslash$} p{2.5cm} <{$}>{\centering\arraybackslash$} p{2.5cm} <{$}>{\centering\arraybackslash$} p{1.6cm} <{$}>{\centering\arraybackslash$} p{1.6cm} <{$}}
\mbox{Arcs} &
\begin{array}{>{\centering\arraybackslash$} p{2cm} <{$}} \mbox{Annealer} \end{array} &  
\begin{array}{>{\centering\arraybackslash$} p{2.9cm} <{$}} \mbox{Successes} \\ \mbox{(standard)} \end{array} &
\begin{array}{>{\centering\arraybackslash$} p{2.9cm} <{$}} \mbox{Successes 95\%} \\ \mbox{(standard)} \end{array} &
\begin{array}{>{\centering\arraybackslash$} p{2.9cm} <{$}} \mbox{Successes} \\ \mbox{(IQPMS)} \end{array} &  
\begin{array}{>{\centering\arraybackslash$} p{2.9cm} <{$}} \mbox{Successes 95\%} \\ \mbox{(IQPMS)} \end{array} & 
\begin{array}{>{\centering\arraybackslash$} p{2cm} <{$}} \mbox{Gain} \end{array} & 
\begin{array}{>{\centering\arraybackslash$} p{2cm} <{$}} \mbox{Gain 95\%} \end{array} \\  \\ 
N = 10 & 
\begin{array}{>{\centering\arraybackslash$} p{2cm} <{$}} \bcv  \mbox{Adv1}  \\ \bcd  \mbox{Adv2} \end{array} &  
\begin{array}{|>{\centering\arraybackslash$} p{2.9cm} <{$}} \bcv 198 \,\,\,(1.24\%) \\ \bcd 330 \,\,\, (2.20\%) \end{array} &
\begin{array}{>{\centering\arraybackslash$} p{2.9cm} <{$}} \bcv 256 \,\,\,(1.60\%) \\ \bcd 463 \,\,\, (2.89\%) \end{array} &
\begin{array}{|>{\centering\arraybackslash$} p{2.9cm} <{$}} \bcv 1366 \,\,\,(8.54\%) \\ \bcd 3940 \,\,\, (23.61\%) \end{array} &
\begin{array}{>{\centering\arraybackslash$} p{2.9cm} <{$}} \bcv 2054 \,\,\,(12.84\%) \\ \bcd 5152 \,\,\, (32.20\%) \end{array} & 
\begin{array}{>{\centering\arraybackslash$} p{2cm} <{$}} \bcv 6.90  \\ \bcd 10.79  \end{array} & 
\begin{array}{>{\centering\arraybackslash$} p{2cm} <{$}} \bcv 8.02  \\ \bcd 11.13  \end{array} \\
N = 12 & 
\begin{array}{>{\centering\arraybackslash$} p{2cm} <{$}}  \rcv \mbox{Adv1} \\ \rcd \mbox{Adv2} \end{array} &  
\begin{array}{|>{\centering\arraybackslash$} p{2.9cm} <{$}} \rcv 19 \,\,\,(0.12 \%) \\ \rcd 63 \,\,\, (0.39 \%) \end{array} &
\begin{array}{>{\centering\arraybackslash$} p{2.9cm} <{$}} \rcv 36 \,\,\,( 0.23\%) \\ \rcd 85 \,\,\, ( 0.53\%) \end{array} &
\begin{array}{|>{\centering\arraybackslash$} p{2.9cm} <{$}} \rcv 1012 \,\,\,( 6.30\%) \\ 2735\rcd  \,\,\, (17.09 \%) \end{array} &
\begin{array}{>{\centering\arraybackslash$} p{2.9cm} <{$}} \rcv 1149 \,\,\,(7.18\%) \\ 3149\rcd  \,\,\, ( 19.68\%) \end{array} & 
\begin{array}{>{\centering\arraybackslash$} p{2cm} <{$}} \rcv  53.26\\ 43.41\rcd   \end{array} & 
\begin{array}{>{\centering\arraybackslash$} p{2cm} <{$}} \rcv   31.92\\ 37.05\rcd   \end{array} \\ 
N = 14 & 
\begin{array}{>{\centering\arraybackslash$} p{2cm} <{$}} \bcv \mbox{Adv1} \\ \bcd \mbox{Adv2} \end{array} &  
\begin{array}{|>{\centering\arraybackslash$} p{2.9cm} <{$}} 6\bcv  \,\,\,( 0.04\%) \\ 29\bcd  \,\,\, (0.18 \%) \end{array} &
\begin{array}{>{\centering\arraybackslash$} p{2.9cm} <{$}} 24 \bcv  \,\,\,( 0.15\%) \\ 42\bcd  \,\,\, ( 0.26\%) \end{array} &
\begin{array}{|>{\centering\arraybackslash$} p{2.9cm} <{$}} 661\bcv  \,\,\,( 4.13\%) \\ 1765\bcd  \,\,\, (11.03 \%) \end{array} &
\begin{array}{>{\centering\arraybackslash$} p{2.9cm} <{$}} \bcv 869  \,\,\,( 5.43\%) \\ 2284\bcd  \,\,\, (14.28 \%) \end{array} & 
\begin{array}{>{\centering\arraybackslash$} p{2cm} <{$}} 110.17\bcv  \\ \bcd 60.86  \end{array} & 
\begin{array}{>{\centering\arraybackslash$} p{2cm} <{$}} 36.21\bcv   \\ 54.38\bcd   \end{array} \\ 
N =  16& 
\begin{array}{>{\centering\arraybackslash$} p{2cm} <{$}} \rcv \mbox{Adv1} \\ \rcd \mbox{Adv2} \end{array} &  
\begin{array}{|>{\centering\arraybackslash$} p{2.9cm} <{$}} 2\rcv  \,\,\,(0.01 \%) \\ 10\rcd  \,\,\, ( 0.06\%) \end{array} &
\begin{array}{>{\centering\arraybackslash$} p{2.9cm} <{$}} 8\rcv  \,\,\,( 0.05\%) \\ \rcd  23\,\,\, ( 0.14\%) \end{array} &
\begin{array}{|>{\centering\arraybackslash$} p{2.9cm} <{$}} 296\rcv  \,\,\,( 1.85\%) \\ 1149\rcd  \,\,\, ( 7.18\%) \end{array} &
\begin{array}{>{\centering\arraybackslash$} p{2.9cm} <{$}} 659\rcv  \,\,\,( 4.12\%) \\ 2015\rcd  \,\,\, ( 12.59\%) \end{array} & 
\begin{array}{>{\centering\arraybackslash$} p{2cm} <{$}} \rcv 148.00 \\ 114.90\rcd   \end{array} & 
\begin{array}{>{\centering\arraybackslash$} p{2cm} <{$}} \rcv  82.38 \\ \rcd 87.61  \end{array} \\ 
N = 18 & 
\begin{array}{>{\centering\arraybackslash$} p{2cm} <{$}} \bcv \mbox{Adv1} \\ \bcd \mbox{Adv2} \end{array} &  
\begin{array}{|>{\centering\arraybackslash$} p{2.9cm} <{$}} \bcv 2  \,\,\,( 0.01\%) \\ 4\bcd  \,\,\, ( 0.03\%) \end{array} &
\begin{array}{>{\centering\arraybackslash$} p{2.9cm} <{$}} \bcv  5\,\,\,( 0.03\%) \\ \bcd  6\,\,\, ( 0.04\%) \end{array} &
\begin{array}{|>{\centering\arraybackslash$} p{2.9cm} <{$}} \bcv  322\,\,\,( 2.01\%) \\ \bcd 735 \,\,\, (4.59 \%) \end{array} &
\begin{array}{>{\centering\arraybackslash$} p{2.9cm} <{$}} \bcv  454\,\,\,(2.84 \%) \\ \bcd 934 \,\,\, (5.84 \%) \end{array} & 
\begin{array}{>{\centering\arraybackslash$} p{2cm} <{$}} \bcv 161.00 \\ \bcd  183.75 \end{array} & 
\begin{array}{>{\centering\arraybackslash$} p{2cm} <{$}} \bcv  90.80 \\ \bcd 155.67  \end{array}
\end{array}
$$
\caption{Number of annealing runs that provided either the optimal solution or an output satisfying the \MPBS constraints and had an objective function within the 95\% of the optimal. 
These data are reported for both annealers, namely Adv1 and Adv2, and the two techniques considered to obtain QUBO realizations of \MPBS, namely the standard and IQPMS methods. 
The number of successes for each given $N$ and annealer corresponds to the sum of the successes obtained among the 4 corresponding instances. Hence, since we performed 4k annealing runs per instance and annealer, the percentages indicated are relative to the total 16k runs performed for each arc-setting and annealer. Finally, the gain columns correspond to the ratios between the successes obtained using the IQPMS methods and the standard method. The same has been done by including those outputs satisfying the \MPBS constraints and had an objective function within the 95\% of the optimal. 
}\label{Dwavestats}
\end{table}

\begin{figure}
\includegraphics[width=0.78\textwidth]{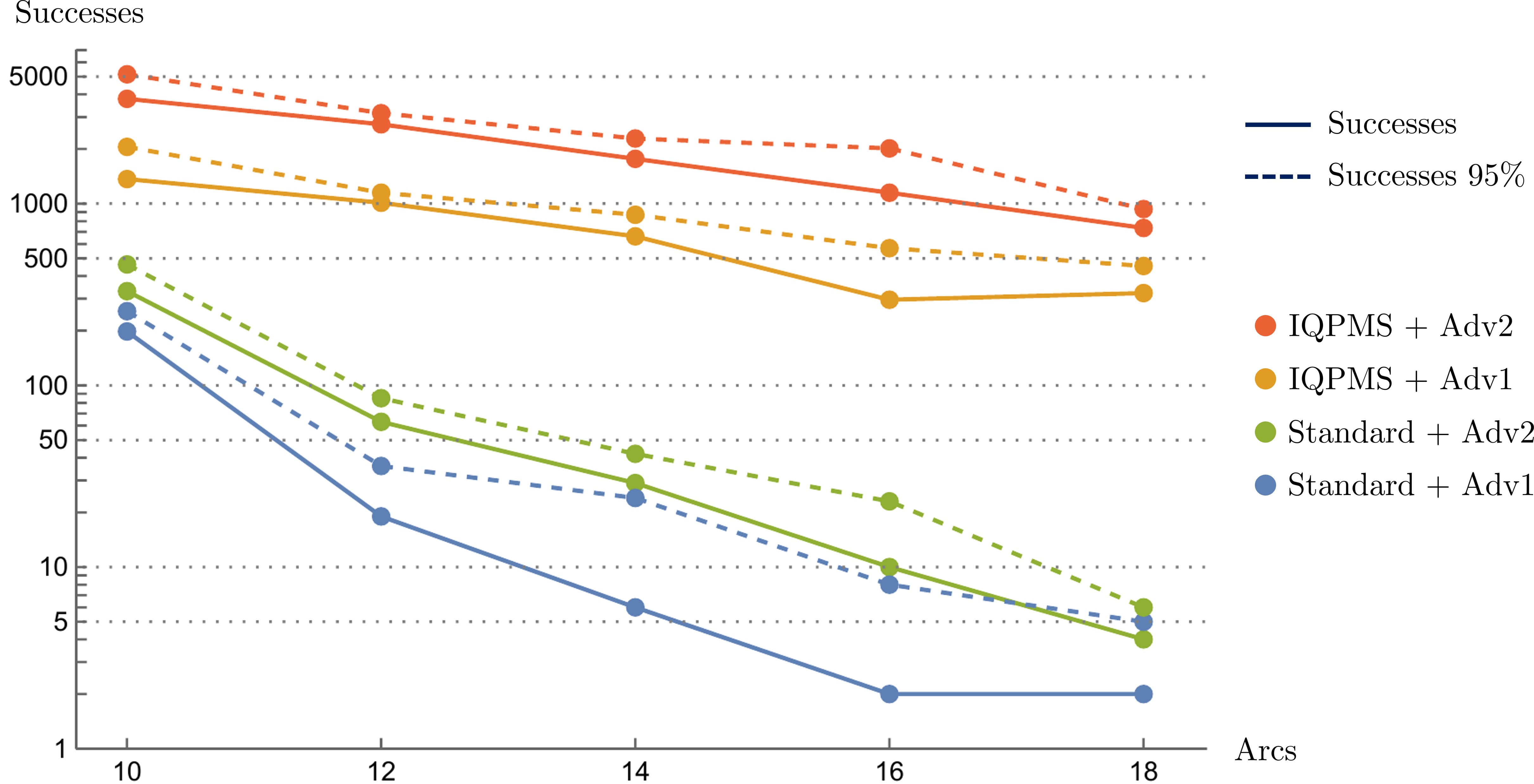}
\caption{Number of runs, in logarithmic scale, where D-wave annealers found the optimal solutions of \MPBS problems with different number of arcs after 16k runs, where their QUBO formulations were either obtained with the standard or the IQPMS methods (dots connected by solid lines). We also show the results obtained if we consider those outputs satisfying the \MPBS constraints and having objective function within the 95\% of the optimal (dots connected by dashed lines). These analyses were performed with both annealers Adv1 and Adv2. }\label{grafico_successi}
\end{figure} 

We show the average number of qubits needed for these embeddings depending on the method and the annealer adopted, for different number of arcs in Figure~\ref{grafico_statistiche} (right). We notice that, when considering the standard method, D-wave embeddings need approximately 10.5 qubits per arc introduced into the graph. Instead, when considering our method, we only have approximately 2 qubits per arc introduced into the graph.
 Several details regarding the embeddings of these problems can be found in Table~\ref{Dwavestatshard}. In particular, depending on the number of arcs in our instance, annealer used and method adopted, we report the average number of qubits employed, the average number of qubits employed per QUBO variable of the instance and the corresponding average maximum chain lengths. The maximum chain length of a given embedding is the maximum number of qubits needed by D-wave to embed a single QUBO variable into the quantum chip. Indeed, since D-wave annealers have limited connectivity, which are 15 for Adv1 and 20 for Adv2, it often happens that more than one qubit is needed to embed a single QUBO variable. Naturally, since the IQPMS methods require far fewer slack variables per node, if compared with the standard method, each binary variable is connected with fewer other variables and therefore our methods allow a much more efficient consumption of qubits. This is one of the main goals achieved by this work, namely providing an encouraging scaling for the qubits deployed by the embedding of locally-constrained QUBO problems into D-wave quantum annealers.

We recorded the number of successes obtained by Adv1 and Adv2 in the individuation of the \MPBS optimal solutions. Moreover, we performed the same analysis by counting how many times the D-wave annealers were able to provide outputs $\mathbf x$ that satisfied the \MPBS constraints and gave a value of the objective function that was at least the $95\%$ of the optimal value $W^* = \sum_i w_i x_i^*$, where $\mathbf x^*$ is the optimal solution, which are referred to as the ``95\% solutions''. We summarize these results in Table~\ref{Dwavestats}, where we also include the ratio between the number of successes obtained with the IQPMS and the standard methods. These gains, both for the optimal and the 95\% solutions, result to be monotonically increasing with the number of arcs in the \MPBS instances. Finally, in Figure~\ref{grafico_successi} we plot the number of successes obtained in the described scenarios.

\section{Discussion}\label{SecDiscussion}
We provided a novel set of techniques to translate optimization problems into a QUBO form. In particular, we showed their potential to drastically reduce the number of slack variables and to increase the sparsity of the generated QUBO problems. The first of the two techniques, the IQP method, is particularly useful whenever the constraints to be enforced are defined over few logical variables and/or the values of the parameters inside inequality constraints are large. This method does not necessary require the locally-constrained graph problem structure. Nonetheless, in this case we can easily adopt the MS method, which provides the simultaneous enforcement of multiple constraints  defined over the same set of binary variables. Moreover, we underline that our method has the same efficiency whether the  constraints considered are equality/inequality linear/non-linear constraints defined with integer/non-integer parameters. Finally, we showed how to tune the corresponding constraint multipliers and we gave a generalization of the MS method for those cases where at least three constraints share the same set of variables.

We followed by considering an NP-hard problem from finance, namely \MPBS, and we showed how to apply our methods to this real-world optimization problem. Noteworthy, we obtained  closed forms for the implementations of its inequality constraints INOUT($u$) and CAPFLOOR($u$) for all the nodes with $N(u)\leq 5$ arcs, with the only exception of CAPFLOOR($u$) when $N(u)= 5$, which we evaluated case by case. Hence, all the constraints for which we achieved a closed form will not need any additional calculation in order to find the corresponding QUBO implementation: the only required operation is the identifications of the node binary variables combinations that violate/satisfy the constraints.

We generated several instances of this problem for different numbers of nodes and edges involved. First, we quantified how better our methods work in terms of slack variables employed and later we showed how these improvements provide a much higher output quality when quantum annealers are the chosen QUBO solver. 
Indeed, we considered two D-wave quantum annealers, namely Advantage\_system4.1 and Advantage2\_prototype1.1, comparing our methods with a standard procedure that provides QUBO forms. In addition, this study provides a performance comparative study between the considered annealers.  

We believe that our approach is particularly useful in the context of quantum computation as all the pre-processing applied is devoted to make the resulting QUBO as light as possible and we do not require additional rounds of computation on the QUBO solver. Indeed, quantum annealers in general provide higher-quality outputs when problems are defined in a more compact form, namely when fewer and less-connected qubits are necessary. Hence, focusing on the refinement of QUBO formulations is clearly one key ingredient to obtain better near-term quantum computation results. Classical computing strategies that aim to obtain solutions for these kind of combinatorial problems within shorter times, usually undertake completely different paths~\cite{power}, avoiding QUBO formulations.

There are several research directions that could be explored starting from our results. First of all, the adoption of IQPMS to solve other combinatorial problems. Possible candidates are the \algo{MultipleKnapsack}, the \algo{Multiple-ConstrainedKnapsack} and the \algo{MultipleSubsetSum} problems. Moreover, given the high performances obtainable in case of locally-constrained optimization problems, its employment could easily fit for real-world combinatorial problems defined over sparse networks (see, e.g.,~\cite{powernetwork}).

Another interesting direction would be given by comparing our methods with those from Ref.~\cite{Djidjev}. Both output quality from quantum annealing and running times could be compared in order to understand under which regimes one could outperform the other. Indeed, while our methods may employ more slack variables, a fair comparison would allow the IQPMS methods to consider much longer quantum computation times. Moreover, it is not clear whether the number of rounds necessary to the iterative strategies are more sensitive to an increase of the input size or the number of constraints. If the latter would be the case, given realistic instances of \MPBS are highly constrained, namely given by sparse graphs with node-by-node constraints, we believe that our methods should provide higher success rates. 
A similar comparison with the techniques presented in Ref.~\cite{powernetwork} would be fruitful as well.
In summary, we expect that, if we compare the already available strategies with our novel methods in the context of solving locally-constrained lowly-connected optimization problems, we should obtain optimal solutions in shorter quantum computation times.

Future applications of the IQPMS methods could be employed with more sophisticated techniques, e.g., the tuning of the annealing times, the adoption of D-wave classical-quantum hybrid QUBO solvers or particular pre/post-processing procedures, e.g., the variable reduction proposed in Ref.~\cite{samplepersistence}, where an iterative algorithm fixes several binary variables to values that have a high probability of being optimal.
Another path is given by the use of different NISQ devices, such as in the context of the Quantum Approximate Optimization Algorithm~\cite{QAOA} or the Quantum Alternating Operator Ansatz~\cite{QAOA2}, which are designed for logical-gate quantum computers. We remind that the aim of this work is to compare in the fairest way our methods with a well-established standard procedure. Hence, the setup considered has not the intent to prove how far our algorithms can be pushed to achieve optimal solutions via quantum annealing for larger and larger instances, but it rather focuses on giving a clear motivation of when and why we should adopt our novelties. Therefore, we are curious to discover up to which input size our methods, in conjunction with other techniques, allow to provide optimal solutions of \algo{MPBS} instances defined via realistic datasets.

\section*{Acknowledgements}
We thank Francesco Gullo and Ilaria Bordino for useful discussions.
Dario De Santis acknowledges
support from the research project ``Dynamics and Information Research Institute - Quantum Information,
Quantum Technologies” within the agreement between
UniCredit Bank and Scuola Normale Superiore di Pisa
(CI14\_UNICREDIT\_MARMI).
Salvatore Tirone acknowledges support from projects PRIN 2017 Taming complexity via Quantum Strategies: a Hybrid Integrated Photonic approach (QUSHIP) Id. 2017SRN-BRK and PRO3 Quantum Pathfinder.
Vittorio Giovannetti acknowledges financial
support by MUR (Ministero dell’Istruzione, dell’Università e
della Ricerca) through the following projects: PNRR MUR
project PE0000023-NQSTI.

\bibliography{bib}

\appendix

\section{The MS method without the IQP method}\label{MSstandard}

We showed how to apply the MS method when constraints are enforced with the IQP method in Section \ref{GQPsatellite}. Despite their combination, namely the IQPMS method, provides various advantages, the MS method can be adopted independently from the IQP method. Here, we show how to employ this method when constraints are enforced in a QUBO form with the standard techniques described in Section \ref{QUBOintro}. This approach could be  useful when the constraints considered are defined over a number of variables that is too large to be handled by the IQP method.

Consider an optimization problem of the generic form (\ref{optprob}). Suppose that, among all the equality and inequality constraints $EC_i$ and $IC_j$, two of them are defined over the same three binary variables. For instance, consider the case where one equality and one inequality constraints are of the form:
\begin{eqnarray}
EC(x_1,x_2,x_3) = x_1 + x_2 +x_3 - 1  \, , \nonumber \\ 
IC(x_1,x_2,x_3) = - 2 x_1 - 2 x_2 +  x_3 + 2 \, ,\nonumber 
\end{eqnarray}
where it is required $EC(x_1,x_2,x_3) = 0$ and $IC(x_1,x_2,x_3)\leq 0$. Our goal is to achieve quadratic forms of $x_1$, $x_2$ and $x_3$ that provides penalties when these constraints are violated. In the context of the MS approach, we are going to construct a penalty function corresponding to the master constraints, which provides a penalty iff it is violated, otherwise it is zero-valued. Instead, the penalty function corresponding to the satellite constraint provides penalties when  $(x_1,x_2,x_3)$ satisfy the master and violates the satellite constraint and is zero-valued when both constraints are satisfied. Hence, we do not impose any requirement to the satellite penalty function when the master constraint is violated.

We pick $EC(x_1,x_2,x_3) = 0$  as master   and  $IC(x_1,x_2,x_3)  \leq  0$ as satellite constraint. Hence,  in order to have a simultaneous reformulation of these constraints within the MS formalism, we start by enforcing $EC(x_1,x_2,x_3) = 0$ with the standard method (see Section \ref{QUBOintro}), hence with the quadratic penalty function $
P^{EC}(x_1,x_2,x_3) =- ( x_1 + x_2 +x_3 - 1)^2 $.

The quadratic penalty function relative to $IC(x_1,x_2,x_3)\leq 0$ has not to be enforced by considering all the possible values of $(x_1, x_2, x_3)$, but solely those combinations already satisfying  $EC(x_1,x_2,x_3) =0$, namely $(x_1, x_2, x_3) = (1,0,0), (0,1,0), (0,0,1)$. Notice that $(x_1, x_2, x_3) = (1,0,0), (0,1,0)$ satisfy $IC(x_1,x_2,x_3) \leq 0$, while $(x_1, x_2, x_3) =(0,0,1)$ leads to a violation. Hence, in order to construct the quantity $S(\mathbf s)$ needed to enforce our inequality/satellite constraint (see Eq. (\ref{S})), we solely have to worry to build a quadratic form $P^{IC}(x_1,x_2,x_3,\mathbf s) = - ( IC(x_1,x_2,x_3) + S(\mathbf s))^2$  such that $\max_{\mathbf s} P^{IC}(x_1,x_2,x_3,\mathbf s) = 0$ for $(x_1, x_2, x_3) = (1,0,0), (0,1,0)$ and $\max_{\mathbf s} P^{IC}(0,0,1,\mathbf s) \leq - 1$. 

If $P^{IC}(x_1,x_2,x_3,\mathbf s)$ would be enforced outside the MS formalism, namely as in Section \ref{QUBOintro}, we would require  $S(\mathbf s)$ to assume all the integer values in the interval $[0,|\min_{\mathbf x} IC(\mathbf x)|]= [0,2]$, where the minimization is performed over all $\mathbf x \in\{ 0,1\}^3$. Hence, in the standard case we would obtain $S(\mathbf s) = s_0 + s_1$: two slack variables would be required. Now, since $IC(x_1,x_2,x_3)\leq 0$ has to be enforced as a satellite of $EC(x_1,x_2,x_3)\leq 0$, we have to ignore those strings violating the master constraint and therefore the same minimization is now performed solely over $(x_1, x_2, x_3) = (1,0,0), (0,1,0), (0,0,1)$, which leads to:
$$
\left|\min_{\mathbf x = \{0,1\}^3 } IC(\mathbf x) \right| = 2 \,\,\,\,\,\, \stackrel{\mbox{MS}}{\longrightarrow} \,\,\,\,\,\, \left|\min_{\mathbf x = (1,0,0), (0,1,0), (0,0,1)} IC(\mathbf x) \right| =   0 .
$$
Hence, by using the MS method, $IC(x_1,x_2,x_3) \leq 0$ can be enforced with a quadratic form without employing any slack variable via $P^{IC}(x_1,x_2,x_3) = - ( - 2 x_1 - 2 x_2 +  x_3 + 2 ) ^2$, which is equal to $0$ iff $\mathbf x$ satisfies the master and the satellite constraints, namely for $(x_1, x_2, x_3) = (1,0,0), (0,1,0)$ and is smaller or equal to $-1$ otherwise. Hence, in this case, there are no accidental incentives and the relative multiplier of the master constraint can be fixed to $\lambda^{EC}=1$ (see Section \ref{tunepenalty}). Notice that $P^{IC}(x_1,x_2,x_3)$ penalises also those combinations of $(x_1,x_2,x_3)$ satisfying the satellite and violating the master constraint. Hence, the sum of the two penalty functions:
$$
P^{EC}(x_1,x_2,x_3)+P^{IC}(x_1,x_2,x_3) = - ( x_1 + x_2 +x_3 - 1)^2  - ( - 2 x_1 - 2 x_2 +  x_3 + 2 ) ^2 \, ,
$$
is equal to zero iff $(x_1,x_2,x_3)$ satisfies both constraints and is smaller or equal to -1 iff one or both constraints are violated, where no slack variables have been employed.

\section{IQPMS solutions for CAP/FLOOR($u$) when $N(u)=4$}\label{FORMULOZZI}
The solution of the linear system~(\ref{SISTEMONE4}) is:
\begin{eqnarray}\nonumber
&P_u^{CF}(\mathbf x) =&(\sigma(1,0,0,1) + \sigma(1,0,1,0) - \sigma(1,0,1,1) + \sigma(1,1,0,0) - \sigma(1,1,0,1) - \sigma(1,1,1,0) + \sigma(1,1,1,1)) x_1 \\ \nonumber && 
+( 2 \sigma(1,0,1,1)  -\sigma(1,0,0,1) - \sigma(1,0,1,0)  + \sigma(1,1,0,1) + \sigma(1,1,1,0) - 2 \sigma(1,1,1,1) )x_2    \\ \nonumber && + 
(1+ \sigma(1,0,1,0) - \sigma(1,0,1,1) - \sigma(1,1,1,0)+ \sigma(1,1,1,1)) x_3 \\ \nonumber && + (-\sigma(1,0,1,0) + \sigma(1,0,1,1) - \sigma(1,1,0,0) + \sigma(1,1,0,1) + 2 \sigma(1,1,1,0)  -  2 \sigma(1,1,1,1))x_4 \\ \nonumber && +   (-\sigma(1,0,1,1) + \sigma(1,1,1,1)) x_1x_2 \\ \nonumber && + (1- \sigma(1,0,0,1)- \sigma(1,0,1,0)+ 2 \sigma(1,0,1,1)- \sigma(1,1,0,0) + \sigma(1,1,0,1) + 2 \sigma(1,1,1,0)- 2 \sigma(1,1,1,1)) x_1x_3 \\ \nonumber && +   (-\sigma(1,1,1,0) + \sigma(1,1,1,1))x_1x_4 + (\sigma(1,0,0,1) - \sigma(1,0,1,1)  - \sigma(1,1,0,1) + \sigma(1,1,1,1)) x_2 x_3 \\ \nonumber && +  (\sigma(1,0,1,0) - \sigma(1,0,1,1) - \sigma(1,1,1,0) +\,  \sigma(1,1,1,1)) x_2 x_4 \\  && +  (\sigma(1,1,0,0)   - \sigma(1,1,0,1) - \sigma(1,1,1,0) + \sigma(1,1,1,1))x_3 x_4 \label{formulozza1} \, .
\end{eqnarray}
The solution of the linear system~(\ref{SISTEMONE42}) is:
\begin{eqnarray} \nonumber
&P_u^{CF}(\mathbf x) =& (1 - \sigma(0,1,0,1) - \sigma(0,1,1,0) + 2 \sigma(0,1,1,1) + \sigma(1,0,1,1) + \sigma(1,1,0,1)+ \sigma(1,1,1,0) - 2 \sigma(1,1,1,1)) x_1 \\ \nonumber && + 
(-1 - \sigma(0,1,1,1)- \sigma(1,0,1,1) + \sigma(1,1,1,1)) x_1 x_2  \\ \nonumber && + 
(\sigma(0,1,0,1)- \sigma(0,1,1,1)- \sigma(1,1,0,1)+ \sigma(1,1,1,1)) x_1 x_3  \\ \nonumber && + 
(\sigma(0,1,1,0) - \sigma(0,1,1,1)- \sigma(1,1,1,0) + \sigma(1,1,1,1)) x_1 x_4  \\ \nonumber && + 
( 1 + \sigma(0,1,1,1)- \sigma(1,0,0,1)- \sigma(1,0,1,0) + 2 \sigma(1,0,1,1)+ \sigma(1,1,0,1)+ \sigma(1,1,1,0) - 2 \sigma(1,1,1,1)) x_2  \\ \nonumber && +
 ( \sigma(1,0,0,1)- \sigma(1,0,1,1)- \sigma(1,1,0,1)+ \sigma(1,1,1,1)) x_2 x_3  \\ \nonumber && + 
(\sigma(1,0,1,0) - \sigma(1,0,1,1)- \sigma(1,1,1,0) + \sigma(1,1,1,1)) x_2 x_4  \\ \nonumber && +
 (-1 + \sigma(0,1,1,0) -  \sigma(0,1,1,1)+ \sigma(1,0,1,0) - \sigma(1,0,1,1)- \sigma(1,1,1,0) + \sigma(1,1,1,1)) x_3  \\ \nonumber && + 
(1 - \sigma(0,1,0,1)- \sigma(0,1,1,0) + 2 \sigma(0,1,1,1)- \sigma(1,0,0,1)- \sigma(1,0,1,0) + 2 \sigma(1,0,1,1) \\ \nonumber && + \sigma(1,1,0,1)+ \sigma(1,1,1,0) - 2 \sigma(1,1,1,1)) x_3 x_4  \\  \label{formulozza2} && +
(-1 + \sigma(0,1,0,1)- \sigma(0,1,1,1)+ \sigma(1,0,0,1)- \sigma(1,0,1,1)- \sigma(1,1,0,1)+ \sigma(1,1,1,1)) x_4 \, .
\end{eqnarray}

\section{Worst-case scenarios for \MPBS penalty multipliers}\label{mpbsmulti}

{\bf Local tuning:} 
 Consider the worst-case scenario where the optimal solution (see Eq.~(\ref{solution})) $\mathbf{x}^*$ is unique and for the node $u$ provides the minimum contribution to the term $\sum_{i\in \mathcal{E}(u)} w_i x_i$, while there is a different combination $\mathbf{x}_{worst}$ that violates  CAP/FLOOR($u$) and provides the maximum contribution to $\sum_{i\in \mathcal{E}(u)} w_i x_i$. Hence, we would have that the binary variables of $\mathbf x^*$ associated to the transactions involving $u$ are $\{x_1^*, \dots,x_{N(u)}^*\}=\{0,\dots,0\}$ and $\{x_{worst,1}, \dots,x_{worst,{N(u)}}\}=\{1,\dots,1\}$. We choose $\lambda_u$ such that $Q_u(\mathcal G, \mathbf x,\mathbf s)$ is larger for $\mathbf x= \mathbf x^*$  than for $\mathbf x= \mathbf x_{worst}$:
\begin{equation}\label{local1}
 \sum_{i\in \mathcal{E}(u)} w_i x_{worst,i} +\lambda_u P^{CF}_u(\mathbf x_{worst},\mathbf s) \leq \sum_{i\in \mathcal{E}(u)} w_i -\lambda_u    \stackrel{!}{<} \sum_{i\in \mathcal{E}(u)} w_i x_i^* + \lambda_u P^{CF}_u(\mathbf x^*,\mathbf s^*)\leq  0 \, , 
\end{equation}
where we used $P^{CF}_u(\mathbf x_{worst},\mathbf s)\leq -1$ because $\{x_{worst,1}, \dots,x_{worst,N(u)}\}=\{1,\dots,1\}$ satisfies the node constraints, 
$\{x_{1}^*, \dots,x_{N(u)}^*\}=\{0,\dots,0\}$ and the fact that $P^{CF}_u(\mathbf x^*,\mathbf s^*)\leq 0$ because $\{x_{1}^*, \dots,x_{N}^*\}$ satisfies CAP/FLOOR($u$) and IN/OUT($u$). The exclamation mark above the inequality symbol underlines that, if satisfied, $\lambda_u$ can be considered large enough in this local setting. 

From Eq.~(\ref{local1}), we derive the lower-bound $\lambda_u> w(u)$, where $w(u)=\sum_{i\in \mathcal{E}(u)} w_i$ is the total amount of the transactions involved with the node $u$. Notice that this is the same result obtained in Eq.~(\ref{lambdalocal}).
Hence, we fix the notation
\begin{equation}\label{lambdaCF}
\lambda_u^{local}=\gamma \, w(u) \hspace{0.4cm} \mbox{ where } \,\, \gamma>1 \, .
\end{equation}

An estimate for $\lambda^{IO}_u$ can be obtained similarly. The worst-case scenario is obtained when the restriction of $\mathbf{x}^*$ on $u$ is $\{0,\dots,0\}$, namely providing a null contribution to $\sum_i w_i x_i$, while $\mathbf{x}_{worst}$ is the combination providing the largest contribution to $\sum_i w_i x_i$ among those violating IN/OUT($u$) and $P^{CF}(\mathbf x_{worst},\mathbf s)$ provides the largest accidental incentive for $\mathbf{x}_{worst}$, namely $P^{CF}_u(\mathbf x_{worst},\mathbf s)= \max_{\mathbf x,\mathbf s} P^{CF}_u(\mathbf x ,\mathbf s) = P^{CF}_{max,u} \geq 0$.
 We choose $\lambda^{IO}_u$ such that the term $\max_{\mathbf{s}} Q_u(\mathcal{G},\mathbf{x},\mathbf{s})$ is larger for $\mathbf x= \mathbf x^*$   than for $\mathbf x=\mathbf x_{worst}$:
\begin{eqnarray}\nonumber
\max_{\mathbf{s}} Q_u(\mathcal{G},\mathbf{x}_{worst},\mathbf{s})&=& \max_{\mathbf{s}} \left( \sum_{i\in \mathcal{E}(u)} w_i x_{worst,i} + \lambda_u \left( P^{CF}_u(\mathbf x_{worst},\mathbf s)+ \lambda^{IO}_u P^{IO}_u(\mathbf x_{worst},\mathbf s) \right) \right) < (1+\gamma\, P^{CF}_{max} (u)) w(u) -\lambda^{IO}(u)  \\ \nonumber
 &  \stackrel{!}{<} & \max_{\mathbf{s}} Q_u(\mathcal{G},\mathbf{x}^*,\mathbf{s}) =\max_{\mathbf{s}} \left( \sum_{i\in \mathcal{E}(u)} w_i x_i^* + \lambda_u \left( P^{CF}_u(\mathbf x^*,\mathbf s)+ \lambda^{IO}_u P^{IO}_u(\mathbf x^*,\mathbf s) \right)\right)= 0 \, .
\end{eqnarray}
The first inequality is justified because $\{1,\dots,1\}$ satisfies IN/OUT($u$) and therefore $\sum_{i\in\mathcal{E}(u)} w_i x_{worst,i} < w(u)$, $\lambda_u=\gamma\,w(u)$,  $P^{CF}_u(\mathbf x_{worst},\mathbf s) = P^{CF}_{max,u}$ and  $P^{IO}_u(\mathbf x,\mathbf s)\leq -1$ for combinations violating IN/OUT($u$). Moreover, we used $\sum_i w_i x_i^*=0$ and the fact that $\mathbf x^*$ satisfies the node constraints. A consequence of this first inequality is that we are imposing a looser condition than previously declared \footnote{A tighter version of $\lambda^{IO}(u)$ could be formulated, but we do not see any particular advantage in performing the extra computation required.}.
Hence, we have the lower-bound $
\lambda^{IO}_u> (1+\lambda\, P^{CF}_{max,u}) / \gamma \, .
$
Since we assumed $ P^{CF}_{max,u} \geq 0$, for the general case we can consider:
\begin{equation}\label{lambdaIO}
\lambda^{IO}_u= 1+ \gamma\max\left\{0,\max_{\mathbf x,\mathbf s}  P^{CF}_{u} (\mathbf x,\mathbf s)  \right\}  \, , \,\,
\mbox{ where $\gamma >1$.}
\end{equation}

{\bf Non-local tuning:} We set some lower-bounds~(\ref{lambdaCF}) and~(\ref{lambdaIO}) for, respectively, $\lambda_u$ and $\lambda^{IO}_u$ by imposing that the sub-QUBO that corresponds to the node $u$ assumes the largest value when $\mathbf x$ satisfies CAP/FLOOR($u$) and IN/OUT($u$). We called it local because we are not considering the possible consequences that a violation of CAP/FLOOR($u$) and/or IN/OUT($u$) may have on $\max_{\mathbf{s}} Q_v(\mathcal{G},\mathbf x,\mathbf s)$ for some node $v\neq u$. Indeed, the possible incentives that these situation may generate could require to set higher values of $\lambda_u$ and $\lambda^{IO}_u$.

Imagine the following unfortunate scenario: $\mathbf x^*$, in order to respect CAP/FLOOR($u$) and IN/OUT($u$), has many $0$s for the variables $x_i$ corresponding to the transactions in $u$. As a consequence, the transactions of the nodes $v$ neighbouring $u$ are not activated too (otherwise CAP/FLOOR($v$) or IN/OUT($v$) may be violated), resulting in many other $x_i=0$ for the arcs belonging to the nodes $v$ sharing at least one arc with $u$. Hence, it may happen that a violation of CAP/FLOOR($u$) and/or IN/OUT($u$) in $u$ allows the activation of many other transactions for the nodes neighbouring $u$, where instead there are no violations of CAP/FLOOR($v$) and IN/OUT($v$). Therefore, there may exists a string that receives a penalty for the violation of CAP/FLOOR($u$) and/or IN/OUT($u$) in $u$, 
but this penalty is overcompensated by the incentives activated by the neighbouring nodes (without penalties).
We define the nodes adjacent to $u$ as follows:
\begin{equation}\label{neigh}
\mbox{neigh}(u)=\{ v \in \mathcal{V} \,|\, \exists (u,v) \lor (v,u)\in \mathcal{E}\} \, .
\end{equation}
Therefore,  to prevent that $ \arg (\max_{\mathbf x,\mathbf s} Q(\mathcal G,\mathbf x,\mathbf s))$ violates CAP/FLOOR($u$) and/or IN/OUT($u$) as described, we consider:
\begin{equation}\label{QUBOneigh}
\max_{\mathbf x,\mathbf s} Q(\mathcal G,\mathbf x,\mathbf s) = \max_{\mathbf x,\mathbf s} \sum_i w_i x_i + \gamma  \sum_{u\in\mathcal{V}} w^{neigh}(u)\left(   P^{CF}_u(\mathbf x, \mathbf s)  + \left(1+ \gamma\max\left\{0,\max_{\mathbf x,\mathbf s}  P^{CF}_{u} (\mathbf x,\mathbf s)  \right\}\right) P^{IO}_u(\mathbf x, \mathbf s)\right)  \, ,
\end{equation}
$$
w^{neigh}(u)=\sum_{v\in \mbox{\scriptsize	 neigh}(u)} w(v) \, ,
$$ where $\gamma>1$ and
 is the  the sum of all the transactions involving $u$ and the neighbouring nodes. Notice that the multipliers involved in Eq.~(\ref{QUBOneigh}) are obtained by replacing $w(u)$ with $w^{neigh}(u)$ from the local formulation, namely we have:
\begin{eqnarray} \label{lambdaCFneigh}
\lambda_u= \gamma \, w^{neigh}(u)  \, , \,\,
\mbox{ where $\gamma >1$.}
\end{eqnarray}
 \\ 

{\bf Global tuning:}  In order to get a global tuning, we consider the following choice of $\lambda_u$:
\begin{eqnarray} 
&&\lambda_u= \gamma \sum_i w_i   \, , \,\,
\mbox{ where $\gamma >1$.}
\end{eqnarray}
and $\sum_i w_i$ is the sum of all the transaction defined on $\mathcal{E}$.

\end{document}